\documentclass[12pt,english]{article}

\renewcommand{\Phi}{\phi}

\newcommand{\yz}{v}

\usepackage[latin9]{inputenc}
\usepackage{geometry}
\usepackage[greek,USenglish]{babel}
\usepackage{verbatim}
\usepackage{prettyref}
\usepackage{amsmath}
\usepackage{amssymb}
\usepackage{graphicx}
\usepackage{setspace}
\usepackage{esint}

\usepackage{color}
\definecolor{darkgreen}{rgb}{0,0.5,0}
\definecolor{darkblue}{rgb}{0,0,0.6}
\definecolor{purple}{rgb}{0.4,.2,0.7}
\usepackage[colorlinks=true,citecolor=darkgreen,linkcolor=purple,urlcolor=purple]{hyperref}

\numberwithin{equation}{section}
\numberwithin{figure}{section}
\numberwithin{table}{section}

\def\CN{{\cal N}}

\def\N{N}

\begin{document}

~
\vskip15mm

\begin{center} {{\huge \textsc{Holographic Vitrification}}

\vskip15mm

Dionysios Anninos${^a}$, Tarek Anous$^{b}$, Frederik Denef$^{c,d}$, Lucas Peeters$^{a}$

\vskip5mm

\it{$^a$ Stanford Institute for Theoretical Physics, Stanford University} 

\it{$^b$ Center for Theoretical Physics, Massachusetts Institute of Technology} 

\it{$^c$ Institute for Theoretical Physics, University of Leuven} }

\it{$^d$ Department of Theoretical Physics, Tata Institute of Fundamental Research}
\vskip5mm

\tt{\quad \quad \quad \quad  danninos@stanford.edu, tanous@mit.edu,}
\newline
\tt{denef@physics.harvard.edu, lpeeters@stanford.edu.}

\end{center}

\vskip15mm

\begin{abstract}

We establish the existence of stable and metastable stationary black hole bound states at finite temperature and chemical potentials in global and planar four-dimensional asymptotically anti-de Sitter space. We determine a number of features of their holographic duals and argue they represent structural glasses. We map out their thermodynamic landscape in the probe approximation, and show their relaxation dynamics exhibits logarithmic aging, with aging rates determined by the distribution of barriers. 

\end{abstract}


\pagebreak 
\pagestyle{plain}

\setcounter{tocdepth}{2}

\tableofcontents


\section{Introduction}

Any liquid, when cooled sufficiently fast, turns into a glass: a peculiar state of matter, disordered like a liquid, yet rigid like a solid \cite{glassperspectives,glasslectures,staticreviews,kineticreviews}. Black hole horizons behave like perfect fluids \cite{Damour,Price:1986yy,Iqbal:2008by,Bredberg:2010ky,Bredberg:2011jq} and have been used extensively to holographically model liquid states of matter \cite{Policastro:2001yc,Policastro:2002se,Kovtun:2004de,Son:2007vk,Bhattacharyya:2008jc,Liu:2009dm,Hartnoll:2009sz,Iqbal:2011ae}. It is natural to hypothesize that such holographic liquids will similarly vitrify upon rapid cooling. But what, then, is a holographic glass? This is is the question we want to address here.

\subsection{The glass problem}

In fact, even disregarding interpretations in terms of black holes and holography, the nature of glass and the glass transition remains shrouded in mystery \cite{glassperspectives,glasslectures,staticreviews,kineticreviews}. Over a century of theoretical, experimental and numerical research have led to many new insights, but so far no fully successful grand unified theory of the glass transition has emerged. The central question in the field is what causes the  dynamical arrest that occurs when a supercooled liquid approaches its glass transition temperature, signaled by an explosive growth in relaxation time scales, leading for example to a dramatic increase in shear viscosity. This happens without any discernible change in spatial structure or order compared to the liquid phase. One of the diverging points of view in the field is whether the transition is essentially thermodynamic or kinetic in nature.  

What is known is that in any case, the glass transition is not described by equilibrium statistical mechanics. Glasses are thermodynamically metastable states failing to find their way to true equilibrium on experimental time scales. As a result, their properties depend on the details of their history, in particular on their age, that is to say on how long it has been since they fell out of equilibrium. This often manifests itself in universal ``aging'' behavior of observables. A concrete example of this is the aging of the conductivity $\sigma$ of electron glasses, which when measured at different times decreases as $\sigma(t) - \sigma(t_0) \sim -\log(t/t_0)$ \cite{amirreview,amirshortreview,ovadyahu2000,ovadyahu2010}. Notice that instead of the usual time translation invariance of exponential relaxation (e.g.\ discharge of a capacitor), we now get time \emph{scale} invariance. Non-exponential relaxation laws have been observed for many other amorphous systems, apparently going back at least to Weber, who noticed it in the relaxation of the silk threads he used to hang his magnets \cite{weber1835,amirshortreview}. 

A qualitative picture for how these and other features may come about is that glasses have extremely rugged free energy landscapes, with exponentially many local minima, in which the system gets hopelessly lost in its attempts to find the true global minimum. It has been known for a long time that this picture can be given a precise meaning in mean field models of spin glasses \cite{spinglassesandbeyond}, in which the equations determining local magnetization densities (the so-called TAP equations \cite{TAP1977}) have exponentially many solutions below the spin glass transition temperature. This is in contrast to the analogous mean field equations for the Ising model, which below the critical temperature has just two solutions for the magnetization density, corresponding to the homogeneous spin up/down equilibrium states. The degrees of freedom governed by TAP-like equations are local order parameters, coarser than the fundamental microscopic degrees of freedom, but finer than the global thermodynamic variables. They have definite expectation values in particular, metastable, macroscopically distinguished states, with small thermal fluctuations around these expectation values. This is analogous to classical backgrounds in field or string theory. Indeed string theory likewise exhibits an exponentially large landscape of solutions, the ``vacua'' of the theory. 

Various incarnations of the landscape-based, essentially static thermodynamic approach, adapted to structural glasses\footnote{\label{spinvsstructural} The standard terminology in the glass literature is that spin glass models are models with quenched disorder (Hamiltonians with randomly frozen couplings), while structural glass models start from simple Hamiltonians (e.g.\ Lennard-Jones), the disorder being spontaneously generated. The systems studied in this work are analogous to structural glasses, existing without any quenched disorder.} have been the leading theories of the glass transition for many years \cite{staticreviews}. More recently however, there has been a renewed focus on real space-time dynamical properties, as opposed to static configuration space properties, due to a large extent to the observation, in experiments and especially numerical simulations, of the ubiquitousness and importance of ``dynamical heterogeneities''. As mentioned earlier, one of the striking features of the glass transition is the complete absence of \emph{static} structural changes accompanying it. The new insight is that this is not true for \emph{dynamic} structural changes. Supercooled liquids turn out to have long-lived, localized regions of high cooperative dynamic activity, mixed with localized regions of almost no cooperative motion whatsoever.\footnote{See \href{http://prx.aps.org/supplemental/PRX/v1/i2/e021013}{http://prx.aps.org/supplemental/PRX/v1/i2/e021013} for some neat movies illustrating the phenomenon.} The size, distribution and evolution of these structures changes significantly when approaching the glass transition. These remarkable kinetic features are hard to explain in the essentially static landscape-based theories, and  theories giving a central role to spacetime trajectories have gained some prominence as a result \cite{kineticreviews}. 

Neither of these different classes of theories is fully satisfactory however, and getting the theories in line with observations often requires patching together elements from different approaches, which are not always logically consistent with each other. There is no universal framework analogous to the framework statistical mechanics provides for equilibrium thermodynamics. One of the theoretical obstacles  making this such a challenging problem is the lack of models that retain all the basic features of a structural glass while also remaining analytically tractable. There is at this point nothing like the hydrogen atom or the Ising model for the glass transition. 

\subsection{Challenges for holographic constructions}

A good model for the glass transition ideally should be able to quantitatively capture the strongly coupled physics of a liquid at all temperatures, allowing analytic study of thermodynamic and transport properties as well as real time relaxation dynamics, making the emergence of dynamical arrest and the salient features of glass formation manifest. Holographic models would therefore seem to be the perfect candidates, as they provide exact solutions to thermodynamic questions (at least for certain large $N$ field theories), and have proven to be particularly powerful exactly in modeling liquid phases of matter and their transport properties, at arbitrary temperatures. Moreover, holography provides direct access to the complete thermodynamic state space, including the landscape of macroscopically distinct stable and metastable states, giving the latter a precise meaning. In other words, it provides a precise analog to the TAP equations mentioned above, in the universal form of the bulk gravity field equations. Explicit solutions can be constructed and probed at will, aided by geometric intuition. The relevant local order parameters are manifest, and it becomes possible to directly deduce whether or not a thermodynamic free energy landscape emerges, and if so what its physical consequences are. In addition, holography automatically incorporates spacetime dynamics, no matter if we are near or far from equilibrium. This makes it possible to study within the same local but macroscopic framework also relaxation dynamics, aging, dynamical heterogeneities and other kinetic features, without having to go back to the microscopic details. Thus, it appears holography would be an ideal platform for theoretical work towards a unified theory of the glass transition.

This brings us back to our question: What, then, is a holographic glass? As mentioned before, a black hole or black brane in AdS behaves as a holographic liquid. Famously, the shear viscosity to entropy density ratio $\eta/s$ has the universal value $\hbar/4\pi k_B$ for any thermodynamic state dual to a black brane at finite temperature \cite{Policastro:2001yc,Kovtun:2004de,Son:2007vk,Iqbal:2008by}. 
The entropy density $s$ goes down with temperature, so for holographic liquids described by black branes, the shear viscosity $\eta$ will decrease with decreasing temperature. In contrast, for ordinary liquids such as water, the shear viscosity goes \emph{up} when  temperature goes down,  dramatically so in the supercooled regime, effectively diverging when approaching the glass transition. Thus a simple homogenous black brane will never behave like a supercooled liquid, let alone a glass.

The holographic dual of a glass should therefore break the translation invariant, homogeneous nature of the black brane. It is known that in certain circumstances, homogeneous bulk geometries such as black branes may indeed become unstable to formation of inhomogeneities \cite{Gregory:1993vy,Gubser:2000ec,Gubser:2000mm,Caldarelli:2008mv}, sometimes giving rise to mildly spatially modulated, ``striped'' phases, vortex lattices or other inhomogeneous structures \cite{Domokos:2007kt,Nakamura:2009tf,Ooguri:2010kt,Bergman:2011rf,Ammon:2011hz,Donos:2011bh,Donos:2011qt,Donos:2012yu,Donos:2013gda,Donos:2013wia,Withers:2013kva,Rozali:2013ama,Maeda:2009vf,Albash:2009iq,Iizuka:2013ag,Cremonini:2012ir,Vegh:2013sk,Bolognesi:2010nb,Sutcliffe:2011sr,Allahbakhshi:2011nh,Bu:2012mq}. These are distinct from holographic lattices that are quenched into the geometry by applying a modulated source on the boundary of AdS such as those studied in \cite{Kachru:2009xf,Horowitz:2012ky,Bao:2013fda,Hartnoll:2012rj}, in the sense that they are spontaneously generated, in the absence of explicit inhomogeneous sources. The works cited above exhibit holographic geometries with lattice-like, periodic order. Not surprisingly, they have crystal-like properties, rather than glass-like. One would expect a glassy geometry to be disordered. Disordered holographic geometries have been studied too \cite{Hartnoll:2008hs,Kachru:2009xf,Fujita:2008rs,Ryu:2011vq,Adams:2011rj,Adams:2012yi,Saremi:2012ji}, but as far as we know, all examples studied to date are quenched by explicit sources. Thus, in the terminology of footnote \ref{spinvsstructural}, they describe holographic ``spin'' (quenched) glasses, rather than holographic structural (spontaneous) glasses. What we are after are holographic structural glasses, with the disordered geometry generated in the absence of explicit disordered boundary sources.

It is possible that in the models of spontaneous breaking of translation symmetry cited above, disordered brane geometries might exist besides the ordered ones studied so far, the reason for the occurrence of order and lattice symmetry being more related to the relative simplicity of constructing symmetric geometries, rather than being intrinsic. Indeed, finding metastable disordered geometries would require finding isolated disordered solutions to a set of coupled nonlinear partial differential equations, arguably not an easy task. To construct simpler examples of disordered geometries in the absence of disordered sources, one might want to try to sprinkle charged massive probe particles onto a charged black brane background. In the dual CFT, this would correspond to adding small matter density inhomogeneities. However, just dropping a charged particle onto a black brane horizon will not generate a metastable structure, as the charge will quickly dissolve into the horizon. More fundamentally, whether we treat the problem perturbatively or use the fully nonlinear Einstein equations, we  are still left with a single, smooth brane horizon, and it is not clear (to us at least) if local inhomogeneities are enough to destroy universal properties such as $\eta/s = 1/4\pi$, which as we have seen is incompatible with what one expects from a supercooled liquid or a glass.

\def\CO{{\cal O}}

\subsection{Black hole bound states as holographic glasses}

\begin{figure}
\centering{
\includegraphics[width=0.45 \textwidth]{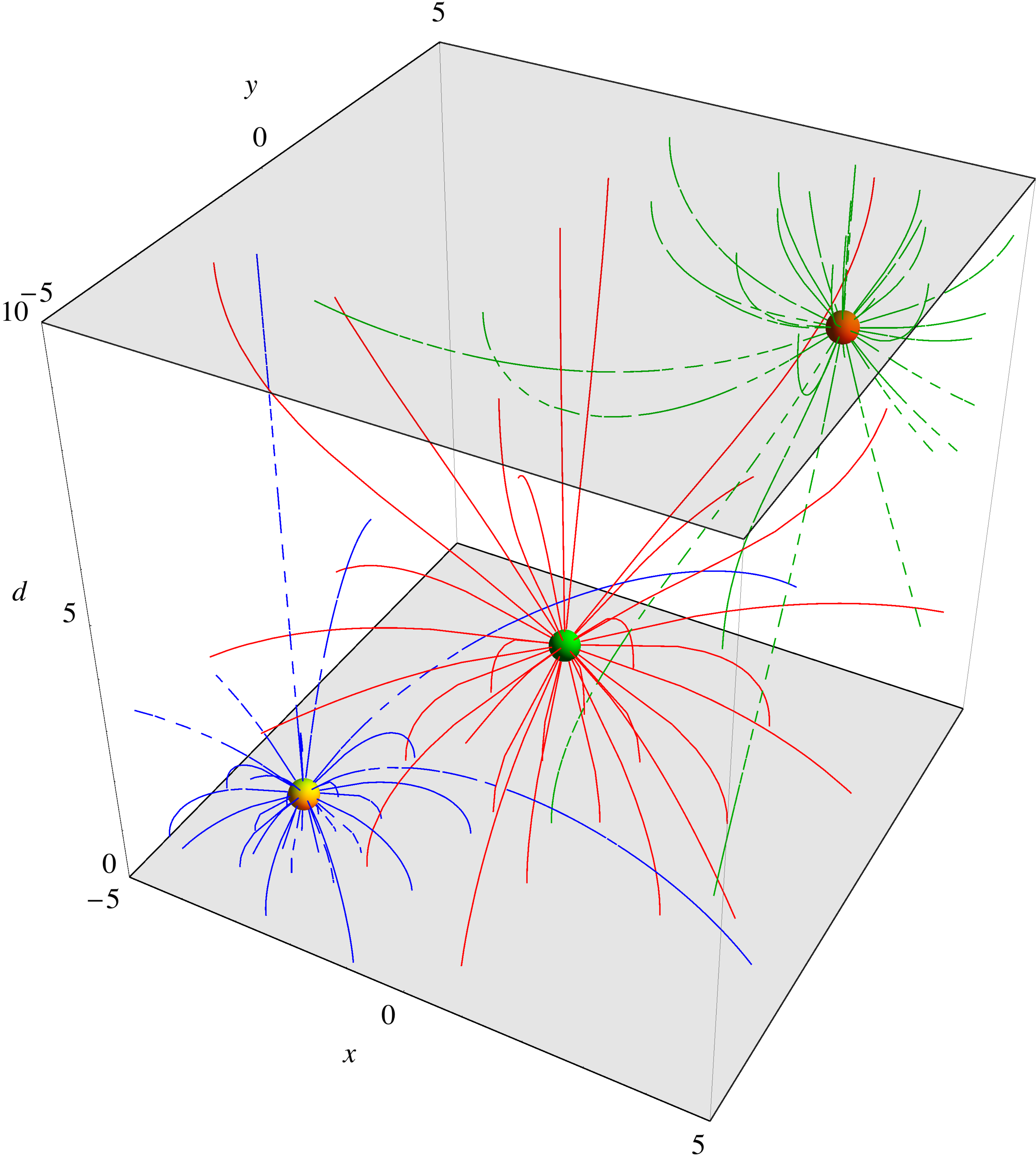}
\includegraphics[width=0.45 \textwidth]{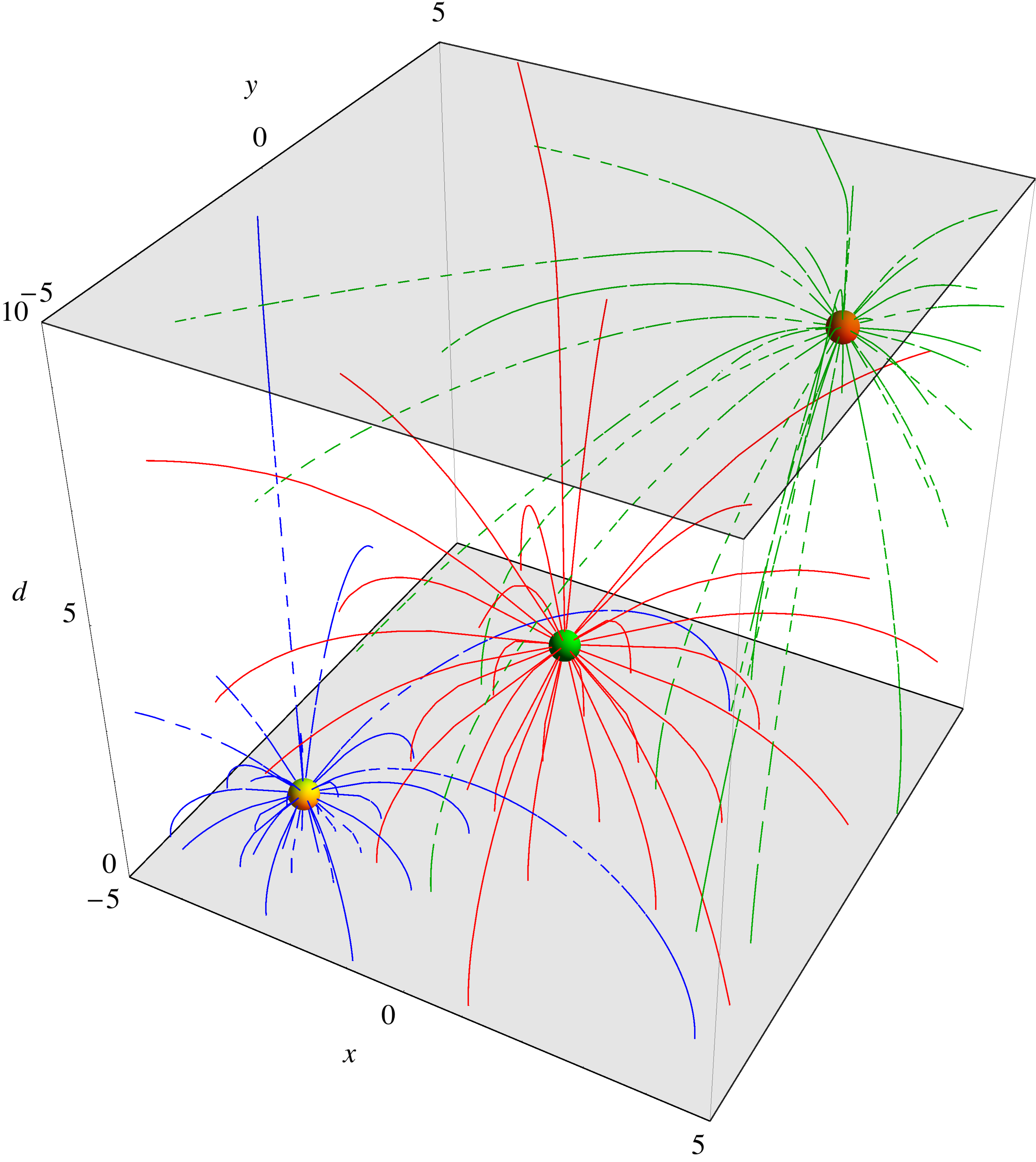}
}
\caption{Electric (left) and magnetic (right) field lines for some bound charges. The bottom plane is the horizon, the top plane is the boundary. The vertical coordinate is the optical distance from the horizon (cf.\ (\ref{opticaldist})).
\label{fig:fields}}
\end{figure}

In this paper, we explore a different idea, continuing on our work in \cite{arXiv:1108.5821}. The idea is that glassy and supercooled liquid phases of matter are holographically dual to disordered geometries with fragmented horizons, or more precisely to metastable  black hole bound states. A zoo of absolutely stable, supersymmetric, stationary black hole bound states has been known in the context of 4d $\CN=2$ supergravity for some time \cite{SabraLust,Denef:2000nb,LopesCardoso:2000qm,Bates:2003vx}, and the bound states we will study in this paper are cousins thereof, lifted to finite temperature and to asymptotically AdS$_4$ space. The generalization to finite temperature, asymptotically flat bound states was the subject of \cite{arXiv:1108.5821} (and independently \cite{Chowdhury:2011qu,Chowdhury:2013ema}), where existence of such bound states was established by considering charged probe black holes in the presence of a nonextremal background black hole. In this work we show that such bound states persist for nonextremal charged black holes in AdS$_4$, including planar ones. The specific bulk gravity theory we consider is the simplest possible natural uplift of the asymptotically flat model studied in \cite{arXiv:1108.5821} to AdS$_4$. It is given by four-dimensional Einstein gravity coupled to two $U(1)$ gauge fields and a non-minimally coupled scalar, with a scalar potential of ``Fayet-Iliopoulos'' gauged supergravity form \cite{de Wit:1983rz,VanProeyen:2004xt,Binetruy:2004hh}, which appears naturally in flux compactifications of string theory. The scalar potential has a negative energy extremum, leading to AdS$_4$ vacuum solutions, and the model has the virtue of having known explicit nonextremal black hole solutions with running scalars, a prerequisite for constructing bound states generalizing those of \cite{arXiv:1108.5821}. The probe particles are characterized by two electric and two magnetic charges, and crucially the existence of stationary bound states requires the charges of probe and background to be mutually nonlocal (i.e.\ in a duality frame in which the background is considered electric, the magnetic charge of the probes must be nonzero in order for a bound state to exist). As a result the bound particles are dynamically trapped by the black hole background, the way electrons get trapped in a magnetic background. 

At sufficiently high temperature, the bound states disappear, melting away into the background. At sufficiently low temperatures, the expulsion of particles from the mother black hole to form bound states lowers the free energy and thus becomes thermodynamically favored. However, at equilibrium, the rates for emission and absorption of charges are exponentially small in the semiclassical ($N \to \infty$) limit in which the bulk gravity picture becomes reliable, leading to a distribution of exponentially large relaxation time scales $\tau_c \sim e^{c N}$ with $c$ broadly distributed. As we will demonstrate, this naturally leads to aging, with a logarithmic aging law of the type mentioned above. 

A number of glassy features of these bound states is evident. A rugged free energy landscape with an extensive configurational entropy is manifest as the space of stationary bound state configurations. This space is large as a result of the fact that the constituents can have a wide range of different possible charges, with frustrated Van der Waals type static and electron-monopole type magnetic interactions between them. In the language of \cite{Anninos:2012gk}, still in the probe approximation, we can think of the local minima in this landscape as equilibrium configurations of ``supergoop'' clouds surrounding the mother black hole. Examples are shown in figs.\ \ref{fig:fields} and further on in \ref{fig:holes}. Although it may well be that there exists one particular ordered cloud configuration minimizing the free energy, there is no straight path that takes the system there starting from say the bare black brane state. Indeed, as we will see, even disregarding interactions between probes, it is in general not true that transitions towards lower free energy states are faster; in fact for a range of charges the opposite is true, with time scales for transitions to the lowest free energy bound states being exponentially much longer than those to higher free energy states. As a result, the system gets trapped in valleys of the free energy landscape which do not continue down to the true lowest free energy states. In this regime (the regime in which bound states are thermodynamically preferred over the bare brane), the probe density will eventually increase to the point that we necessarily exit the range of validity of the probe approximation. At higher temperatures, all bound states have higher free energy than the bare brane, and are thus metastable. In this case, time evolution preserves the validity of the probe approximation, and we can follow the evolution of the system for arbitrarily long times. For example, after quenching some initial cloud configuration with order 1 occupation numbers, we can see how it relaxes back to the bare brane configuration with an exponentially dilute cloud, which is the equilibrium state. As already mentioned, we observe characteristic $\CO(t_2)-\CO(t_1) \sim \log(t_2/t_1)$ aging behavior for such relaxation processes, similar to aging laws in other amorphous systems. 

The main goals of this paper are to establish the existence of AdS black hole bound states at finite temperature, to map out their detailed phase diagram, and to exhibit quantitatively the crude glassy features we just sketched. In addition, we set up the basic holographic dictionary between black hole bound states and their CFT dual thermal states. In particular we determine the features to which the localized bulk probe particles map in the boundary CFT --- for instance localized electric charge maps to a localized excess of matter density given by the normal component of the electric field strength at the boundary, while localized magnetic charge maps to a localized current loop (or spin) given by the cross product of the unit normal and the magnetic field at the boundary. The electric and magnetic fields of some probes are shown in fig.\ \ref{fig:fields} and CFT charge and current densities are illustrated for actual examples in figures \ref{fig:vortices}, \ref{fig:densplots} and \ref{fig:densplots2}.

To understand the glass transition itself in this context, to extract detailed transport properties and to place these models in the larger framework of theories of supercooled liquids and the glass transition, more work will be needed. In the final section of this paper, we give a detailed discussion of the gaps in our present analysis, offering some speculation and the general outlook we have on further progress. In a nutshell, our current speculative view is as follows. Bound states get highly populated during a cooling quench via classical horizon charge clumping instabilities, thus generating a finite density of local structures that are not in equilibrium but nevertheless metastable with exponentially long lifetimes. The bound charges backreact onto the brane, and because they are necessarily magnetically charged, they will act as magnetic brakes onto charged horizon currents, generating eddy current friction due to Lenz's law. This will obstruct charge transport, leading to a dramatic decrease in diffusion rates. Shear viscosity on the other hand, which is transversal momentum conductivity, will get greatly enhanced with respect to the bare black brane, as momentum can now efficiently be conducted through the supergoop cloud covering the brane. Indeed, due to the mutual electric-magnetic nonlocality of the charges in the cloud, implying that pairs of charges produce an intrinsic angular momentum stored in their electromagnetic field, the cloud can be thought of as a network of gyroscopes, resisting shearing and dynamically rigidifying the system, for any one of its configurations. As a result, the viscosity will greatly increase. Finally, dynamical heterogeneities are seeded by the bound charges, and their hierarchical dynamics as observed in simulations such as \cite{Keys2011} finds a natural geometrization in terms of the hierarchy of layers of particles bound at different radii.

\subsection{Outline}

The structure of the paper is as follows. In section \ref{setup} we introduce the model. In section \ref{sec:solution} we review the  bulk background solutions we will use, and settle on a convenient parametrization.  In section \ref{sec:stability} we discuss in detail the thermodynamic phase diagram of the background black holes, including the identification of possible charge clumping instabilities. We also study the small and planar black hole limits, and hyperscaling-violating limits. In section \ref{hhads} we exhibit the existence of probe bound states and map out their phase diagram. We give special attention to the planar case, and briefly discuss the opposite limit, ``AdS supergoop'', which is a possible endstate in which the horizon has completely fragmented into small black holes. In \ref{sec:relaxationdynamics} we study the relaxation dynamics of probe clouds in the ideal gas approximation, demonstrating the appearance of logarithmic aging and relating the aging coefficients to barrier distributions. In section \ref{sec:holointerpret} we give the holographic interpretation of our bulk constructions. Finally in section \ref{outlook} we summarize our conclusions, point out the gaps in our analysis, provide some speculations and give an outlook on future directions. The appendices provide details of some results which may be of independent interest.

\section{The Model}\label{setup}

\subsection{Qualitative features and motivation}

The bulk gravity theory we consider consists of four-dimensional Einstein gravity coupled to two $U(1)$ gauge fields and a non-minimally coupled scalar. In addition it has a scalar potential with an AdS$_4$ vacuum solution, with tunable parameters controlling the scalar vev and the AdS curvature scale. The specific Lagrangian we start from is given below in (\ref{modelLag}) and the part we will actually use in this paper is given in (\ref{modelLagsimple}). It can be viewed as a bosonic truncation of the simplest possible $\CN=2$ gauged supergravity theory, sometimes called Fayet-Iliopoulos (FI) gauged supergravity \cite{de Wit:1983rz,VanProeyen:2004xt,Binetruy:2004hh} (for a concise review with black holes in mind see \cite{Hristov:2012bk}). 

In the flat space limit (vanishing scalar potential), the model reduces to the one considered in \cite{arXiv:1108.5821}, which was obtained there as a universal consistent truncation of any Calabi-Yau compactification of type IIA string theory. One motivation for our choice of model is that this flat space limit is known to have  stationary BPS black hole bound states of arbitrary complexity \cite{SabraLust,Denef:2000nb,LopesCardoso:2000qm,Bates:2003vx,Denef:2007vg}, which persist at finite temperatures \cite{arXiv:1108.5821,Chowdhury:2011qu}. Hence by continuity we are guaranteed that black hole bound states will also exist in the present model, at least in the limit in which the size of the black holes is much smaller than the AdS radius. Another motivation is that the asymptotically flat background black hole solutions used in \cite{arXiv:1108.5821} have explicit asymptotically AdS counterparts \cite{Hristov:2012bk,hep-th/9901149}. This allows us to copy the probe strategy followed in \cite{arXiv:1108.5821}, making manifest the specific new features induced by the lift to AdS$_4$. The final motivation is the plausibility that this model has a suitable (stable) embedding in string theory, possibly with a holographic dual description as a three-dimensional conformal field theory. 

The string theory embedding will have at least one important imprint on the low energy physics which is not determined by the 4d bulk Lagrangian (\ref{modelLag}) itself, namely the spectrum of charged particles. To stay as close as possible to \cite{arXiv:1108.5821}, we will assume the charged particles in the model are all much heavier than the AdS curvature scale. This allows treating them as well-localized probes. As detailed in section \ref{sec:probeaction} below,  we will infer their mass by thinking of them as black holes much smaller than any of the length scales of the background.

\subsection{Bulk action} \label{sec:action}

Our notation is chosen to parallel that of \cite{arXiv:1108.5821}, the asymptotically flat limit of the model.\footnote{To conform to more standard conventions, we will however change the normalization of the gauge fields by a factor $-\frac{1}{2}$: $A_\mu^{\rm new} = -\frac{1}{2} A_\mu^{\rm old}$.} The light field content consists of the metric $g_{\mu\nu}$, a complex scalar $z \equiv x + i y$ and two $U(1)$ gauge fields $A^I_\mu$, $I=0,1$, with field strengths $F_{\mu\nu}^I \equiv \partial_\mu A^I_\nu - \partial_\nu A^I_\mu$. The four-dimensional bulk action is taken to be the bosonic sector of Fayet-Iliopoulos $\CN=2$ gauged supergravity with cubic prepotential: $S=\frac{1}{8 \pi} \int d^4 x \, \sqrt{-g} \, {\cal L}$ with
\begin{equation} \label{modelLag}
 {\cal L}= \frac{1}{2\ell_p^2} R - \frac{3}{4 \ell_p^2}  \frac{(\partial x)^2 + (\partial y)^2 }{y^2}- V_g(x,y) 
  - G_{IJ} F^I_{\mu\nu} F^{J\mu\nu} +   \Theta_{IJ} F^I_{\mu\nu} \tilde{F}^{J\mu\nu}  \, ,
\end{equation}
where $\tilde{F}_{\mu\nu} \equiv \tfrac{1}{2}\epsilon_{\mu\nu\rho\sigma}F^{\rho\sigma}$, with $\epsilon_{0123}=+\sqrt{-g}$.
The scalar is neutral but is non-minimally coupled to the electromangnetic field strengths through the coupling and theta angle matrices
\begin{equation}
 G = \begin{pmatrix} 
   \frac{1}{6} y^3 + \frac{1}{2} x^2 y& -\frac{1}{2} xy \\
 -\frac{1}{2} xy & \frac{1}{2} y
 \end{pmatrix}\, , \qquad \Theta = \begin{pmatrix} 
   \frac{1}{3} x^3 & -\frac{1}{2} x^2 \\
 -\frac{1}{2} x^2 & x
 \end{pmatrix}.
\end{equation}
The scalar potential $V_g$ for $\CN=2$ Fayet-Iliopoulos-gauged supergravity is schematically of the form $V_g = |DW|^2 - 3|W|^2$ where $W \sim \frac{1}{y^{3/2}} \bigl( -g_{p_1} \frac{z^2}{2} + g_{q_0} \bigr)$, which is also of the form of Gukov-Vafa-Witten-type $\CN=1$ superpotentials arising from flux compactifications \cite{Gukov:1999ya}. It leads to the following potential:
\begin{equation}
V_g (x,y) = - \frac{3}{2 \ell_p^4} g_{p_1}  \bigl( g_{p_1} y  + g_{q_0} \frac{1}{y}  + g_{p_1} \frac{x^2}{y}  \bigr).
\end{equation}
In the context of flux compactifications, the constants $g_{p_1}$ and $g_{q_0}$ would be fixed by the choice of fluxes supporting the compactification, and by values of moduli we are taking to be frozen here.

We will take $(g_{p_1}, g_{q_0})$ to be arbitrarily tunable but fixed real valued parameters of the theory. If they have the same sign, which we assume from now on, the potential is extremized at a negative local maximum $z=z_0$, giving rise to an AdS$_4$ vacuum with AdS length $\ell$, with 
\begin{equation} \label{y0def}
 x_0=0\, , \quad \yz = \sqrt{\frac{g_{q_0}}{g_{p_1}}} \, , \qquad  
  V_{g} = -\frac{3}{\ell_p^4} \sqrt{g_{p_1}^{3} g_{q_0}} =- \frac{3}{\ell_p^2\ell^2} \, .
\end{equation} 
In this vacuum the scalar has the conformally coupled value $m^2 = - 2/\ell^2$, above the Breitenlohner-Freedman AdS tachyon bound \cite{Breitenlohner:1982jf}, which for AdS$_4$ is $m^2_{BF} = -2.25/\ell^2$.

For the background black hole solutions which we consider, it is consistent to put $x \equiv 0$, in which case the coupling matrix $G$ becomes diagonal and the theta angle matrix $\Theta$ is zero. Putting furthermore $y \equiv \yz \, e^{\chi}$, the Lagrangian (\ref{modelLag}) then simplifies to
\begin{equation} \label{modelLagsimple}
 {\cal L}= \frac{1}{2\ell_p^2} \bigl( R - \frac{3}{2}  (\partial \chi)^2 + \frac{6}{\ell^2}  \cosh \chi  \bigr)
  -\frac{1}{6} \yz^3 \, e^{3 \chi}   \, F_0^2 -\frac{1}{2}  \yz \, e^\chi  \, F_1^2   \, .
\end{equation}

Without making a commitment to any stringy interpretation at this point, we reparametrize the $g_i$ by constants $k$ and $\N$ as follows
\begin{equation}
 g_{q_0} \equiv \frac{1}{k} \, \qquad g_{p_1} \equiv \frac{1}{\N} \, .
\end{equation}
Then we have
\begin{equation} \label{abjmlike}
 \yz = \sqrt{\frac{\N}{k}} \, , \qquad \frac{\ell^2}{\ell_p^2} =  \frac{\N^2}{\yz}  =  \sqrt{k \N^3} \, .
\end{equation}
If the gravity theory has a CFT dual, its central charge is proportional to the second quantity, 
the AdS radius squared in four dimensional Planck units (see e.g.\ \cite{Kovtun:2008kx} for a general discussion). This will also be evident from the scaling of various thermodynamic quantities in  (\ref{rescaledext}) further down. In ABJM theory \cite{Aharony:2008ug}, a Chern-Simons-matter CFT proposed to be dual to type IIA string theory on AdS$_4 \times {\mathbb C}P^3$ with $k$ units of magnetic RR 2-form flux and $\N$ units of magnetic 6-form flux turned on in the ${\mathbb C}P^3$, the central charge is of the same form, with $\N$ interpreted as the rank of the gauge group, and $k$ as the inverse coupling constant of the Chern-Simons theory. The quantity $\yz^2 = \N/k$ is identified with the 't Hooft coupling $\lambda$ in this setting, and $\ell_s = \ell/\sqrt{\yz}$ with the string length. Further down we will see that  other quantities such as particle mass spectra have ABJM-like scalings with $k$ and $\N$. 

However, the model we are considering is \emph{not} the low energy effective action of the ABJM AdS$_4 \times {\mathbb C}P^3$ compactification, as in this theory one of the $U(1)$s is actually massive, Higgsed by a charged scalar (the universal 4d axion) with D0- and D4- charges proportional to  $(g_{q_0}^{-1},g_{p_1}^{-1}) \propto (k,N)$ \cite{Aharony:2008ug}. One of the consequences of this is that D2 and D6 charges will come with strings attached and that one of the two electrostatic forces will fall off exponentially rather than by the usual Coulomb law.

\subsection{Probe action} \label{sec:probeaction}

Since our model has two $U(1)$s, the electromagnetic fields couple to two magnetic charges $p^I$ and two electric charges $q_I$, $I=0,1$. The $q_I$ couple electrically to the $A^I$, while the $p^I$ couple electrically to the \emph{dual} gauge potentials $B_I$, defined as
\begin{equation} \label{Gmagndual}
dB_I = G_I = G_{IJ} \tilde{F}^J - \Theta_{IJ} F^J~.
\end{equation} 
The equations of motion for $F^I$ are the Bianchi identities for $G_I$ and vice versa. With these dual gauge fields one can conveniently write down a general expression for the action of a probe particle in a general background. For a probe charge $\gamma = (p^0,p^1,q_0,q_1)$ this is \cite{Denef:2000nb,Billo:1999ip} 
\begin{equation}\label{probeaction}
 S_\gamma =  - \int m_\gamma(z) \, ds - \int  q_I A^I - p^I B_I  \, .
\end{equation}
We 
will take probe charges to be quantized in units of order 1, roughly thinking of them as wrapped D6, D4, D2 and D0 brane charges in a type IIA compactification. The mass $m(p,q;z)$ depends on the charges and the local background scalar value $z=x+i y$. We will consider probe black holes which are much smaller than the AdS radius as well as much smaller than the background black hole, albeit at the same temperature. As argued in \cite{arXiv:1108.5821} and as we will check again in section \ref{sec:probevalidity} below, in this limit, the background becomes effectively cold from the point of view of the probe, in the sense that the thermal contribution to its mass becomes negligible. Hence the probe acquires the properties of an extremal black hole in asymptotically flat space. Extremal asymptotically flat black holes in $\CN=2$ supergravity may be BPS or non-BPS. In the first case, their mass is given by the absolute value of the central charge of the asymptotically flat $\CN=2$ supersymmetry algebra, which for our model is
\begin{equation} \label{centralcharge}
m_\gamma(z)=\frac{1}{\ell_p} \sqrt{\frac{3}{4 y^3}} \biggl| \frac{1}{6} p^0 z^3-\frac{1}{2} p^1
   z^2+q_1 z +q_0 \biggr| \, .
\end{equation}
In the second case, the mass is strictly greater than this. As in \cite{arXiv:1108.5821}, we restrict ourselves to probe charges that are in fact BPS. Besides being the simplest to analyze systematically, BPS probes are also the most stable. Although non strictly supersymmetric in AdS, the phase space for decay of these nearly-BPS probes will generically be much smaller than for probe charges which have a non-BPS flat space limit. 

When $x=0$, (\ref{centralcharge}) reduces to $m_\gamma =  \frac{\sqrt{3}}{2 \ell_p} \left[ (\frac{1}{6} p^0 y^{3/2} - q_1 y^{-1/2})^2 + (\tfrac{1}{2} p^1
   y^{1/2} +q_0 y^{-3/2})^2 \right]^{1/2}.$ Since we work with normalization conventions in which charges are integrally quantized, we can read off the orders of magnitude of the masses of various types of charge. Expressed in terms of the AdS scale $\ell$ and the the parameters $\yz$, $\N$ and $k$ introduced in (\ref{y0def}) and (\ref{abjmlike}), these are: 
\begin{equation} \label{Dprobemasses}
 \ell \, m_{\rm D0} \sim \frac{\N}{\yz^2} = k \, , \quad
 \ell \, m_{\rm D2} \sim \frac{\N}{\yz} = \sqrt{\N k} \, , \quad
 \ell \, m_{\rm D4} \sim \N \, , \quad
 \ell \, m_{\rm D6} \sim \N \yz = \sqrt{\frac{\N^3}{k}} \, . 
\end{equation}
Notice that this agrees with the masses of wrapped D0- and D4-branes in ABJM theory \cite{Aharony:2008ug} (D2- and D6-branes carry magnetic charge for the massive $U(1)$ in ABJM, and as a result would come with additional magnetic flux strings attached to them). The condition that all charged particles be much heavier than the AdS scale is thus
$\frac{1}{\N} \ll \yz \ll \sqrt{\N}$, or equivalently $\N^3 \gg k \gg 1$. 

As in \cite{arXiv:1108.5821,Denef:2007vg}, we may parametrize the charges as 
\begin{equation} \label{bnparametrization}
 \gamma=(p^0,p^1,q_1,q_0)=p^0(1,\kappa,-b+\frac{\kappa^2}{2},n+b\kappa-\frac{\kappa^3}{6}).
\end{equation} 
The parameter $\kappa$ can be thought of as proportional to $U(1)$ worldvolume flux on the wrapped D-brane; switching it on effectively shifts $z \to z - \kappa$ in (\ref{centralcharge}). The (flat) BPS black hole entropy is independent of $\kappa$ and given by $S_\gamma =  \pi (p^0)^2 \sqrt{\tfrac{8}{9} b^3 - n^2}$ \cite{Shmakova:1996nz}. For charges $\gamma=p^1(0,1,\kappa,n'-\frac{{\kappa'}^2}{2})$, this becomes $S_\gamma = \pi (p^1)^2  \sqrt{\frac{2}{3} n'}$. Evidently the quantities under the square root must be positive for the black hole to exist. 
We should note however that not all BPS particles have a realization as a single centered black hole in supergravity, even when we allow singular limits in which the horizon goes to zero size. Some BPS states are realized as multi-centered bound states \cite{Denef:2000nb}. A notable example is a pure wrapped D4-brane, which has a \emph{negative} worldvolume curvature-induced D0-charge $q_0 = -p_1^3/24$, and is realized as a two particle bound state of charges $(1,\frac{p_1}{2},\frac{p_1^2}{8},-\frac{p_1^3}{48})$ and $(-1,\frac{p_1}{2},-\frac{p_1^2}{8},-\frac{p_1^3}{48})$ \cite{Denef:2007vg}. However for our purposes it will be sufficient to consider single centered probes, and so we will require $\tfrac{8}{9} b^3 - n^2 \geq 0$.


\section{Background solution} \label{sec:solution}

We consider a spherically symmetric nonextremal charged black hole metric of the form
\begin{equation} \label{metricansatz}
 ds^{2}=-V(r) \, dt^{2}+\frac{1}{V(r)} dr^{2}+ \, W(r) \, d\Omega_{2}^{2} \, .
\end{equation}
The scalar $z$ is assumed to only depend on the radial coordinate $r$. Note that $r$ is in principle not the Schwarzschild radial coordinate; namely because it can go negative. In general the black hole may have arbitrary electric and magnetic charges $Q_I$ and $P^I$, but as in \cite{arXiv:1108.5821} we limit ourselves to a setup with $P^0=0$ and $Q_1=0$, in which case we can consistently set $x=0$ throughout, and the field strengths  
\begin{equation} \label{F01sol}
 F^0 = Q_0  \, \frac{3}{y(r)^3} \frac{dt \wedge dr}{W(r)} \, , \qquad F^1 = -\frac{1}{2} \, P_1 \, \sin \theta \, d\theta \wedge d\phi \, 
\end{equation}
automatically solve the Bianchi identities and equations of motion \cite{arXiv:1108.5821}.

Exact solutions satisfying this ansatz, for arbitrary charges $P_1$, $Q_0$ and mass $M$, were constructed in \cite{hep-th/9901149} (related solutions were considered in
\cite{Caldarelli:1998hg,Behrndt:1998jd,Sabra:1999ux,Chamseddine:2000bk,Cucu:2003yk,Chong:2004na,Cacciatori:2009iz,Kimura:2010xe,Goldstein:2010aw,Chow:2010fw,Dall'Agata:2010gj,Hristov:2010ri,Toldo:2012ec,Klemm:2012yg,Klemm:2012vm,Gnecchi:2012kb}). These solutions will be the starting point for our analysis.\footnote{This is not the most general set of solutions compatible with the ansatz. Indeed in the neutral limit, it reduces to the standard hairless AdS-Schwarzschild solution, while it is known  that there also exist hairy solutions with the same boundary conditions \cite{Hertog:2004dr} (for a recent discussion see \cite{Amsel:2012ir}). 
The (numerically constructed) hairy neutral black hole is thermodynamically disfavored compared to the hairless one \cite{Hertog:2004dr}, and thus by continuity the same will be true for at least a finite range of charged black holes, for which this restriction will not invalidate the thermodynamic analysis. It is not known however if this continues to hold for arbitrary charges. In principle it should be possible to address this question by (numerically) analyzing the reduced equations of motion obtained e.g.\ in \cite{Dall'Agata:2010gj}.}

\subsection{Metric, scalar and gauge potentials} \label{sec:msgp}

For any given mass $M$ and charges $P_1$, $Q_0$, the solution of \cite{hep-th/9901149} can be written in the form (\ref{metricansatz}) with $x=0$ and
\begin{equation}\label{bhsoln}
 V(r) = \frac{1}{W}\left(r^2-c^2+\frac{1}{\ell^2}W^2\right), \qquad
 W(r)=\sqrt{f_0 f_1^3} \, , \qquad y(r)=\yz \sqrt{\frac{f_0}{f_1}}\,,
\end{equation}
where the $f_i$ are functions linear in $r$:
\begin{equation}
 f_0(r)=r+a_0 \, , \quad
 f_1(r)=r+a_1 \, ,
\end{equation}
the AdS length $\ell$ and asymptotic scalar $\yz = y|_{r=\infty}$ are fixed by $g_{q_0}$ and $g_{p_1}$ as in (\ref{y0def}), and $c$, $a_0$ and $a_1$ are positive constants determined by the mass $M$ and charges $Q_0$ and $P_1$ of the black hole:
\begin{equation} \label{MQfromac1}
a_0=\sqrt{c^2+\tfrac{12}{\yz^3} \ell_p^2 Q_0^2},\qquad a_1=\sqrt{c^2+\tfrac{\yz}{3} \ell_p^2 P_1^2} \, , 
\end{equation}
with $c=c(M,Q_0,P_1)$ the unique positive solution to
\begin{equation} \label{MQfromac2}
  M \ell_p^2 =  \tfrac{1}{4}a_0 + \tfrac{3}{4} a_1 = \tfrac{1}{4}\sqrt{c^2+\tfrac{12}{\yz^3} \ell_p^2 Q_0^2} + \tfrac{3}{4}\sqrt{c^2+\tfrac{\yz}{3} \ell_p^2 P_1^2} \, .
\end{equation}
The definition and computation of the mass $M$ is subtle due to the presence of the $m^2<0$ scalar. We computed it as in \cite{Hertog:2004dr,Balasubramanian:1999re}. The parameter $c$ is a measure for the deviation from extremality, as in the asymptotically flat case studied in \cite{arXiv:1108.5821}. However in the case at hand the point $c=0$ is  not physically reachable: extremality occurs at some nonzero value of $c$, as will be clear from the discussion further down. Notice that when $a_0 = a_1$, i.e. when $|Q_0| = \yz^2 |P_1|/6$, the profile of the scalar field becomes constant everywhere and the metric becomes that of the ordinary Reissner-Nordstrom-AdS black hole.

We denote the radial location of the outer horizon by $r_+$. It satisfies $V(r_+)=0$, that is:
\begin{equation} \label{rplusdef}
 r_+^2 - c^2 + \frac{1}{\ell^2} (r_+ + a_0)(r_+ + a_1)^3 = 0 \, , 
\end{equation}
and in addition $W(r)>0$ and $V(r) > 0$ for all $r>r_+$. 

The gauge potentials $A^I$ and their magnetic duals $B_I$ are obtained by integrating the field strengths $F^I$ and $G_I$ specified by (\ref{F01sol}) and (\ref{Gmagndual}):
\begin{align} \label{ABexpr}
  A^0 &=  \biggl(\frac{3}{\yz^3}\frac{Q_0}{r+a_0} - \Phi_0\biggr) \, dt \, , \qquad &
  A^1 &= \frac{1}{2} P_1 \bigl(\cos \theta \pm 1\bigl) \, d\phi \, , \\
  B_0 &= \frac{1}{2} Q_0 \bigr(\cos \theta \pm 1 \bigr) \, d\phi  \, , \qquad &
  B_1 &=  -\biggl( \frac{\yz}{4}\frac{P_1}{r+a_1} - \Phi_1 \biggr) \, dt \, .
\end{align}
We choose the integration constants $\Phi_0$ and $\Phi_1$ such that the electric potentials vanish at the black hole horizon $r=r_+$. This guarantees regularity of the gauge connection after Euclidean continuation of the solution, and fixes
\begin{equation} \label{phivalsreg}
 \Phi_0 = \frac{3}{\yz^3} \frac{Q_0}{r_+ + a_0} \, , \qquad 
 \Phi_1 = \frac{\yz}{4} \frac{P_1}{r_+ + a_1} \, ,
\end{equation}

The asymptotic scalar profile in the standard Schwarzschild radial coordinate $r_s = \sqrt{W(r)}$ is given by $\log y(r_s) = \frac{\alpha}{r_s} + \frac{\beta}{r_s^2} + \cdots$, where 
$\alpha = (a_0-a_1)/2$ and $\beta = - \alpha^2/2$. Thus all solutions found in \cite{hep-th/9901149} obey the generalized conformally invariant boundary condition $\beta = f \alpha^2$ of \cite{Hertog:2004dr}, for a specific value of $f$ (which depends on the normalization of the scalar).\footnote{These generalize the ``standard'' Dirichlet ($\alpha=0$) and ``alternate'' Neumann ($\beta=0$) zero source boundary conditions. In language of the dual CFT, the $\alpha=0$ boundary conditions corresponds to a CFT where the operator ${\cal O}$ dual to the scalar has dimension $\Delta=2$, while $\beta=0$ boundary conditions correspond to a CFT where this operator has dimension $\Delta=1$. The $\alpha=0$ CFT is the IR fixed point of a relevant double trace deformation $\Delta {\cal L}_{\rm CFT} \propto {\cal O}^2$ of the $\beta=0$ CFT, while the $\beta + \frac{\alpha^2}{2} = 0$ CFT is obtained from the $\beta=0$ one by an approximately marginal triple trace deformation $\Delta {\cal L}_{\rm CFT} \propto {\cal O}^3$.}

\subsection{Parametrization} \label{sec:param}

We found it most convenient to parametrize the vacua by $\ell$ and $\yz$ and the black hole solutions by $r_+$, $u_0$ and $u_1$, where we define
\begin{equation}
 u_I \equiv r_+ + a_I.  
\end{equation}
The parameters $c,a_0,a_1$ appearing in the solution as given above can be written in terms of $(r_+,u_I)$ as: 
\begin{equation} \label{cauI}
 c = \sqrt{r_+^2+\frac{1}{\ell^2} u_0 u_1^3} \, , \qquad a_I = u_I - r_+ \, ,
\end{equation}
and thus the conserved quantitities $Q_0$, $P_1$ and $M$ are obtained using the relations (\ref{MQfromac1})-(\ref{MQfromac2}). Explicitly:
\begin{equation}  \label{QPMurpnotilde} \small
 \ell_p|Q_0| = \sqrt{\frac{\yz^3}{12}} \sqrt{u_0(u_0-2r_+) - \frac{u_0 u_1^3}{\ell^2}}, \quad \ell_p|P_1| = \sqrt{\frac{3}{\yz}} \sqrt{u_1(u_1-2r_+) - \frac{u_0 u_1^3}{\ell^2}},  \\
 \quad \ell_p^2M=\frac{1}{4}(u_0+3u_1) - r_+ \, .
\end{equation}
The AdS-Reissner-Nordstrom limit corresponds to $u_0=u_1\equiv u$, while the neutral AdS-Schwarzschild limit has $r_+=\frac{1}{2}(u-\frac{1}{\ell^2} u^3)$, with $M =\frac{1}{2}(u+\frac{1}{\ell^2} u^3)/\ell_p^2$.

\subsection{Entropy and temperature}

The black hole entropy is one quarter of the horizon area, which in our parametrization takes the simple form
\begin{equation}\label{entropy}
S= \frac{\pi\sqrt{u_0 u_1^3}}{\ell_p^2} \, .
\end{equation}
Its temperature $T$ is obtained in the standard way by requiring regularity of the Euclidean continuation at $r=r_+$ by imposing Euclidean time periodicity $1/T$, giving
\begin{equation}\label{bhtemp}
 T=\frac{V'(r_+)}{4 \pi}=\frac{2r_+ +  u_1^2(3u_0+u_1)/\ell^2}{4\pi\sqrt{u_0 u_1^3}}\,.
\end{equation}
Notice that in the flat space limit, the BPS black holes would have $r_+ = 0$ and are thus connected to finite temperature black holes in AdS where we do not take the strict $\ell \to \infty$ limit.

\subsection{Physical region of parameter space}

The physical parameter range is given by the values of $(r_+, u_0, u_1)$ for which the constants $a_I$ and $c$ appearing in the metric are all positive, and for which $T>0$ and $\Phi_I \in {\mathbb R}$. This implies in particular that $u_I>0$, as can be seen by making use of (\ref{MQfromac1}) and (\ref{cauI}). The horizon radial position can be either positive or negative: for example a large neutral AdS-Schwarzschild black hole has $r_+<0$ while a small neutral black hole has $r_+>0$. 

To obtain all possible black hole solutions for a given $(T,\Phi_0,\Phi_1)$, we solve numerically for $(r_+,u_0,u_1)$ and retain the solutions with $u_0,u_1>0$. This guarantees the solution is physical and that $r_+$ is indeed the outer horizon, i.e.\ $V(r)>0$, $W(r)>0$ for all $r>r_+$.\footnote{To see this, express $V$ and $W$ in terms of $(r_+,u_0,u_1)$ and $x \equiv r-r_+$. Then $W=\sqrt{(u_0+x)(u_1+x)^3}$, which is manifestly positive for $x>0$, since $u_I>0$. Furthermore $W V=\left(2 r_+ +\frac{1}{\ell^2} u_1^2 (3 u_0+u_1)\right) x + \left(1+\frac{3}{\ell^2} u_1 (u_0+u_1)\right) x^2 +\frac{1}{\ell^2}(u_0+3 u_1) x^3+\frac{1}{\ell^2} x^4$, which is also manifestly positive, since the coefficient of $x$ equals $4 S T >0$, and $u_I>0$.}

\subsection{Scaling symmetries and invariant parametrization} \label{scalingsymm}

We have parametrized the solutions by a total of 5 parameters $(\yz,\ell,r_+,u_0,u_1)$, with the first two fixing the AdS vacuum and the last three parametrizing the black hole solutions within a given vacuum. However, as in the asymptotically flat case \cite{arXiv:1108.5821}, there are two scaling symmetries trivially relating different solutions. They act as
$X \to \lambda_1^{n_1} \lambda_2^{n_2} X$, $\lambda_i \in {\mathbb R}^+$, on the various quantities $X$ defined so far, with the exponents $(n_1,n_2)$ indicated in the first two lines of this table:
\begin{center}
\begin{tabular}{|c|cccc|ccc|cccc|ccc|c|} 
 \hline 
   & $\ell$ & $\yz$ & $k$ & $\N$ & $r_+$ & $u_0$ & $u_1$ & $M$ & $Q_0$ & $P_1$ & $S$ & $T$ & $\Phi_0$ & $\Phi_1$ & $r$ \\
  \hline
 $n_1$  & 1  & 0 & 1 & 1  & 1 & $1$ & $1$  &  1 & 1 & 1 & 2 & $-1$ & 0 & 0 & 1  \\
 $n_2$  & $0$  & 1 & $-\frac{3}{2}$ & $\frac{1}{2}$ & 0 & 0 & 0  &  0 & $\frac{3}{2}$ & $-\frac{1}{2}$ & 0 & $0$ & $-\frac{3}{2}$ & $\frac{1}{2}$ & $0$ \\
 $\delta$ & $-1$ & 0 & 0 & 0 & $-1$ & $-1$ & $-1$ & 1 & 0 & 0 & 0  & 1 & 1 & 1 & $-1$ \\
 $\N^\#$ & 0 & 0 & 1 & 1 & 0 & 0 & 0 & 2 & 1 & 1 & 2 & 0 & 1 & 1 & 0 \\
 $\yz^\#$ & 0 & 1 & $-2$ & 0 & 0 & 0 & 0 & $-1$ & 1 & $-1$ & $-1$ & 0 & $-2$ & 0 & 0 \\
    \hline
\end{tabular}
\end{center}
The third line shows the mass dimension $\delta$. Physical observables will depend only on invariant combinations of the parameters, up to an overall factor determined by the scaling properties of the observable. 
Specifically, we will express any quantity $X$ of mass dimension $\delta$ and scaling exponent $(n_1,n_2)$ in terms of a dimensionless, scaling invariant $\tilde{X}$, as follows:
\begin{equation}\label{eq:scalingsymm}
X= \ell^{-\delta} \N^{n_1+\delta} \, \yz^{n_2-(n_1+\delta)/2} \tilde{X}~  \, .
\end{equation}
The quantities $\N$ and $k$ were introduced in (\ref{abjmlike}). The last two lines of the table indicate the powers of $\N$ and $\yz$ appearing in various quantities. We will display our phase diagrams as functions of the  rescaled intensive variables $\bigl(\tilde{T},\tilde{\phi}_0,\tilde{\phi}_1\bigr)$ related to the original ones by
\begin{equation} \label{rescaledpotentials}
T = \frac{1}{\ell} \, \tilde{T} \, , \qquad \Phi_0 =  \frac{\N}{\yz^2 \ell} \, \tilde{\Phi}_0\, , \qquad \Phi_1 = \frac{\N}{\ell} \, \tilde{\Phi}_1 \, .
\end{equation}
The extensive variables (\ref{QPMurpnotilde}) and (\ref{entropy}) are related to their invariant counterparts by
\begin{equation} \label{rescaledext}
 Q_0 = \N \yz \, \tilde{Q}_0 \, , \qquad P_1 = \frac{\N}{\yz} \,\tilde{P}^1, \qquad
 M = \frac{\N^2}{\yz \ell} \, \tilde{M}  \, , 
 \qquad  S = \frac{\N^2}{\yz} \,  \tilde{S} \, .
\end{equation}
Working consistently with the rescaled variables instead of the original ones effectively sets 
\begin{equation} \label{rescaledeffect}
  \qquad \ell_p \equiv 1 , \qquad \ell \equiv 1 \, \qquad \yz \equiv 1 \, 
\end{equation} 
in the expressions of the previous sections. In what follows we will always use rescaled variables, and to avoid cluttering we will therefore drop the tildes, keeping in mind that in order to get the actual physical quantities, we need to rescale as indicated above. 

Finally note that besides the obvious charge conjugation symmetry $(P_1,Q_0) \to (-P_1,-Q_0)$, the background metric and scalar profile are also invariant under $(P_1,Q_0)\to(P_1,-Q_0)$. This descends from an enhanced ${\mathbb Z}_2$ symmetry of the action that exists only when the pseudoscalar $x$ is zero. 




\section{Background thermodynamics}\label{sec:stability}

Before moving on to examine probe black holes in the black hole background, we analyze the phase structure of the background itself, which is already quite interesting. This comes down to a generalization of the classic work \cite{Chamblin:1999tk} on phases of pure AdS-Reissner-Nordstrom black holes to the case with running scalars, with the former retrieved in our setup as the special case $u_0=u_1$, which indeed has $y(r)=\yz$ constant. The presence of running scalars leads to a considerably more intricate structure.


\subsection{Thermodynamic equilibrium and stability}

We will mostly work in a thermodynamic ensemble with fixed temperature $T$ and chemical potentials $\Phi_0$, $\Phi_1$ dual to the charges $Q_0$ and $P_1$, and fixed charges $P^0=0$, $Q_1=0$. That is to say, if we imagine coupling the system to a reservoir at fixed temperature $T$ and potentials $\Phi_I$, the total (system plus reservoir) entropy will change as $\Delta S_{\rm tot} = \Delta S - \frac{1}{T} \Delta E + \frac{\Phi_0}{T} \Delta Q_0 + \frac{\Phi_1}{T} \Delta P_1 = -\Delta F /T$, where $\Delta E$, $\Delta Q_0$, $\Delta P_1$ and $\Delta S$ refer to the system, and we have defined  
\begin{equation} \label{freeenergywe}
 F \equiv E-T\, S -  \Phi_0 Q_0 - \Phi_1 P_1.
\end{equation}
Stable equilibrium with the reservoir requires $S_{\rm tot}$ to be maximized, or equivalently $F$ to be minimized under variations of energy and charges; locally this requires
\begin{equation} \label{stableeq}
 F' = 0 \, , \qquad F'' > 0 \, .
\end{equation}
The derivatives are understood to be with respect to the system's extensive variables, at fixed, externally tuned values of $T$, $\Phi_0$ and $\Phi_1$. The parametrization of the extensive variables can be arbitrary. We will work with the black hole metric parameters $(u_0,u_1,r_+)$ defined in \ref{sec:param}. Thus, using (\ref{QPMurpnotilde}) and (\ref{entropy}) keeping in mind (\ref{rescaledeffect}),
\begin{equation} \label{Furp}
 F = \frac{1}{4} u_0+ \frac{3}{4} u_1 - r_+ - \pi T \sqrt{u_0 u_1^3} - 
 \frac{\Phi_0}{2\sqrt{3}} \sqrt{u_0^2-2 u_0 r_+ - u_0 u_1^3} -
 \sqrt{3} \, \Phi_1 \sqrt{u_1^2-2 u_1 r_+ - u_0 u_1^3}.
\end{equation}
Solving $F'=0$ in (\ref{stableeq}) at fixed $(T,\Phi_0,\Phi_1)$ then provides the local equilibrium relation between $(T,\phi_0,\phi_1)$ and $(r_+,u_0,u_1)$: 
\begin{equation} \label{Tphi}
 T = \frac{2 r_+ + 3u_0 u_1^2+u_1^3}{4\pi\sqrt{u_0 u_1^3}} \, , \quad
 \phi_0 = \frac{\sqrt{3}}{2} \frac{\sqrt{u_0^2-2 u_0 r_+ - u_0 u_1^3}}{u_0}  \, , \quad
  \phi_1 = \frac{\sqrt{3}}{4} \frac{\sqrt{u_1^2-2u_1 r_+ - u_0 u_1^3}}{u_1} \, ,
\end{equation}
in agreement with the values obtained earlier in (\ref{phivalsreg}) and (\ref{bhtemp})  by requiring regularity of the Euclidean continuation. The corresponding equilibrium free energy is remarkably simple:
\begin{equation}
\label{Frplusrel}
 F_{\rm eq}=\frac{r_+}{2} \,. 
\end{equation} 
This can also be obtained as the on shell Euclidean action $I_E=F/T$, provided the action is defined with the appropriate boundary counterterms, as in \cite{Hertog:2004dr}. Note that this simple expression suggests a nice interpretation of the radial coordinate $r$. Roughly, it is to free energy what the Schwarzschild radial coordinate is to entropy. We can also give a more physical interpretation to the parameters $u_0$, $u_1$ by noticing that at equilibrium $u_0 = 3 Q_0/\phi_0$, $u_1 = P^1/4 \phi_1$. This shows that $u_0$ and $u_1$ can be thought of as the black hole's D0- and D4-charge susceptibilities. 

For the system-reservoir equilibrium to be stable under small fluctuations, we need a positive definite Hessian, that is $F''>0$ at fixed $T,$ and $\phi_I$. Stability under arbitrarily large fluctuations requires the minimum to be global.  

Note that although we are analyzing stability in this (partial) grand canonical ensemble, this does not necessarily mean we are actually considering a physical situation in which the system is truly coupled to a reservoir. Indeed, in the case of global AdS black holes (dual to thermal states of a CFT$_3$ living on a 2-sphere), it is physically most natural to consider the physical system to be isolated, since there is no natural ``outside'' environment for the 2-sphere. However even for isolated systems, a grand canonical stability analysis provides information. More specifically, an instability in the grand canonical ensemble will, for sufficiently large isolated systems, indicate a thermodynamic tendency  towards the formation of inhomogeneities in the distribution of the energy and charge. Essentially, for a subsystem small compared to the complete system, this is because the remainder of the system acts as a reservoir. In view of the fact that instabilities towards the formation of inhomogeneities is exactly what we want to investigate in this paper, this is therefore an appropriate ensemble to consider.\footnote{By the same token, it would actually have been even more appropriate for us to consider the ensemble where all charges are allowed to fluctuate, including $P^0$ and $Q_1$. Unfortunately this is obstructed by the lack of explicit black holes solutions for the general charge case.}

\subsection{Schwarzschild illustration}

\begin{figure}
\centering{
\includegraphics[height=50mm]{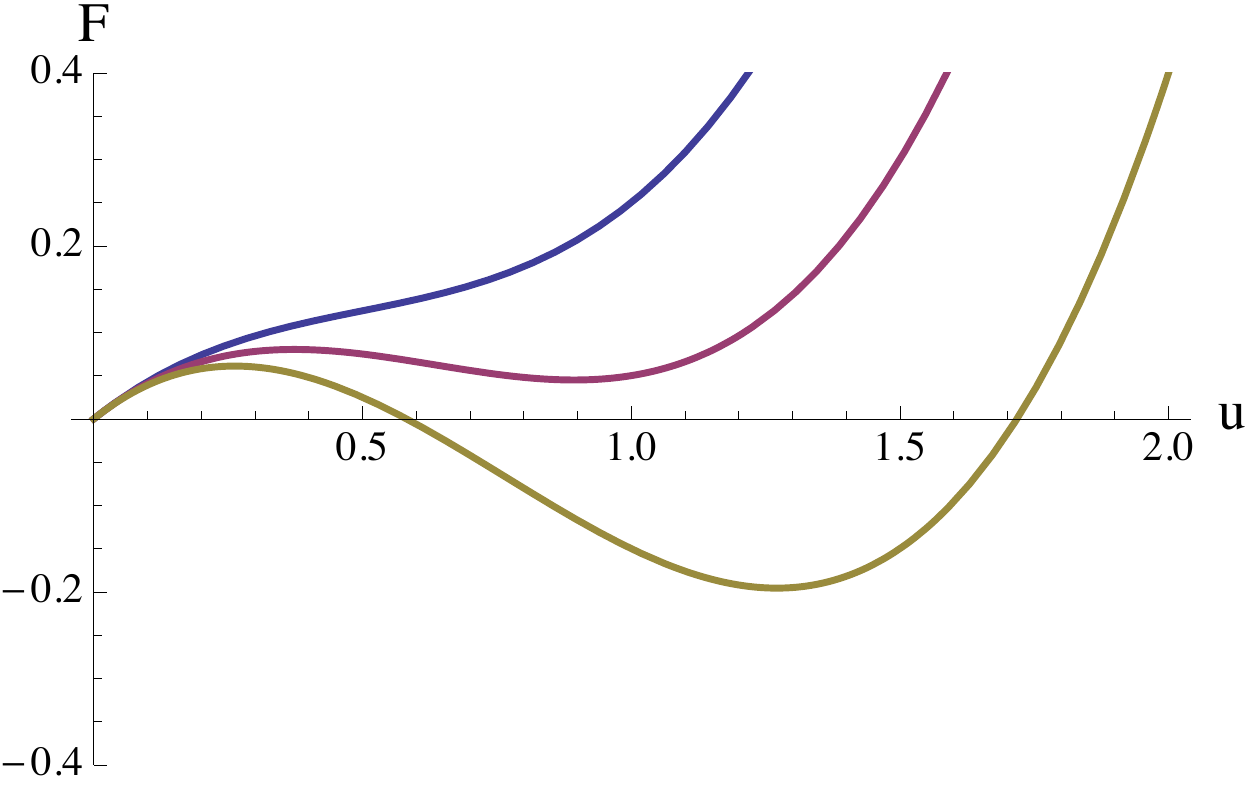}
}
\caption{AdS-Schwarzschild free energy $F$ for a black hole of size $u$ coupled to a heat bath at temperatures (from left to right) $\pi T = 0.75, 0.95, 1.15$. A local minimum corresponds to a perturbatively stable black hole, which is globally stable if it is negative. A local maximum  corresponds to a perturbatively unstable black hole. 
\label{fig:AdSSchw}}
\end{figure}

As a simple check and illustration of the above discussion, consider first the AdS-Schwarzschild black hole (fig.\ \ref{fig:AdSSchw}). 
This amounts to setting $u_0=u_1 \equiv u$ and $r_+=\frac{1}{2}(u-u^3)$, so $S=\pi u^2$, $M=\frac{1}{2}(u+u^3)$, and:
\begin{equation} \label{FSchw}
F=M-T S  = \frac{1}{2}(u+u^3) - \pi T u^2 \, .
\end{equation}
The local equilibrium condition (\ref{stableeq}) is $\partial_u F=\frac{1}{2} + \frac{3}{2} u^2 - 2 \pi T u=0$ and $\partial_u^2 F=3 u - 2 \pi T > 0$. The first equation expresses the equilibrium temperature in terms of $u$: $T_{\rm eq}(u)=\frac{1}{4 \pi}(u^{-1}+3u)$. Plugging this value for $T$ into (\ref{FSchw}) gives $F_{\rm eq} = \frac{1}{4}(u-u^3) = \frac{1}{2} r_+$, confirming (\ref{Frplusrel}). The minimum value of $T_{\rm eq}(u)$, reached at $u=1/\sqrt{3}$, is $T_{\rm min}=\sqrt{3}/2\pi$; there are no black holes at temperatures below this. For any given $T > T_{\rm min}$, there are two solutions $u$ to the equilibrium equation, hence two black hole solutions. The larger one will be at a local minimum of $F(u)$ ($F''(u)>0$), the smaller one at a local maximum.  
The local minimum of $F(u)$ is not necessarily a global minimum. To verify global minimality, we also have to compare to the free energy at the boundary points of state space, in this case at $u=0$. From the third expression in (\ref{FSchw}), it follows that for any value of $T$, we have $F=0$ when $u=0$.\footnote{This is true in the classical gravity approximation $N \to \infty$ where $N$ was defined in (\ref{eq:scalingsymm}). At one loop, there will be a contribution from thermal fluctuations, capturing the free energy of an ideal thermal gas in global AdS, but this will be of order $1$ in a large $N$ expansion, and hence negligible to leading order.} Therefore global stability requires $F_{\rm eq}<0$. This is the case if and only if $u>1$. Hence a first order phase transition occurs at $u=1$, where $T_{\rm eq}=1/\pi$. This was first pointed out by Hawking and Page \cite{Hawking:1982dh}. The transition is accompanied by a macroscopic jump in mass and entropy in the large $N$ limit and can thus be considered to be a first order phase transition. In the context of the AdS-CFT correspondence, it can be interpreted as a confinement-deconfinement phase transition occurring on the sphere at a temperature of the order of the inverse curvature radius \cite{Witten:1998zw}.

\subsection{Background phase diagram} \label{sec:phasediag}

\begin{figure}[t]
\centering{
\includegraphics[height=80mm]{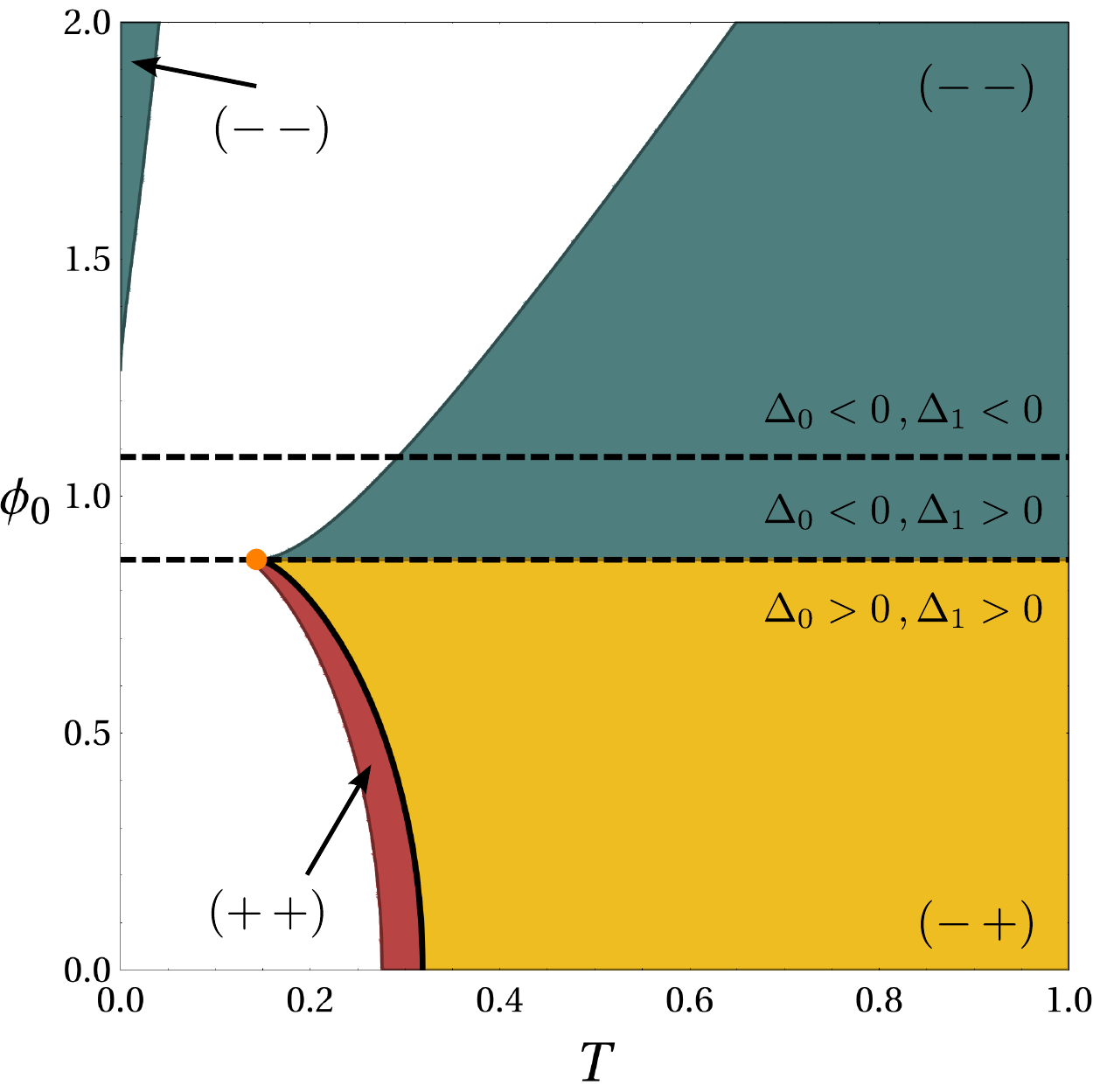}
\includegraphics[height=80mm]{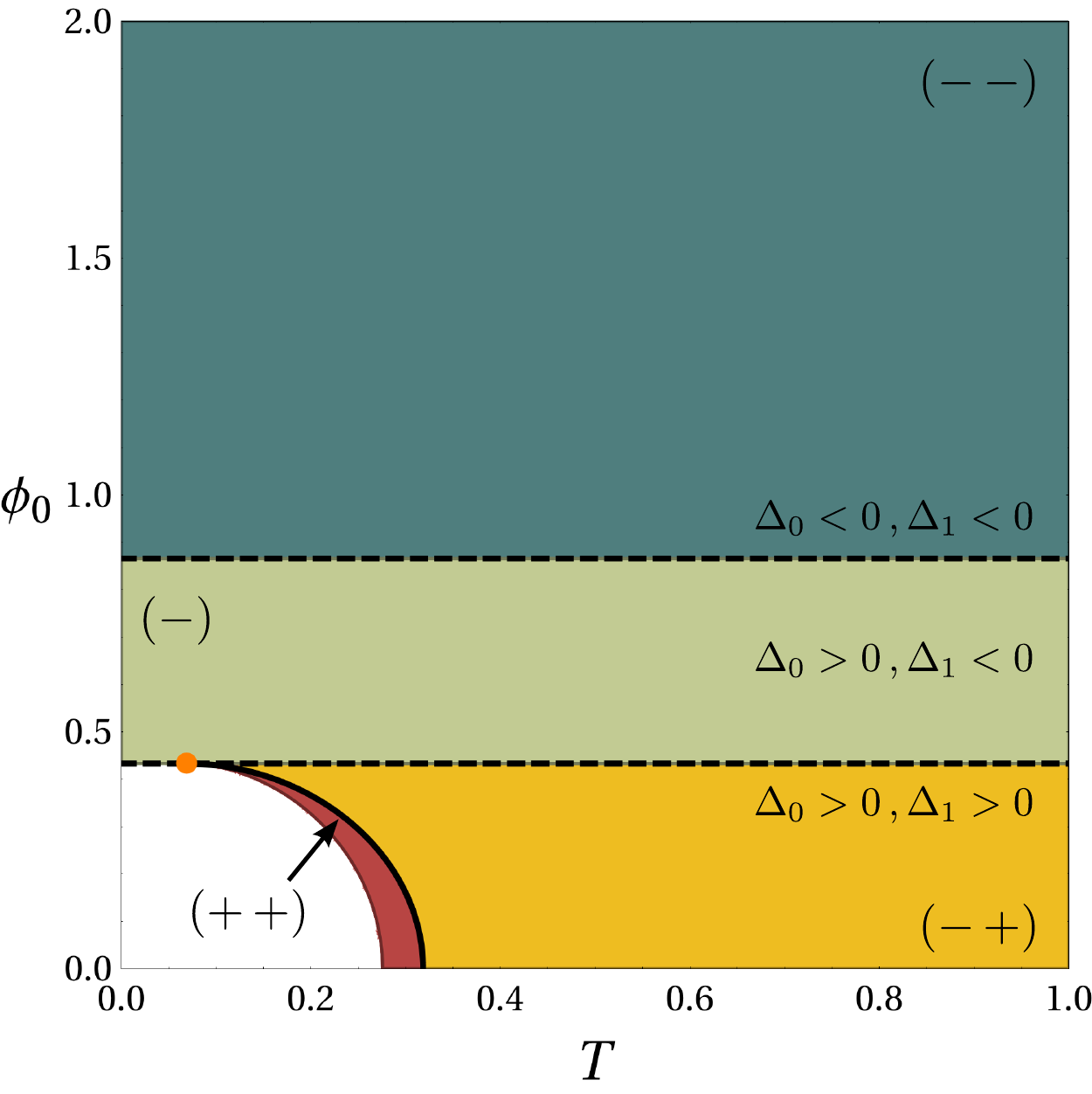}
}
\caption{Phase diagrams for the black hole background. On the left we have $\Phi_1=0.4 \,\Phi_0$ and on the right $\Phi_1=\Phi_0$. The different regions are labeled by a the signs of the free energies of the black hole solutions in the region.  For example $(-\,+)$ is a region with two black holes, one with negative and one with positive free energy, while $(-)$ indicates a region with just one black hole, with negative free energy. Across the dotted lines either $\Delta_0$ or $\Delta_1$ changes sign. 
The white regions represent configurations where no black holes exist. The Hawking-Page transition occurs at the thick black line, terminating in the orange dot. \label{fig:BHPhasediag}}
\end{figure}

\newcommand{\bp}{\bar{\phi}} 

Figure~\ref{fig:BHPhasediag} shows the phase diagrams in the $(\Phi_0, T)$ plane, for two different fixed $\Phi_1/\Phi_0$ ratios. The diagrams are obtained by solving (\ref{Tphi}) for $r_+$, $u_0$ and $u_1$. For the $\phi_0$ and $\phi_1$ equations this can be done in a relatively simple closed form:
\begin{equation} \label{rplusu0sol}
 r_+ = \frac{u_1}{2} \frac{\Delta_0 \Delta_1-u_1^4}{
   \Delta_0+ u_1^2 } \, ,  \quad  
 u_0 = u_1 \, \frac{\Delta_1+u_1^2 }{\Delta_0+ u_1^2} \, , \quad \mbox{where} \quad
 \Delta_0 \equiv 1 - \frac{4}{3} \phi_0^2, \quad \Delta_1 \equiv 1 - \frac{16}{3} \phi_1^2 \, ,
\end{equation}
The remaining relation to be inverted is 
\begin{equation} \label{Teq}
 T = \frac{\Delta_0 + 3 u_1^2}{4 \pi u_1} \sqrt{\frac{\Delta_1 + u_1^2 }{\Delta_0+ u_1^2 }}\, .
\end{equation}
This can be reduced to finding the roots of a cubic polynomial but as usual the explicit expression for the solutions is not illuminating. The charges and entropy in terms of $u_1$ and $\phi_0$, $\phi_1$ are
\begin{equation}
 Q_0 = \frac{u_1 \phi_0}{3} \frac{\Delta_1+ u_1^2 }{\Delta_0 + u_1^2 } \, ,
\quad
 P_1 = 4 \, u_1 \phi_1 \, , 
\quad
 S = \pi u_1^2 \sqrt{\frac{\Delta_1+ u_1^2 }{\Delta_0+ u_1^2 }} = \pi \frac{\sqrt{Q_0 P_1^3}}{\sqrt{\phi_0 \phi_1^3}} \, ,
\end{equation}
the free energy is $F=r_+/2$ with $r_+$ as in (\ref{rplusu0sol}), and the energy is 
\begin{equation} \label{massformulageneral}
 M = \frac{u_1}{4} \, \frac{4 u_1^2 + 2 u_1^4+ 3 \Delta_0 + \Delta_1 -2 \Delta_0 \Delta_1}{\Delta_0+ u_1^2 } \, .
\end{equation} 
Recall that the Reissner-Nordstrom limit corresponds to $u_0=u_1$, which implies $\Delta_0=\Delta_1$, or $\phi_1 = \phi_0/2$. 

We list some notable features: 
\begin{enumerate}

\item The temperature (\ref{Teq}) diverges for $u_1 \to \infty$, so at high temperatures there will always be at least one solution, with negative free energy. It is continuously connected to the large AdS-Schwarzschild black hole by tuning $\phi_0$ and $\phi_1$ to zero. As long as $\Delta_0$ and $\Delta_1$ are positive (corresponding to the region below the lower dotted line in the figure), the temperature goes infinite again when $u_1 \to 0$, providing a second high temperature solution branch. This solution is continuously connected to the small Schwarzschild black hole. It ceases to exist when crossing over to $\Delta_0<0$ or $\Delta_1<0$ (from below to above the (lower) dotted line in the figure), as the quantity under the square root then becomes negative for $u_1 \to 0$. When $\Delta_0>0$ and $\Delta_1<0$ (region between the dotted lines), there is only one high temperature solution. However when $\Delta_0 < 0$ (region above the (upper) dotted line), a new high temperature branch emerges for values of $u_1$ approaching the zero of the denominator, i.e.\ for $u_1^2 \to - \Delta_0$. In contrast to the small $u_1$ branch, it has negative free energy. 

\item In regions with two black holes, the one with the lowest free energy is locally stable ($F''>0$), the other one unstable. When there is a unique black hole solution, it is locally stable.
When crossing the dotted lines (corresponding to sign changes of the $\Delta_I$), the stable black hole always continues smoothly, whereas the unstable black hole becomes singular. Consider for example the case $\Delta_1>0$ with $\Delta_0$ small and negative. Putting $u_1=w\sqrt{-\Delta_0}$ and dropping subleading terms turns (\ref{Teq}) into $T \approx \frac{\sqrt{\Delta_1}(3 w^2-1)}{4 \pi w \sqrt{w^2-1}}$, which relates a finite fixed $w$ to a finite fixed $T$. Sending $\Delta_0$ up to zero at fixed $w$ thus corresponds to a black hole with $u_1 \to 0$, $u_0 \approx \frac{w \Delta_1}{\sqrt{-\Delta_0}(v^2-1)} \to \infty$, $r_+ \sim - \sqrt{-\Delta_0}  \to 0-$, $Q_0 \sim 1/\sqrt{-\Delta_0} \to \infty$, and $M \sim 1/\sqrt{-\Delta_0} \to \infty$. The scalar profile and geometry becomes singular in this limit; for instance at the horizon we have $y=\sqrt{u_0/u_1} \sim -1/\Delta_0 \to \infty$. 
 \label{itemcont}

\item The white gaps in the plot occur when the black hole free energy at fixed reservoir temperature and potentials fails to have a local extremum as a function of the extensive variables $(r_+,u_0,u_1)$, the analog of the upper curve in fig.\ \ref{fig:AdSSchw}. In this case none of the family of black holes we consider can exist in equilibrium with the reservoir. When crossing over into a white gap a stable and an unstable saddle point of the free energy coalesce and disappear. At the boundary the Hessian $F''$ develops a zeromode and $\det F'' = 0$. It can be checked that $\det F'' \propto (3 \, u_0 + u_1 - 4 \,r_+)(2 \, u_0 u_1^5+r_+(u_0-u_1) u_1^2 - 2 \, r_+^2)$, up to factors that remain positive throughout; this provides the boundaries of the white gaps.

\item For the white gaps below the dotted line ($\Delta_0, \Delta_1 > 0$), a Hawking-Page transition occurs before reaching the  gap. This is indicated by the thick line forming the boundary between the yellow and red regions. In the red region the free energy still has a local minimum, but it is positive, so the black holes we consider are thermodynamically disfavored compared to a thermal gas in empty AdS. This is the analog of the middle curve in fig.\ \ref{fig:AdSSchw}. The transition temperature $T_{\rm HP}$ is obtained by solving $F=\frac{1}{2} r_+=0$, which gives $u_1=(\Delta_0 \Delta_1)^{1/4}$ and 
\begin{equation}
 T_{\rm{HP}}=\frac{\sqrt{\Delta_0}+3\sqrt{\Delta_1}}{4\pi} \, .
\end{equation} 
which is real if $\Delta_0,\Delta_1>0$. On the Reissner-Nordstrom locus, we have $\Delta_0=\Delta_1$ and this becomes $T_{\rm HP} = \sqrt{\Delta_0}/\pi$, reproducing \cite{Chamblin:1999tk}. For neutral black holes we have $\Delta_0=\Delta_1=1$ and $T_{\rm HP}=1/\pi$, reproducing \cite{Hawking:1982dh}.

\item When  $\Delta_0 < \Delta_1$, as is the case in the figure on the left, there is also a white gap above the dotted line, i.e.\ for $\Delta_0<0$. The instability associated to it is of a very different nature than the Hawking-Page instability. It is still true that the disappearance of black hole solutions is due to the coalescence and then disappearance of a pair of saddle points of the free energy (\ref{Furp}) (one locally stable, the other one unstable), but now this happens for saddle points at a \emph{negative} value of $F$, so the thermodynamically preferred state cannot possibly be that of a thermal gas in empty AdS (which has $F=0$). Indeed there is a much more violent instability in this regime: whenever $\phi_0>\frac{\sqrt{3}}{2}$, the free energy (\ref{Furp}) is unbounded below, with a runway in the large $u_0$ direction. To see this, it is convenient to first eliminate $r_+$ in favor of the charge $P_1=\partial_{\phi_1} F = \sqrt{3} \sqrt{u_1(u_1-2r_+) - u_0 u_1^3}$, in terms of which
\begin{equation}
 F = \frac{u_0}{4} \bigl(1 + 2 u_1^2 - \frac{2 \phi_0}{\sqrt{3}} \bigl[\bigl(1-\frac{u_1}{u_0}\bigr)\bigl(1+u_1^2\bigr) + \frac{P_1^2}{3 u_0 u_1} \bigr]^{1/2} \bigr) - \pi T \sqrt{u_0 u_1^3} - \phi_1 P_1 + \frac{P_1^2}{6 u_1} + \frac{u_1}{4} \, .
\end{equation}
In the large $u_0$ limit at fixed $u_1$ and $P_1$, the leading term is linear in $u_0$, with coefficient proportional to $1+ 2 u_1^2 - \frac{2 \phi_0}{\sqrt{3}} \sqrt{1+u_1^2}$. When $\phi_0 > \frac{\sqrt{3}}{2}$, this becomes negative for a range of $u_1$ values, implying the free energy is unbounded below in this regime. When brought in contact with an infinite reservoir, the system will soak up $Q_0$-charge without bound. For large systems in isolation, one expects a corresponding instability to formation of clumps with large $Q_0$ densities. In the limit of an infinitely large system (the planar limit, which will be detailed in section \ref{sec:planarlimit}), the system acts as an infinite reservoir for finite subsystems, and there again appears to be no limit on how large the charge accumulation can get. This would appear rather unphysical. However, in this limit the solution becomes singular, with the scalar $y$ and curvature growing without bound towards the black hole, outside the regime of validity of the 4d (truncated) supergravity approximation. Presumably, assuming the model has a UV completion, the runaway will therefore be cured by degrees of freedom beyond those considered in our setup. \label{upperwhitegap}.

\item The limit $\Delta_1 \to \Delta_0$ is subtle when $\Delta_0<0$. Naively, (\ref{rplusu0sol}) would seem to imply that the limiting solution is just the $u_0=u_1$ AdS-Reissner-Nordstrom black hole with constant scalar profile. This is indeed one of the limiting solutions, but it misses the solution branch with $u_1^2$ approaching $-\Delta_0$: From (\ref{rplusu0sol}) and (\ref{Teq}) it follows that with $\Delta_1-\Delta_0 \equiv \delta$ and $u_1^2+\Delta_0 \equiv \epsilon$ both small, we have $2 \pi T \approx \sqrt{-\Delta_0 (1+\delta/\epsilon)}$ and $u_0/u_1 \approx 1 + \delta/\epsilon \approx -(2 \pi T)^2/\Delta_0$. This is different from 1 in general so the limiting black hole will not be the RN solution and in particular it will have a nontrivial scalar profile. For $T<\frac{\sqrt{-\Delta_0}}{2 \pi}$, this black hole has lower free energy than the AdS-RN solution, for $T>\frac{\sqrt{-\Delta_0}}{2 \pi}$ it has higher free energy. When $T=\frac{\sqrt{-\Delta_0}}{2 \pi}$ the two solutions coincide with $u_0=u_1=\sqrt{-\Delta_0}$, and the Hessian degenerates. This is also the location where the white gap begins to open up when $\Delta_0<\Delta_1$.

\item The orange dot in the figure corresponds to the singular point $u_1 \to 0$ with either $\Delta_0=0$ and $T=\frac{3 \sqrt{\Delta_1}}{4 \pi}$ (as in the left panel of the figure) or $\Delta_1=0$ and $T=\frac{\sqrt{\Delta_0}}{4 \pi}$ (as in the right panel). When $\Delta_1=0$, $Q_0/P_1$ diverges, and when $\Delta_0=0$, $P_1/Q_0$ diverges. This results in singular limiting solutions, similar to the other degenerations we discussed.

 \end{enumerate}

\subsection{The flat space / small black hole limit} \label{sec:flatlimit}

The asymptotically flat space limit (analyzed in \cite{arXiv:1108.5821}) corresponds to taking $\N \propto \ell/\ell_p \to \infty$ keeping the original, unrescaled $Q_0$, $P_1$ and $M \ell_p$ fixed. From (\ref{rescaledext}) it can be seen that in terms of the rescaled variables we are working with here (which were indicated by tildes in (\ref{rescaledext})), this means we take $(Q_0,P_1,M) \sim \frac{1}{\N} \to 0$, or equivalently $(r_+,u_0,u_1) \sim \frac{1}{\N} \to 0$. From (\ref{Tphi}) it follows that for generic nonextremal black holes in this scaling limit we have $T \to \infty$ while the $\phi_I$ remain finite. This is evident as well from (\ref{rescaledpotentials})), as we are taking the limit with fixed physical temperature and potentials in Planck units. At any rate, since we can now drop terms of higher order in $u_1$ in expressions such as (\ref{Teq}), it becomes easy to invert the relations between intensive and extensive variables; in particular $Q_0=\frac{\phi_0}{12 \, \pi \, T} \sqrt{\frac{\Delta_1^3}{\Delta_0}}$, $P_1=\frac{\phi_1}{\pi T} \sqrt{\Delta_0 \Delta_1}$, $M=\frac{\sqrt{\Delta_0 \Delta_1^3}}{16 \pi T}(\frac{1}{\Delta_0}+\frac{3}{\Delta_1}-2)$, $S=\frac{\sqrt{\Delta_0 \Delta_1^3}}{16 \pi T^2}$, and $F =\frac{\sqrt{\Delta_0 \Delta_1^3}}{16 \pi T}>0$.  

From these expressions we see there is another limit which sends the extensive quantities to zero in the appropriate way, namely taking $(\Delta_0,\Delta_1) \sim \frac{1}{N} \to 0$ (hence $|\phi_0| \to \frac{\sqrt{3}}{2}$, $|\phi_1| \to \frac{\sqrt{3}}{4}$), keeping $T$, the physical temperature in AdS units, fixed. Curiously, from the flat space point of view, this is in fact an extremal limit, since the temperature goes to zero in Planck units: $T \ell_p \sim 1/N$. Indeed in this limit the entropy becomes $S=\pi \sqrt{\frac{2}{3} |Q_0 P_1^3|}$, reproducing the well-known flat space extremal D4-D0 entropy formula.


\subsection{The planar / large black hole limit} \label{sec:planarlimit}

\begin{figure}[t!]
\centering{
\includegraphics[width=0.99 \textwidth]{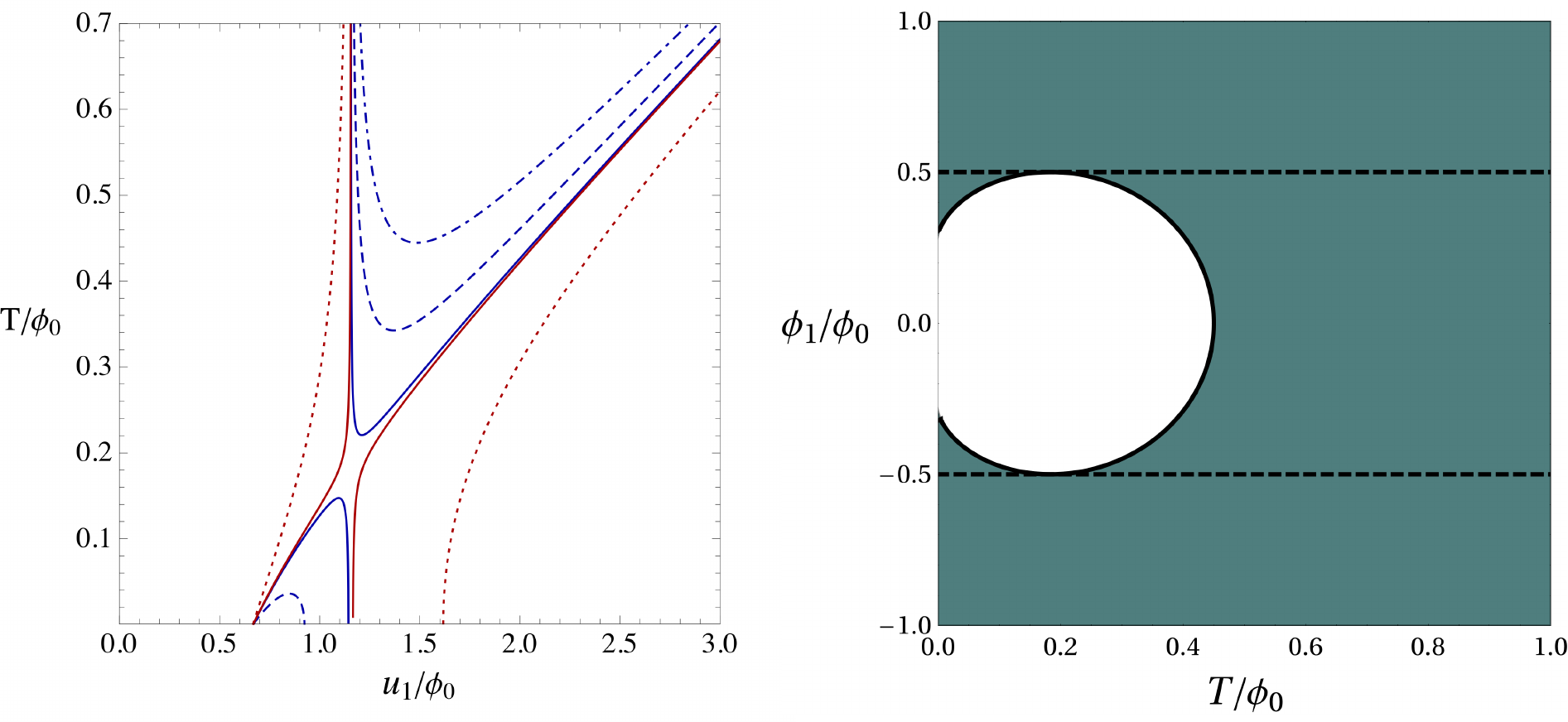}
}
\caption{{\bf Left}: Planar black hole temperature $T/\phi_0$ as a function of $u_1/\phi_0$, for $\phi_1/\phi_0 = 0.1, 0.4, 0.495, 0.505, 0.7$, corresponding respectively to the dash-dotted, dashed and solid blue curves, and to the solid and dotted red curves. Lines of constant $T/\phi_0$ intersect the curves in two points or not at all, illustrating that for given intensive variables, there are always either two black hole solutions or none at all. {\bf Right}: Planar black hole phase diagram. The colored region has two black holes, the white has none. It corresponds to the gaps in accessible temperatures for the curves on the left. The dotted lines denote the Reissner-Nordstrom locus, where one of the planar solutions has no scalar hair. In the white gap, the background becomes unstable to soaking up $Q_0$ charge as discussed in remark~\ref{upperwhitegap} in the previous section.\label{fig:planphasediag}}
\end{figure}

It is often simpler to work in a limit in which we can effectively replace the spherical $S^2$ black hole geometry by an ${\mathbb R}^2$ planar one. This is achieved by zooming in on a small solid angle of the geometry, say around the north pole, while simultaneously scaling up all extensive quantities. In the dual CFT this limit can be thought of as a thermodynamic limit in which the system of interest is living on a flat two-dimensional plane and in contact with a heat reservoir with which it can exchange energy and charge, through a far away boundary. 

The required scalings parallel those used in \cite{Chamblin:1999tk} in the RN case. Introducing a new radial coordinate $\rho>0$ related to the old one $r$ by $r = r_+ + \rho$, we put:
\begin{equation}\label{eq:planarscalings}
  u_0=\lambda\, \bar{u}_0 \, , \quad u_1=\lambda\, \bar{u}_1 \, , \quad r_+=\lambda^3\,\bar{r}_+ \, , \quad
   \rho = \lambda\,\bar{\rho} \, , \quad t = \bar{t}/\lambda \, , \quad \theta = \bar{\theta}/\lambda \, , \quad \phi = \bar{\phi} \, ,
\end{equation}
sending $\lambda \to \infty$ while keeping the barred quantities fixed. For the conformal boundary metric we thus get $d\Omega_2^2 = d\theta^2 + \sin^2 \theta d\phi^2 \to (d\bar{\theta}^2 + \bar{\theta}^2 d \bar{\phi}^2)/\lambda^2$. The quantity in brackets is the flat planar metric in polar coordinates; let $\bar{x},\bar{y}$ be the corresponding Cartesian coordinates.  Then in the limit $\lambda \to \infty$ the metric and scalar (\ref{metricansatz}) become\footnote{Explicit factors of $\ell$ or $\yz$ do not appear here because we are still working in the rescaled invariant coordinates of section \ref{scalingsymm}, including for the metric and coordinates.} 
\begin{equation} \label{planarmetricansatz}
 ds^{2}=-\bar{V} \, d\bar{t}^{2}+\frac{1}{\bar{V}} d\bar{\rho}^{2}+ \, \bar{W} (d\bar{x}^2 + d\bar{y}^2) \, ,\qquad y = \sqrt{\frac{\bar u_0+\bar \rho}{\bar u_1+\bar \rho}} \, ,
\end{equation}
where 
\begin{equation} \label{planarwarp}
 \bar{V} = \frac{2 \, \bar{r}_+ \bar{\rho} + (\bar{u}_0 + \bar{\rho})(\bar{u}_1 + \bar{\rho})^3  - \bar{u}_0 \bar{u}_1^3 }{\sqrt{(\bar{u}_0 + \bar{\rho})(\bar{u}_1 + \bar{\rho})^3}}\, , \quad
  \bar{W} = \sqrt{(\bar{u}_0 + \bar{\rho})(\bar{u}_1 + \bar{\rho})^3} \, .
\end{equation}
The gauge potentials (\ref{ABexpr}) remain unchanged, apart from the small $\theta$ expansion:
\begin{align} \label{ABexprplanar}
  \bar{A}^0 &=  \bar{\phi}_0 \bigl(\frac{\bar u_0}{\bar u_0 + \bar\rho} - 1 \bigr)  d\bar t &
  \bar{A}^1 &= -\frac{1}{4} \bar P_1 \bar{\theta}^2  d \bar \phi \, , \\
  \bar{B}_0 &= -\frac{1}{4} \bar{Q}_0 \bar{\theta}^2  d \bar \phi \, , &
  \bar{B}_1 &=  -\bar\phi_1 \bigl( \frac{\bar u_1}{\bar u_1+\bar\rho} - 1 \bigr) d\bar{t} \, .
\end{align}
Here we used the relations (\ref{phivalsreg}), $Q_0=u_0 \phi_0/3$ and $P_1=4 u_1 \phi_1$.
In fact the original spherical solution differs from this one only in that we have dropped a term $\bar{\rho}^2/\lambda^2$ in the numerator of $\bar{V}$. 
Under this scaling we have $M\sim \lambda^3$, $Q_0 \sim \lambda^2$, $P_1\sim\lambda^2$, $\phi_I \sim \lambda$, $T \sim \lambda$. In the global phase diagram discussed in section \ref{sec:phasediag}, the planar limit thus corresponds to going along diagonal rays out to infinity. 
Analogous to (\ref{eq:planarscalings}) we can introduce barred quantities for these physical variables. These satisfy largely the same relations as the unbarred quantities in section \ref{sec:phasediag}, except that the constant +1 drops out in the relation between $\phi_I$ and $\Delta_I$ in (\ref{rplusu0sol}), and that the lower order terms drop out in the expression for the mass in (\ref{massformulageneral}), so that in fact $\bar{M} = -\bar{r}_+ = - 2 \bar{F}$. Since the mass must be positive, the free energy of planar black holes must be negative. Similarly, in (\ref{Tphi}), the quadratic terms $u_0^2$ and $u_1^2$ under the square roots in the expressions for the potentials drop out in the planar limit. Due to the rescalings, we should consider $\bar M$, $\bar P^1$ and $\bar Q_0$ to be energy and mass densities per unit area.

In what follows we will drop the bars in the notation for the rescaled planar variables; whenever planar black holes are considered, all quantities are understood to be rescaled as indicated above. \def\bar{{}}

For later convenience, let us recapitulate. The energy and mass densities are given in terms of the parameters of the black hole solutions by
\begin{equation} \label{planarextensive}
 \bar M=-\bar r_+ \, , \quad 
 \bar Q_0 = \frac{1}{2 \sqrt{3}} \sqrt{-2\bar u_0 \bar r_+-\bar u_0 \bar u_1^3} \, , \quad
 \bar P^1 = \sqrt{3} \sqrt{-2\bar u_1 \bar r_+ - \bar u_0 \bar u_1^3} \, , \quad
 \bar S = \pi \sqrt{\bar u_0 \bar u_1^3} \, .
\end{equation}
The equilibrium values of the intensive quantities are given by
\begin{equation} \label{Tphi2}
 \bar T = \frac{2 \bar r_+ + 3\bar u_0 \bar u_1^2+\bar u_1^3}{4\pi\sqrt{\bar u_0 \bar u_1^3}} \, , \quad
 \bar \phi_0 = \frac{\sqrt{3}}{2} \frac{\sqrt{-2 \bar u_0 \bar r_+ - \bar u_0 \bar u_1^3}}{\bar u_0}  \, , \quad
  \bar \phi_1 = \frac{\sqrt{3}}{4} \frac{\sqrt{-2\bar u_1 \bar r_+ - \bar u_0 \bar u_1^3}}{\bar u_1} \, .
\end{equation}
The energy and charge densities can be obtained from the temperature and potentials  by eliminating $r_+$, $u_0$ and $u_1$ from the above equations. This can be reduced to solving $\bar T = \frac{\bar \Delta_0 + 3 \bar u_1^2}{4 \pi \bar u_1} \sqrt{\frac{\bar \Delta_1 + \bar u_1^2 }{\bar \Delta_0+ \bar u_1^2 }}$ for $\bar u_1$, where $\bar \Delta_0\equiv - \frac{4}{3} \bar \phi_0^2$ and $\bar \Delta_1\equiv - \frac{16}{3} \bar \phi_1^2$. This equation has zero or two solutions $\bar u_1$, from which we then get the extensive variables:
\begin{equation}
 \bar M = \frac{
\bar u_1}{2} \, \frac{\bar u_1^4-\bar \Delta_0 \bar \Delta_1}{\bar \Delta_0+ \bar u_1^2 } \, ,
 \quad \bar Q_0 = \frac{\bar u_1 \bar \phi_0}{3} \frac{\bar \Delta_1+ \bar u_1^2 }{\bar \Delta_0 + \bar u_1^2 } \, ,
  \quad \bar P_1 = 4 \, \bar u_1 \bar \phi_1 \, ,\quad 
  \bar S = \pi \bar u_1^2 \sqrt{\frac{\bar \Delta_1+ \bar u_1^2 }{\bar \Delta_0+ \bar u_1^2 }} \, .
\end{equation}
It is possible to write a polynomial relation between entropy, energy and charge densities, which can be viewed as the black brane equation of state:
\begin{equation}
 \bigl(\bar S^2+12 \, \pi^2 \bar Q_0^2\bigr) \bigl(3 \, \bar S^2 + \pi^2 \bar P_1^2\bigr)^3 = 
 432 \,\pi^6  \, \bar M^4 \, \bar S^2 \, .
\end{equation}
Notice that we get (by construction) an additional scaling symmetry $\bar{X} \to \lambda^{n_3} \bar{X}$ besides those listed in section \ref{scalingsymm}, with scaling exponents $k$ given by
\begin{center}
\begin{tabular}{|c|cc|ccc|cccc|ccc|c|} 
 \hline 
  & $\yz$ & $\ell$ & $\bar{r}_+$ & $\bar{u}_0$ & $\bar{u}_1$ & $\bar{M}$ & $\bar{Q}_0$ & $\bar{P}^1$ & $\bar{S}$ & $\bar{T}$ & $\bar{\Phi}_0$ & $\bar{\Phi}_1$ & $\bar{\rho}$ \\
  \hline
 $n_3$  & 0  & 0 &  3 & $1$ & $1$  &  3 & 2 & 2 & 2 & $1$ & 1 & 1 & 1  \\
    \hline
\end{tabular}
\end{center}
This scaling is that of a CFT in a 2d box of fixed size $L$, in the limit that $T$ and the $\phi_I$ are all much larger than the IR cutoff $1/L$ imposed by the box. Thermodynamic quantities will only depend nontrivially on scale invariant ratios. This allows us to plot the full planar phase diagram in terms of the two scale-invariant variables, for example $T/\phi_0$ and $\phi_1/\phi_0$ as shown in the panel on the right of figure~\ref{fig:planphasediag}. 
 
\begin{figure}[t!]
\centering{
\includegraphics[width=0.47 \textwidth]{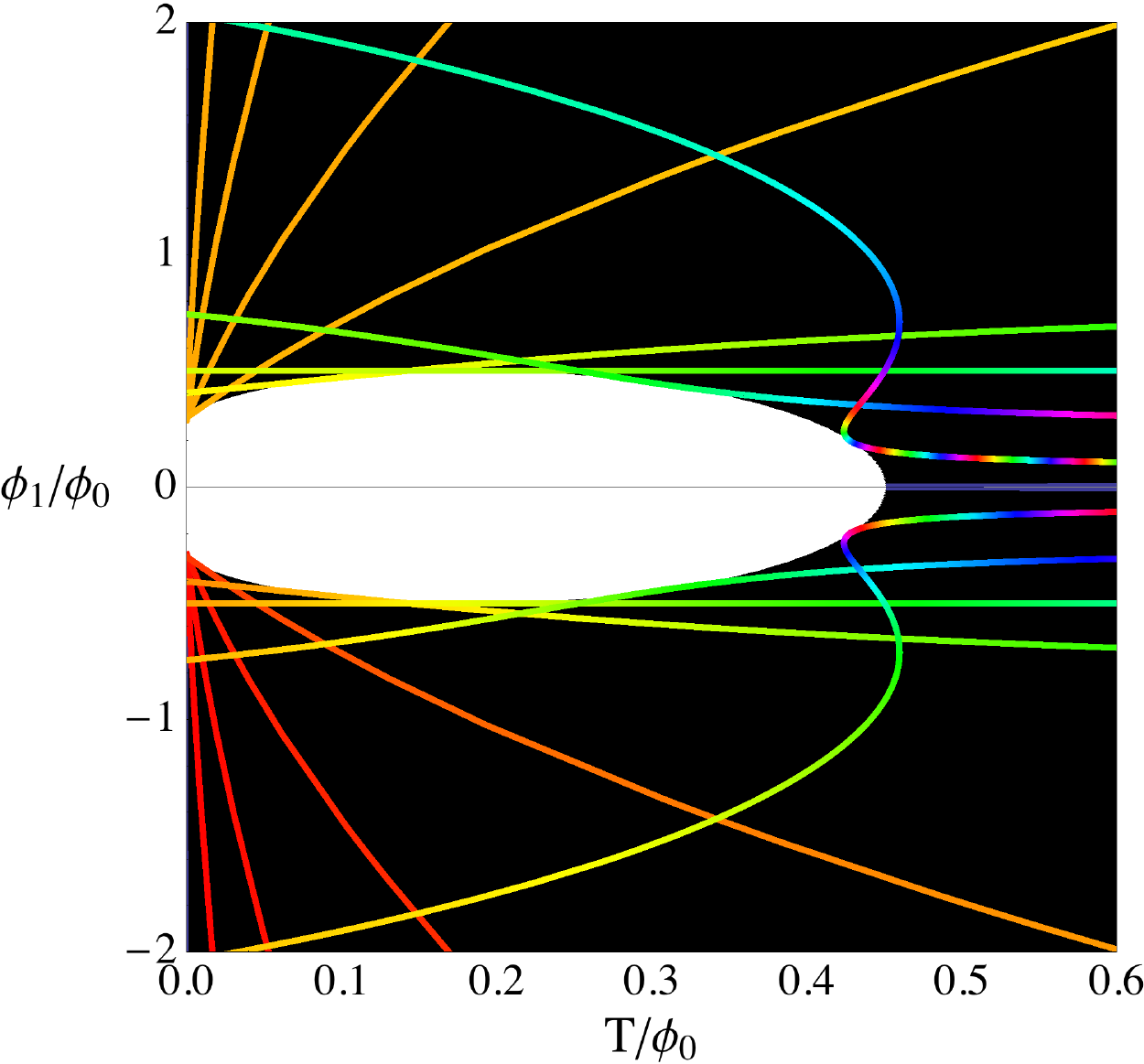}
\includegraphics[width=0.49 \textwidth]{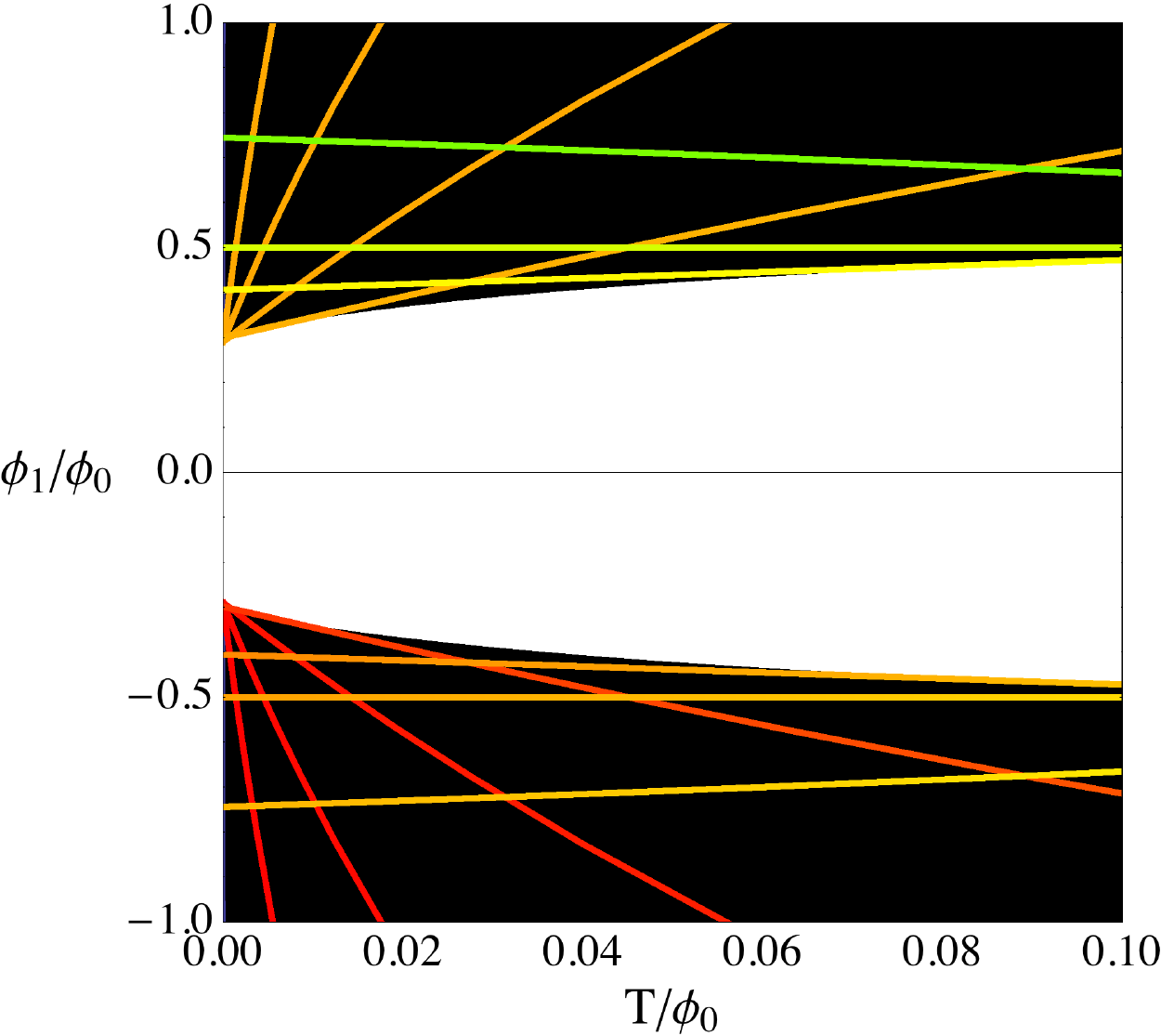}
}
\caption{Lines of constant charge for $\bar{P}^1=1$, $\pm \bar{Q}_0 = 10^{-5},$ $10^{-4}$, $10^{-3}$, $10^{-2}$, $10^{-1}$, $1/6$, $0.316$, $1$, with the larger values of $|\bar{Q}_0|$ being closest to the $\phi_1=0$ axis at high temperatures. The value $\bar{Q}_0 = 1/6$ corresponds to the Reissner-Nordstrom solution. In the lower half of the plane, the hue of the lines goes up according to entropy (going up in red to yellow direction), while in the upper half of the plane, the mass (=free energy) is indicated in this way. The lower values of $|\bar{Q}_0|$ have the lower free energy and entropy. The stable and unstable branches connect at the boundary of the white gap. \label{fig:equichargelines}}
\end{figure}

\subsection{Hyperscaling violating limits} \label{sec:hsvl}

Upon setting $P_1/Q_0$ or $Q_0/P_1$ to zero, as was the case for most degenerations discussed in section \ref{sec:phasediag}, our planar backgrounds reduce to the hyperscaling violating geometries studied in \cite{Ogawa:2011bz,Huijse:2011ef,Iizuka:2012pn,Shaghoulian:2011aa,Dong:2012se} and other recent works. These are characterized in general by a dynamic critical exponent $z$ and a hyperscaling violation exponent $\theta$, parametrizing the radial scaling behavior of the metric (cf.\ eq.\ (1.1) of \cite{Huijse:2011ef}). 

To see this, we fix the temperature $T$ and use (\ref{Tphi}) to write $r_+ = -\frac{3}{2} u_0 u_1^2 - \frac{1}{2} u_1^3 + 2 \pi T \sqrt{u_0 u_1^3}$, and obtain from (\ref{planarwarp})
\begin{equation}
 V =  \frac{4 \pi T \sqrt{u_0 u_1^3} \, \rho + 3 (u_0u_1+u_1^2) \, \rho^2 + (u_0+3 u_1) \, \rho^3 + \rho^4}{W} \, , \qquad W = (\rho+u_0)^{1/2} (\rho+u_1)^{3/2} \, .
\end{equation}
For finite nonzero $u_0$ and $u_1$, the solution is regular; in particular when $T = 0$ it has an AdS$_2 \times {\mathbb R}^2$ near horizon geometry. However if we send $u_1 \to 0$ then for $\rho \ll u_0$:
\begin{equation}
ds^2 = - u_0^{1/2}  \rho^{3/2} dt^2 + \frac{d\rho^2}{ u_0^{1/2}  \rho^{3/2}} + u_0^{1/2} \rho^{3/2} \left( dx^2 + dy^2 \right)~, \qquad y = \sqrt{u_0/\rho}~.
\end{equation}
This is a hyperscaling violating geometry with $\theta = -1$ and $z=1$. Similarly, if we send $u_0 \to 0$ then for $\rho \ll u_1$:
\begin{equation}\label{hsP1}
ds^2 = - 3 u_1^{1/2}  \rho^{3/2} dt^2 + \frac{d\rho^2}{ 3 u_1^{1/2}  \rho^{3/2}} + u_1^{3/2} \rho^{1/2} \left( dx^2 + dy^2 \right)~, \qquad y = \sqrt{\rho/u_1}~.
\end{equation}
This is a hyperscaling violating geometry with $\theta$ and $z$ tending to infinity with the ratio  $\eta \equiv -z/\theta = 1$ fixed. Notice that the above metric (\ref{hsP1}) is conformal to AdS$_2 \times \mathbb{R}^2$. These geometries were studied in the context of the $U(1)^4$ truncation of $\CN=8$ gauged supergravity in \cite{Donos:2012yi}.

To see what this limit corresponds to in our phase diagram, we use the various relations summarized in the previous section, obtaining $\phi_0 = \frac{\sqrt{3}}{2} \sqrt{3 u_1^2- 4\pi T \sqrt{u_1^3/u_0}}$ and $\phi_1 = \frac{\sqrt{3}}{4} \sqrt{u_1^2+2 u_0 u_1 - 4\pi T \sqrt{u_0 u_1}}$, and from this the charges $Q_0 = u_0 \phi_0/3$ and $P_1=4 u_1 \phi_1$. Notice that for these expressions to be real, and therefore the solution to be physical, $T$ must be bounded above for a given $u_0,u_1$. Specifically when $u_1 \to 0$, we need $4 \pi T<\sqrt{u_0 u_1} \to 0$ and when $u_0 \to 0$, we need $4 \pi T < 3 \sqrt{u_0 u_1} \to 0$.

Thus, when $u_0 \to 0$ (metric (\ref{hsP1})), we get $\phi_0 \propto u_1$, $\phi_1 \propto u_1$ and $T \propto \sqrt{u_0 u_1}$, implying $T/\phi_0 \to 0$ while $\phi_1/\phi_0$ remains finite and tunable to any desired value satisfying $|\phi_1/\phi_0|>1/\sqrt{12}$. Hence this limit corresponds to the zero temperature boundary in the phase diagram fig.\ \ref{fig:planphasediag}. The charge ratio in this limit is $P_1/Q_0 \propto u_1/u_0 \to \infty$, that is the black hole becomes purely D4-charged in this limit.  

Similarly, when $u_1 \to 0$, we get $\phi_1/\phi_0 \to \infty$, while $T/\phi_1$ remains finite; this is the boundary at infinity in fig.\ \ref{fig:planphasediag}. The charge ratio is $P_1/Q_0 \propto \sqrt{u_1/u_0} \to 0$; the black hole becomes purely D0-charged in this limit.

Besides the $u_0 \to 0$ solutions we just described, there are also regular $T=0$ solutions with $u_0$ and $u_1$ finite that have AdS$_2 \times {\mathbb R}^2$ near-horizon geometries. Their free energy is $F=-\frac{4}{3} \phi_0 \phi_1^2+\frac{1}{27} \phi_0^3$, whereas the free energy of the $u_0=0$ solution is $F=-\frac{16}{3\sqrt{3}} \phi_1^3$. Away from the boundary point $\phi_1/\phi_0 = 1/\sqrt{12}$, the latter is always lower than the former, so the hyperscaling-violating geometry is always thermodynamically preferred. At the boundary point, the two solutions coincide. 

The entropy $S=\pi \sqrt{u_0 u_1^3}$ vanishes when $u_0=0$ or $u_1=0$. Hence we conclude that at $T=0$, the system under study has vanishing entropy in its thermodynamically preferred state; it does not suffer from the entropy anomaly typical for Einstein-Maxwell setups with scalar-independent couplings.

\subsection{Clumping instability} \label{sec:D0clumping}

The grand canonical ensemble has an instability for all values of the parameters, perturbative for some, nonperturbative for others, because at fixed temperature and potentials, the free grand canonical free energy $F=M-TS-\phi_0 Q_0-\phi_1 P^1$ is unbounded from below. The unbounded direction corresponds to infinite D0-charge density $Q_0$ keeping the D4-charge density $P^1$ fixed. This limit can be reached  e.g.\ by letting $u_0 \to \infty$ with $u_1 = c \, u_0^{-1/3}$ and $r_+ = -\frac{3c^2}{2}\, u_0^{1/3}$. Using (\ref{planarextensive}) this gives $Q_0 \approx \frac{c}{2} u_0^{2/3}$, $P^1 \sim \, c^{3/2}$, $M = - r_+ \sim P^1 \sqrt{|Q_0|}$, $S \sim c^{3/2}$, and (dropping irrelevant numerical factors):
\begin{equation} \label{FapproxlargeQ0}
  F \sim P^1 \sqrt{Q_0} - T P^1 - \phi_0 \, Q_0 - \phi_1 \, P^1 \, .
\end{equation}  
For $\phi_0 \neq 0$, this is unbounded below when $Q_0 \to \infty$. Thus, once the D0-density is sufficiently large, the system will be able to lower its free energy without bound by sucking in D0-charge from the reservoir. The local (wannabe) equilibrium values of the temperature and potentials scale as $T \sim P_1^2/Q_0^{3/2}$,  $\phi_1 \sim \sqrt{Q_0}$, $\phi_0 \sim P^1/\sqrt{Q_0}$. So we see that the local equilibrium chemical potential $\phi_0$ in fact \emph{decreases} with $Q_0$, in other words we get a negative capacitance, hence the runaway instability.\footnote{Somewhat different limits can be considered to produce different asymptotics of the temperature and potentials, but the feature of having $d\phi_0/dQ_0 < 0$ persists.} 

On the other hand, as suggested by (\ref{FapproxlargeQ0}) and as can be checked more generally, for smaller values of $Q_0$, the free energy slope goes the other way, towards zero $Q_0$. So for sufficiently small values of $Q_0$, the instability is nonperturbative; a free energy density barrier must be overcome before the runaway regime is reached.

Notice that for the D4-charge density there is no such instability. If we similarly take $P^1 \to \infty$ while keeping $Q_0$ fixed (which requires scaling $u_0 \sim u_1^{-3}$ and $r_+ \sim - u_1^3$), we obtain $S \sim 1$, $M \sim |P_1|^{3/2}$, so $F \sim |P_1|^{3/2} - \phi_1 P^1+ \cdots$, which is bounded and stable. The crucial difference between the two is the asymptotic growth of the energy with the charge, which has an exponent $1/2 < 1$ for the D0 and $3/2 > 1$ for the D4 charge.

To have a physically more stable setup, we could therefore work for example at fixed temperature and fixed charges rather than at fixed temperature and fixed chemical potentials. This will eliminate the runaway charge transfer from the reservoir, but nevertheless there will be a remnant in the form of a clumping instability, i.e.\ towards formation of inhomogeneities. This is because we can consider any finite subregion of our black brane horizon to be a system held at fixed potentials, with the remainder of the brane playing the role of reservoir. The D0-charge accumulation instability will now correspond to a thermodynamic instability towards accretion of D0-charge in the subregion. 

This can be seen more directly. At fixed temperature and fixed total brane charges, the thermodynamically preferred equilibrium state is the state that minimizes the canonical free energy $F_c = M-TS$. This includes minimization over possible inhomogeneities in the charge and energy densities. Now imagine concentrating a total amount of charge $Q_{0,\rm tot}$ in some finite area $A$, giving a charge density $Q_0=Q_{0,\rm tot}/A$, and let us assume we are in the high D0-charge density regime described earlier. Keeping the D4-charge density $P^1$ fixed, we thus obtain a region with a canonical free energy density $F_c \sim \sqrt{Q_0} \sim \sqrt{Q_{0,\rm tot}}/\sqrt{A}$. The contribution to the \emph{total} free energy of this region is therefore $\Delta F_{c,\rm tot} \sim \sqrt{Q_{0,\rm tot}} \sqrt{A}$, which becomes smaller when $A \to 0$. It is thus thermodynamically favored to concentrate the charge $Q_{0,\rm tot}$ into an ever smaller area $A \to 0$, since in addition also the surrounding region will decrease its free energy in this way (as it lowers its D0-charge density). Thus, according to this simple-minded thermodynamic picture, the initially homogeneous D0-charge will tend to implode into point-like chunks. 

This analysis is of course rather crude. We have not taken into account possible quantum or stringy corrections, which become important in the singular limit under consideration, and may well regulate the singularities. We have not taken into account density gradient contributions to the energy, which would give rise to bubble wall tensions and may also regulate singularities. Finally, even within these approximations, we have fixed by hand the D4-charge density, but in general this density will also run. This may lead for instance to a complete separation of charge, with pure D0-dots inside a pure D4-sea. In any case, to determine the true final state, a more detailed analysis is clearly in order, but this falls outside the scope of this paper.

\section{Bound states}\label{hhads}

We now proceed to establish the existence of bound states of these black holes with suitably charged probes. The probes are assumed much heavier than the AdS scale, and in particular they can be black holes themselves, as long as they are much smaller than the length scales set by the background solution. We compute the probe potentials from (\ref{probeaction}); a local minimum indicates a bound state. We take the probe potential to be zero at the horizon, so negative/positive values of the potential energy indicate stable/metastable bound states. On the other hand, since the probes are massive, an escape to infinity would require an infinite amount of energy; the global AdS metric acts as a confining box. This is a significant difference with the asymptotically flat case studied in \cite{arXiv:1108.5821}. 

Most of our analysis is numerical. We provide some analytic results in the planar zero temperature limit in section \ref{sec:anal}. 

\subsection{Probe potential and validity of the approximation} \label{sec:probevalidity}

Consider a probe with (D6,D4,D2,D0)-brane charge $(p^0,p^1,q_1,q_0)$. In the spirit of section \ref{scalingsymm} it will be convenient to introduce rescaled charges
\begin{equation} \label{probehats}
 \hat{p}^0 = \yz^2 \, \frac{p^0}{g} \, , \quad 
 \hat{p}^1 = \yz \, \frac{p^1}{g} \, , \quad 
 \hat{q}_1 = \frac{q_1}{g}  \, , \quad 
 \hat{q}_0 = \frac{1}{\yz} \, \frac{q_0}{g} \, , 
\end{equation}
with $g$ an at this point arbitrary constant. This differs from the rescaling used for the background black hole charges (\ref{rescaledext}) in that there is no factor of $\N$ involved here; in its place we now have $g$, which we can think of as parametrizing the order of magnitude of the probe charges. We do this because we want to keep the quantized probe charges fixed and finite while taking the $\N \to \infty$ limit. Notice that since charge is quantized in order 1 units in our conventions, the hatted probe charges are quantized in units given by the above scaling factors. At fixed finite $\yz$, these can be made arbitrarily small by taking $g$ large,  making the rescaled charges effectively continuous. Furthermore, ratios of probe to background charges, masses and length scales will involve the rescaled variables (tilde-variables for the background, hatted variables for the probes) and a universal overall factor $\frac{g}{\N}$. For example $\frac{q_0}{Q_0} = \frac{g}{\N} \frac{\hat{q}_0}{\tilde{Q}_0}$ and, using (\ref{Dprobemasses}), $\frac{m_{p_0 D6}}{M} \sim \frac{g}{\N} \frac{\hat{p}^0}{\tilde{M}}$. The discussion in section \ref{sec:probeaction} implies that for order 1 rescaled probe charges, the probe black hole entropy will be of order $g^2 \yz^{-1}$, 
 hence the ratio of its linear size over the AdS length scale will be of order $g \yz^{-1/2} \ell_p/\ell = \frac{g}{\N}$. Thus, for finite rescaled variables, the probe approximation will be justified provided $g \ll \N$. 

The static potential $V_p$ obtained from (\ref{probeaction}) and the solutions described in section \ref{sec:msgp} consists of two parts, a gravitational part $V_{\rm grav}(r)= \sqrt{V(r)} \, m_\gamma(y(r))$ and an electromagnetic part $V_{\rm em} = q_I A^I - p^I B_I$. Explicitly
\begin{equation} \label{VpVp}
  V_p=\frac{g \N}{\ell \yz} \hat{V}_p \, , \qquad \hat{V}_p = \hat{V}_{\rm grav} + \hat{V}_{\rm em} \, ,
\end{equation}  
with:
\begin{equation} \label{probepotdef}
 \hat{V}_{\rm grav} = \frac{\sqrt{3}}{2} \sqrt{\left( \rho(\rho+2 r_+) + f_0 f_1^3-u_0 u_1^3 \right)\biggl[
  \biggl(\frac{\hat{p}_1}{2 f_1}  + \frac{\hat{q}_0}{f_0} \biggr)^2 + \frac{f_0}{f_1} \biggl(\frac{\hat{p}_0}{6\,f_1}   -\frac{\hat{q}_1}{f_0} \biggr)^2
 \biggr] } \, ~,
\end{equation}
and 
\begin{equation}\label{probepotdefel}
 \hat{V}_{\rm em} =  -\frac{\phi_0 \hat{q}_0 \rho}{f_0} - \frac{\phi_1 \hat{p}^1\rho}{f_1}    \, ,
\end{equation}
where as before
\begin{equation}
f_0 =  \rho + u_0 ~, \quad f_1 = \rho + u_1  \, , \quad \rho \equiv r - r_+ \, .
\end{equation}
The radial coordinate $\rho$ vanishes at the horizon. In the above expressions, the background variables are understood to be rescaled as in section \ref{scalingsymm}, but we have suppressed the tildes here.

In contrast to the background metric and scalar, the probe potential is qualitatively altered when flipping the sign of $Q_0$ or $P^1$. Because of this we have to consider both possible signs of $\Phi_1/\Phi_0$ separately.  Notice however that we still have the following symmetry:
\begin{equation} \label{D6D2symmetry}
 (\hat{p}^0,\hat{p}^1,\hat{q}_1,\hat{q}_0) \to (-\hat{p}^0,+\hat{p}^1,-\hat{q}_1,+\hat{q}_0) \, .
\end{equation}
This allows us to assume $\hat{p}^0 \geq 0$ without loss of generality. 

Finally let us check the claim made in section \ref{sec:probeaction} that from the probe point of view the background temperature is effectively zero. The fraction of the probe's energy that is thermal when it has the same temperature as the background is, for order 1 values of the rescaled variables, $T s_\gamma / m_\gamma \sim (g^2 \yz^{-1}) / (g \N \yz^{-1}) = g/\N$, so again if $g \ll \N$, the probe will effectively be extremal. 

In what follows we will mostly drop the hats (and tildes) in our notation, which is equivalent to setting $\ell \equiv 1$, $N \equiv 1$, $v \equiv 1$, $g \equiv 1$. To restore the factors $\ell$, $N$, $v$ and $g$ in equations, one should keep in mind the following scaling weights: $[\ell]=(1,0,-1,0)$, $[v] = (0,1,0,0)$, $[N] = (1,\frac{1}{2},0,0)$,  
$[g] = (0,\frac{1}{2},0,1)$.
The first two entries correspond to the weights $(n_1,n_2)$ for background quantities given in section \ref{scalingsymm}, the third one is the mass dimension, and the fourth one indicates nonzero only for quantities involving the probe; it indicates the scaling with the overall size (charge/mass) of the probe. For example the weights of the probe potential are $[V_p]=(0,0,-1,1)$, hence $V_p \propto g N/v \ell$. We will restore the original factors in the concluding sections.

\subsection{Thermodynamic interpretation} \label{sec:thermointerpr}

When a small probe charge is expelled from a black hole, the black hole entropy changes by an amount
\begin{equation}
 \delta S_{\rm BH} = \frac{1}{T} \delta E_{\rm BH} - \frac{\phi_0}{T} \delta Q_{0,\rm BH} - \frac{\phi_1}{T} \delta P^1_{\rm BH} \, .
\end{equation}
Here we used the microcanonical definitions of temperature and chemical potentials, taking into account that the potentials for D2 and D6 charge are zero. Conservation of charge implies $\delta Q_{0,\rm BH} = - q_0$ and $\delta P^1_{\rm BH} = - p^1$. Conservation of energy implies $\delta E_{\rm BH} = -E_p^{\rm tot}$, where $E_p^{\rm tot}$ is the sum of the probe's rest mass energy plus the binding energy due to the probe-black hole interaction. Up to an additive constant $E_0$ this equals the natural total energy $E_p$ obtained from the probe action given in section \ref{sec:probeaction}: 
\begin{equation} \label{probeE}
 E_p^{\rm tot} = E_p + E_0 \, , \qquad 
 E_p \equiv V_p + E^{\rm kin}_p \, ,
\end{equation} 
where $V_p$ is the probe potential derived there, and $E^{\rm kin}_p$ is the probe kinetic energy. The additive constant $E_0$ is easily obtained by considering a probe at rest asymptotically far away from the black hole. In this case there is no binding energy so $E_p^{\rm tot}$ is just the probe's gravitational rest mass energy $V_{\rm grav}$, defined in (\ref{probepotdef}). On the other hand in this situation we have $E_p=V_p=V_{\rm grav} -{q}_0 \phi_0 - {p}^1 \phi_1$,  as can be seen from (\ref{probepotdefel}). Hence $E_0 = q_0 \phi_0 + p^1 \phi_1$. Putting everything together, the constant term cancels with the other potential dependent terms in $\delta S_{\rm BH}$, leaving us with the simple result
\begin{equation} \label{deltaSBH}
 \delta S_{\rm BH} = -\frac{E_p}{T} \, , 
\end{equation}
where $E_p = V_p(\rho) + E^{\rm kin}_p$. The change in the total microcanonical entropy of the system for a given final state $|\alpha\rangle$ of the  probe viewed as a particle (here $\alpha$ is a one particle state label which includes charge and energy $E_p$) is thus 
\begin{equation} \label{deltaSnlm}
 \delta S|_{\alpha} = S_p-\frac{E_p}{T} \equiv -\frac{F_p}{T} \, ,
\end{equation}
where $S_p$ is the probe's internal entropy. Recall that $V_p/T \propto \frac{g \N}{v}$ while $S_p \propto \frac{g^2}{v}$, so 
in the probe limit $g \ll \N$, the probe's internal entropy contribution to $F_p$ is generically subleading. 

In the planar limit it is also natural to take the system ${\cal S}$ of interest to correspond to a finite (but parametrically large) part of the $xy$-plane, with the remainder of the plane viewed as the reservoir. In this case, by definition, $\delta S_{\rm tot} = -\delta F_{\cal S}/T$, and (\ref{deltaSnlm}) reduces to 
\begin{equation} \label{deltaFFp}
 \delta F_{\cal S} = F_p \, .
\end{equation}
We can now take the system size to infinity, and view this as a formula for the change of total free energy in the grand canonical ensemble. 

Thus, in equilibrium, the probability of finding a single probe in a given state $\alpha$ relative to the probability of having no probes is $e^{-F_\alpha/T}$. In particular we see that if the minimum of the probe potential is negative, ejecting such probes is thermodynamically  preferred at large $N$, while if it is positive, swallowing them is preferred. If $F_\alpha$ is positive for all possible probe charges, we get a cold, exponentially dilute gas in the large $N$ limit (so interactions can be neglected), with average occupation number of the 1-particle state $|\alpha\rangle$ given by
\begin{equation} \label{Nellm}
 \langle N_{\alpha} \rangle = e^{-F_\alpha/T} \, . 
\end{equation}
Alternatively these occupation numbers can be obtained by considering  the thermal atmosphere of the black hole as a statistical mechanical system in the grand canonical ensemble, with the black hole acting as a reservoir. We do not distinguish between Bose or Fermi statistics here because the gas is dilute (the average occupation number is $e^{-N}$ suppressed). 

The average number of probe particles of a given charge $\gamma$ is obtained (still in the dilute gas approximation) by summing this over all fixed charge 1-particle states, or semiclassically by integrating over the relevant phase space volume.\footnote{If extended all the way to the horizon $\rho=0$, this phase space volume is actually infinite due to the infinite redshift. Similarly, with Dirichlet boundary conditions at the horizon, the naive sum over quantum states is infinite, for the same reasons. At the same time, and related to this, the dilute gas approximation breaks down near the horizon, since $V_p(\rho) \to 0$ when $\rho \to 0$. Thus we can only make reliable statements for the average number of probes at separations larger than some IR cutoff $\rho_*$.} This is detailed in appendix \ref{app:degstates}. The final result for the semiclassical spatial number density of particles of charge $\gamma$ is (equation (\ref{mgammadensity}) in the appendix):
\begin{equation} \label{mgammadensity2}
 \langle n_\gamma(\vec x) \rangle =  \frac{1}{(2 \pi)^3} \, \frac{4 \pi W}{V^2} \, \sqrt{\frac{\pi}{2}}
  \left(V_{\rm grav,\gamma} T\right)^{3/2} \, \Omega(\gamma) \, e^{-V_{p,\gamma}/T} \, ,
\end{equation}
where $\Omega(\gamma) = 1$ if the probe is a structureless particle and $\Omega_p(\gamma) = e^{S_p(\gamma)}$ if the probe is a black hole. The expected total number of probes of charge $\gamma$ in a spatial region ${\cal R}$ is then given by $\int_{\cal R} d^3x \, \langle n_\gamma(\vec x) \rangle$.

\subsection{Probe bound states for spherical black holes}

\begin{figure}[h]
\centering{
\includegraphics[height=0.3\textwidth]{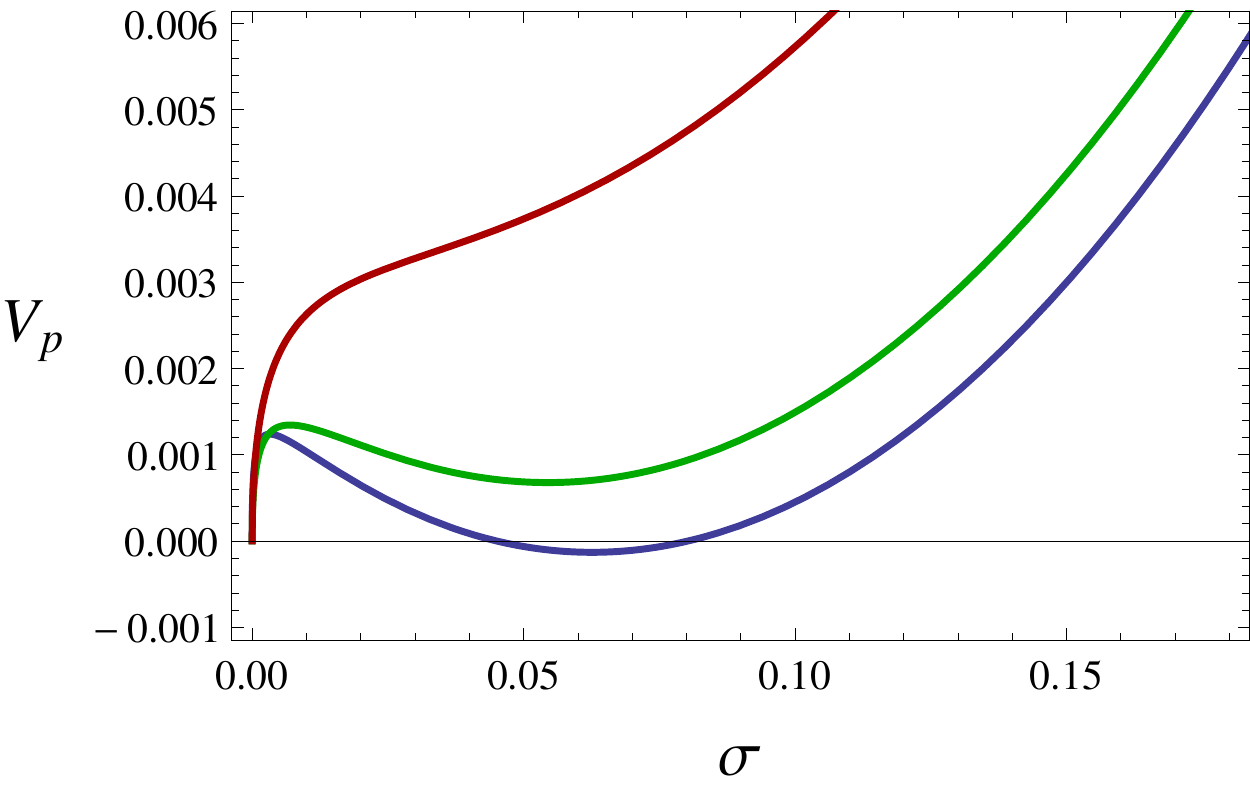}
\includegraphics[height=0.3\textwidth]{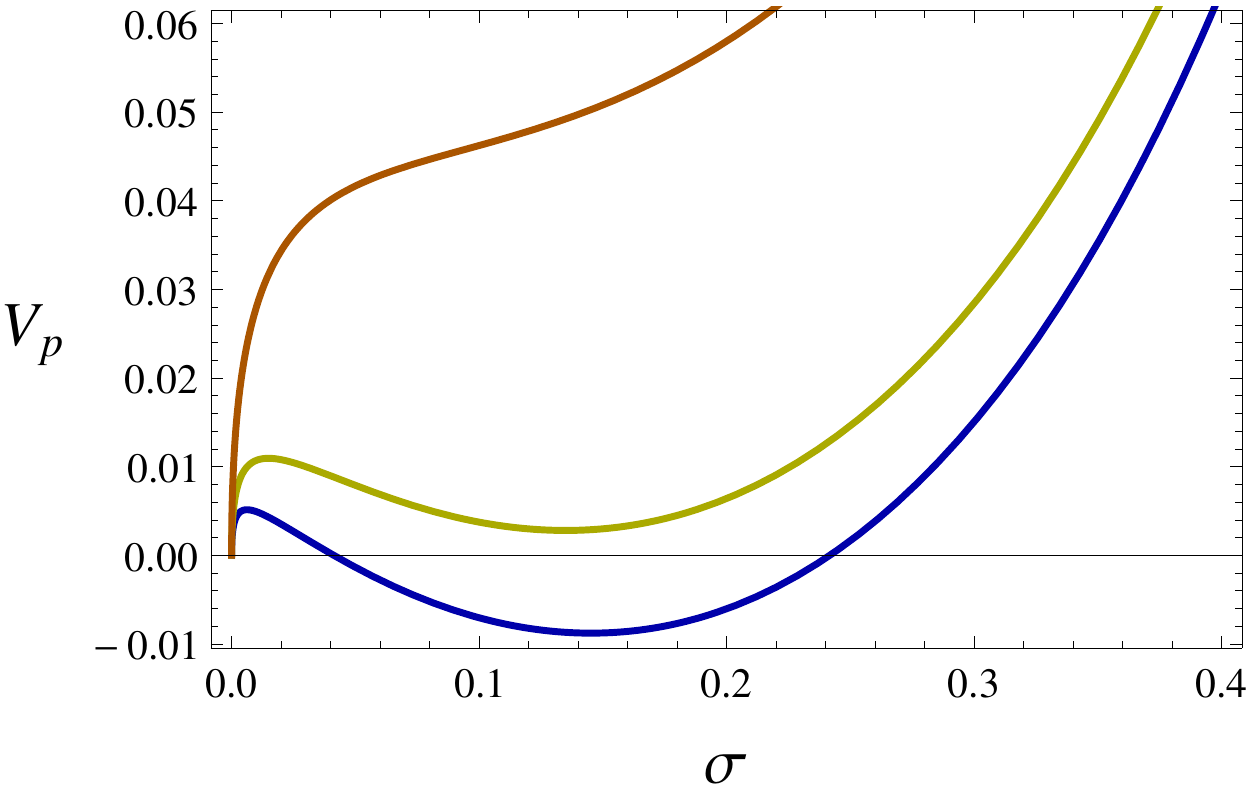}
}
\caption{Probe potentials at different temperatures for $\phi_1/\phi_0 = -0.49$ and $\phi_0 = 1.15$. The coordinate $\sigma$ used here is defined as $\sigma \equiv \rho/(1+\rho)$. The plots are made for pure fluxed D6 probes. \textbf{Left}: $\kappa=0.2908$ and the probe potential is plotted for $T=0.01,0.02,$ and $0.04$ for probes around the stable background. \textbf{Right}: $\kappa=1.0566$ and the probe potential is plotted for $T=0.01,0.02$ and $0.06$ for probes around the unstable background. 
 \label{fig:probepot}}
\end{figure}

We will focus in particular on bound states with ``pure fluxed D6'' probes --- these are probes with charges $\hat{\gamma}=(\hat{p}^0,\hat{p}^1,\hat{q}_1,\hat{q}_0)$ defined by expanding $e^{\kappa x} = 1 + \hat{p}^1 x + \hat{q}_1 x^2 - \hat{q}_0 x^3 + {\cal O}(x^4)$, i.e.:  
\begin{equation} \label{fluxD6}
 \hat{\gamma} = \bigl(1,\kappa,\frac{\kappa^2}{2},-\frac{\kappa^3}{6} \bigr) \qquad \leadsto \qquad \gamma =  \frac{g}{v^2} \bigl(1,\kappa v,\frac{(\kappa v)^2}{2},-\frac{(\kappa v)^3}{6} \bigr) \, ,
\end{equation}
in other words $b=n=0$ in the parametrization introduced at the end of section \ref{sec:probeaction}. 
Such probes can be thought of as wrapped D6-branes with worldvolume flux $F_2 \propto \kappa v$ turned on, which lift to smooth, locally Taub-NUT ``bubbling'' geometries in M-theory \cite{benawarner3,miranda}. The motivation for this restriction is in part simplifying the search for bound states, and in part the observation made in \cite{arXiv:1108.5821} that in the asymptotically flat case, at least in a large part of parameter space, these charges form bound states more easily than any other charge which has a single centered realization. Numerical explorations in the present setup confirm this, although we do not investigate this exhaustively. 

The search for bound states proceeds by looking for local minima of $V_p=V_{\rm grav}+V_{\rm em}$ defined in equations~(\ref{probepotdef}) and~(\ref{probepotdefel}), for all possible values of $\kappa$. This is done numerically. Note that $V_p=0$ at the event horizon and therefore probe bound states with $V_p<0$ are thermodynamically favorable configurations as explained in section \ref{sec:thermointerpr}. Thus, such bound states are stable, and conversely, local minima of the probe potential such that $V_p>0$ are metastable to tunneling into the black hole. Some examples are shown in figure \ref{fig:probepot}. 

A universal feature we observe is that for any given $(\phi_0,\phi_1)$ all bound states with fixed charges disappear at sufficiently high temperatures (depending on the probe charge). Intuitively the reason is clear: when the temperature is increased, black holes gain mass rather than charge, the gravitational pull becomes stronger, and eventually gravitational collapse is inevitable --- the probe is pulled into the black hole.

\begin{figure}
\centering{
\includegraphics[width=0.32\textwidth]{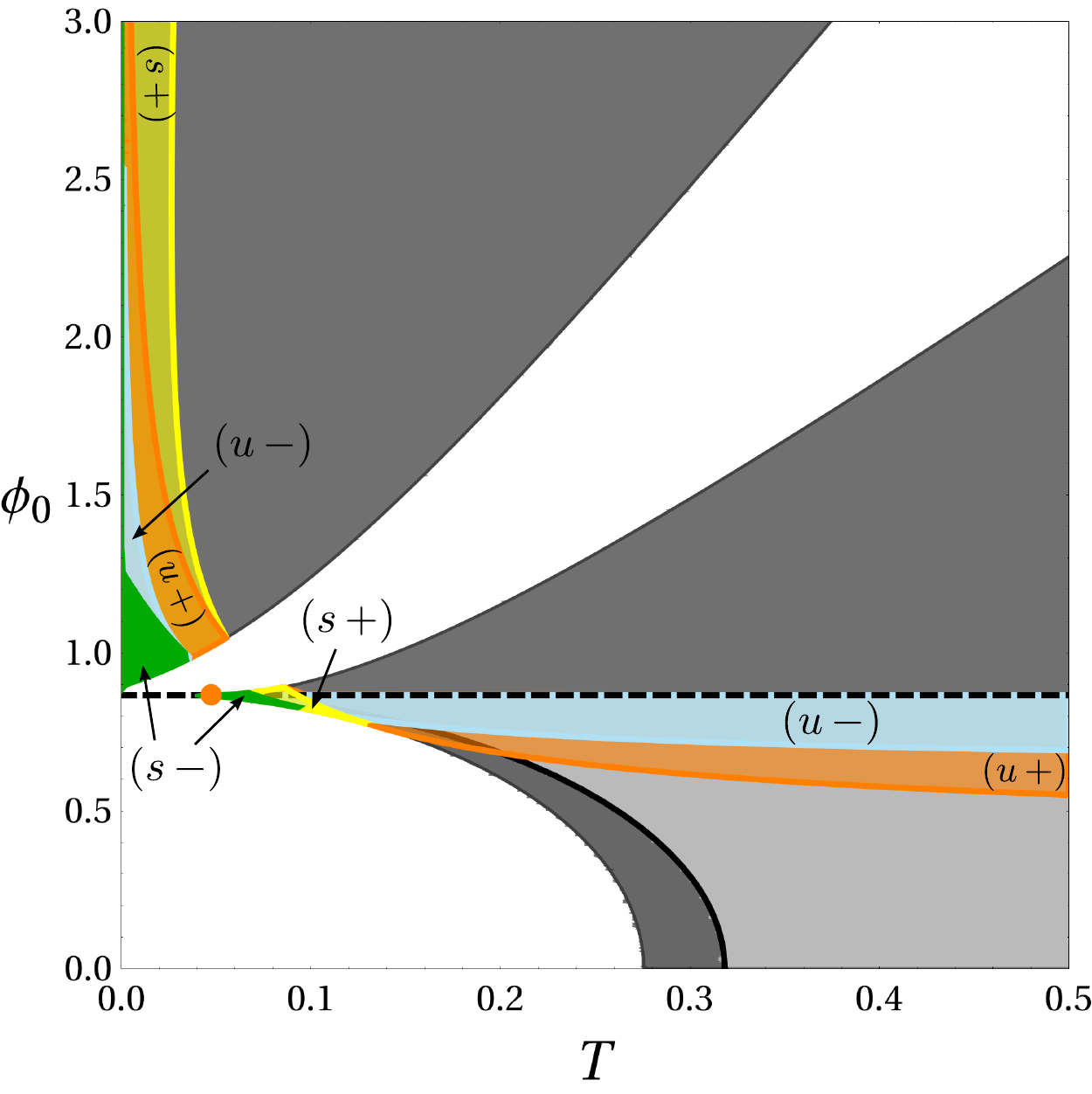}
\includegraphics[width=0.32\textwidth]{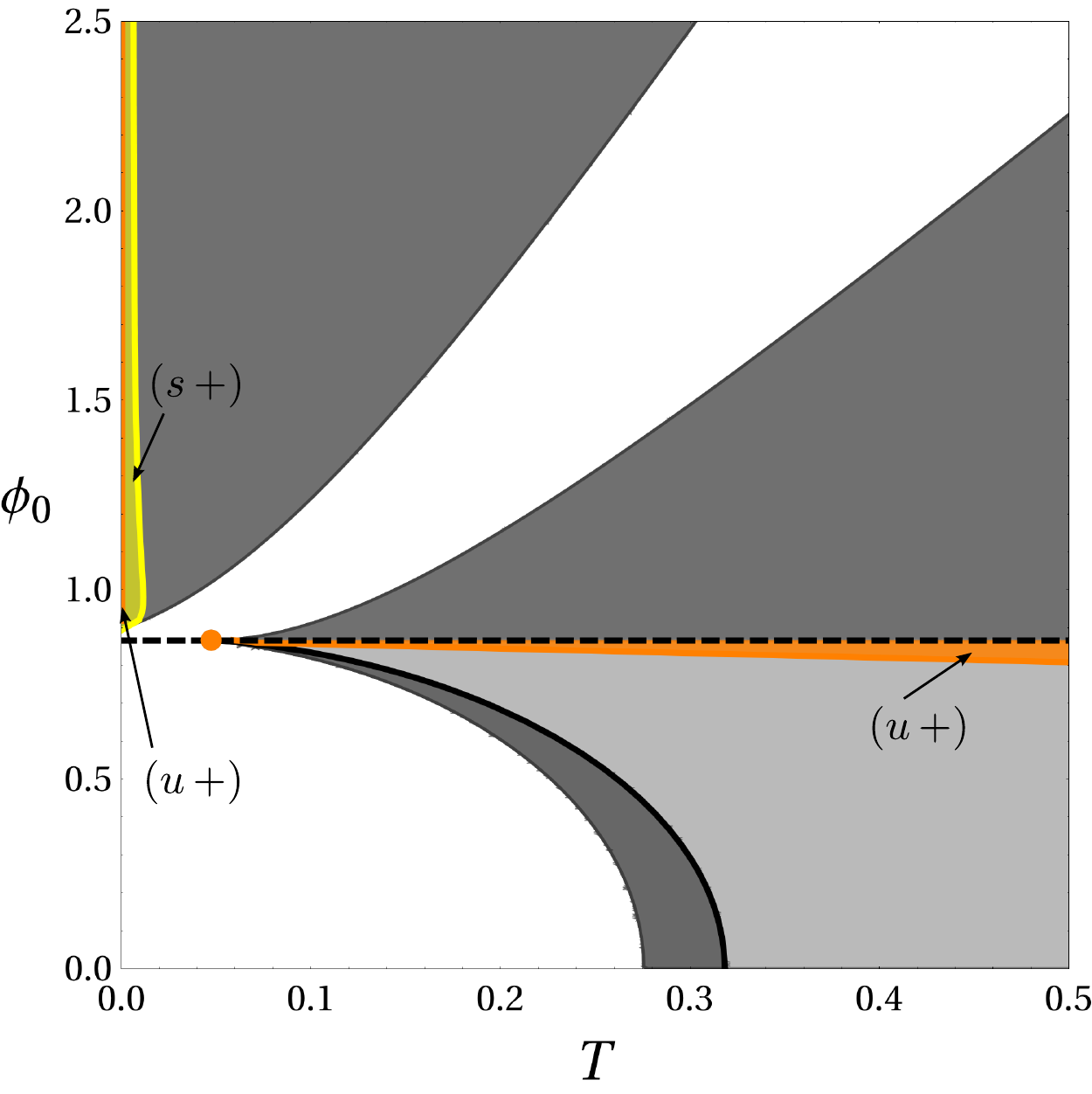}
\includegraphics[width=0.32\textwidth]{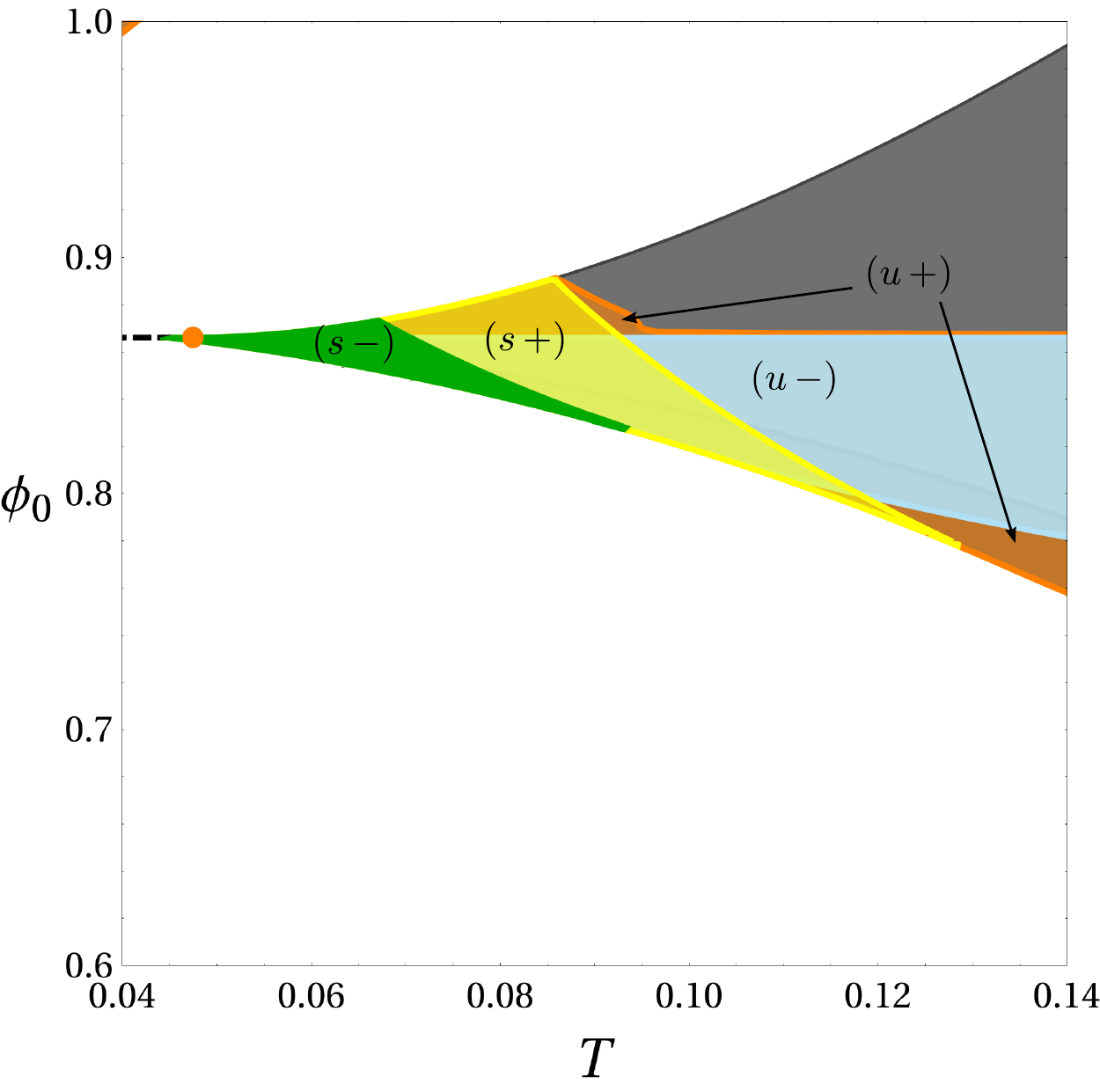}
}
\caption{Existence regions for probe bound states with background potentials set at $\phi_1/\,\phi_0=\pm0.49$, with the negative ratio in the left column. We label bound state regions with $(s/u\,\pm)$. A bound state region labeled $s$ means it forms around the stable black hole and similarly, $u$ regions represents probes bound to the unstable black hole. The $\pm$ denote whether the bound state has positive resp.\ negative potential energy. States with positive potential energy are unstable to tunneling into the black hole. The grayscale background echoes the background phase diagrams of section \ref{sec:phasediag}. The rightmost panel shows a close-up near the orange dot cusp for $\phi_1/\,\phi_0=-0.49$. Notice that the top corner of the $(s\,+)$ region smoothly connects to the top corner of the $(u\,+)$ region. The top of the $(s\,-)$ region connects to the $(u\,-)$ region in the same way. This can be understood simply from continuity in the extensive variable $u_1$. \label{fig:existdiag049}
\label{fig:existdiag049zoom}}   
\end{figure}

\begin{figure}
\centering{
\includegraphics[width=0.43\textwidth]{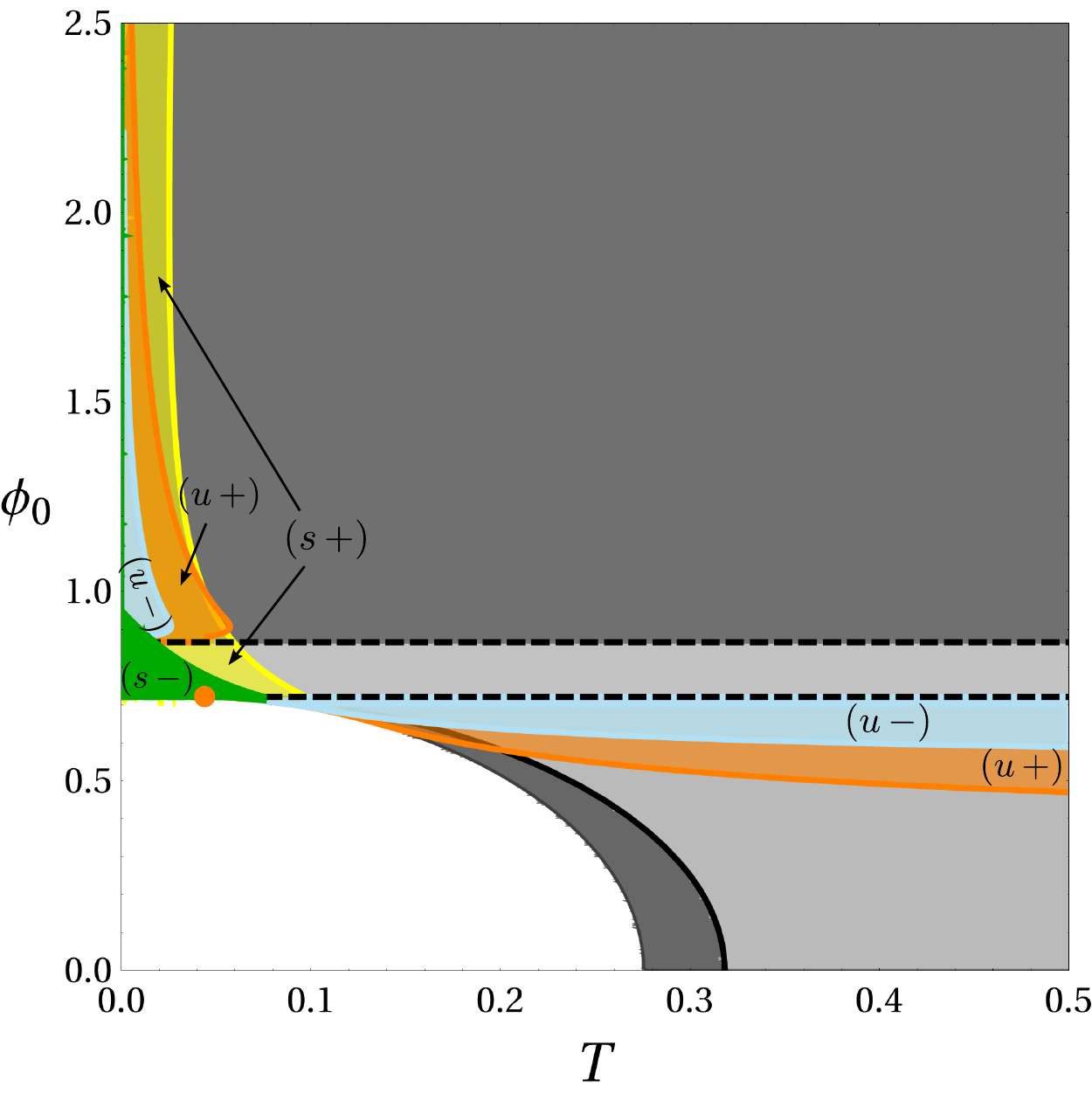}
\hspace{1cm}
\includegraphics[width=0.43\textwidth]{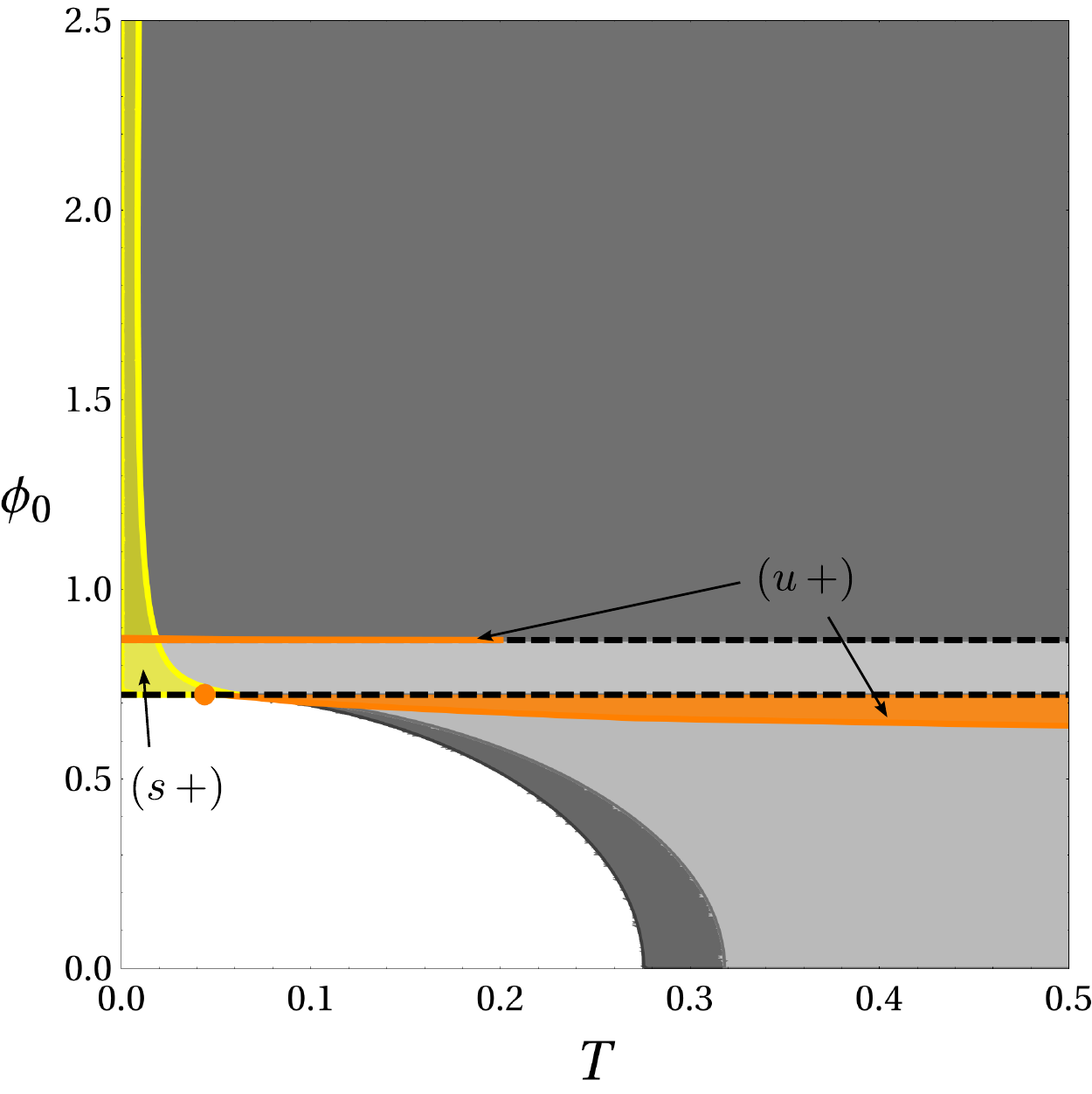}
}
\caption{Existence regions for probe bound states with $\phi_1/\,\phi_0=\pm0.6$, with the negative ratio being in the left column. \label{fig:existdiag06}}
\end{figure}

\begin{figure}
\centering{
\includegraphics[width=0.43\textwidth]{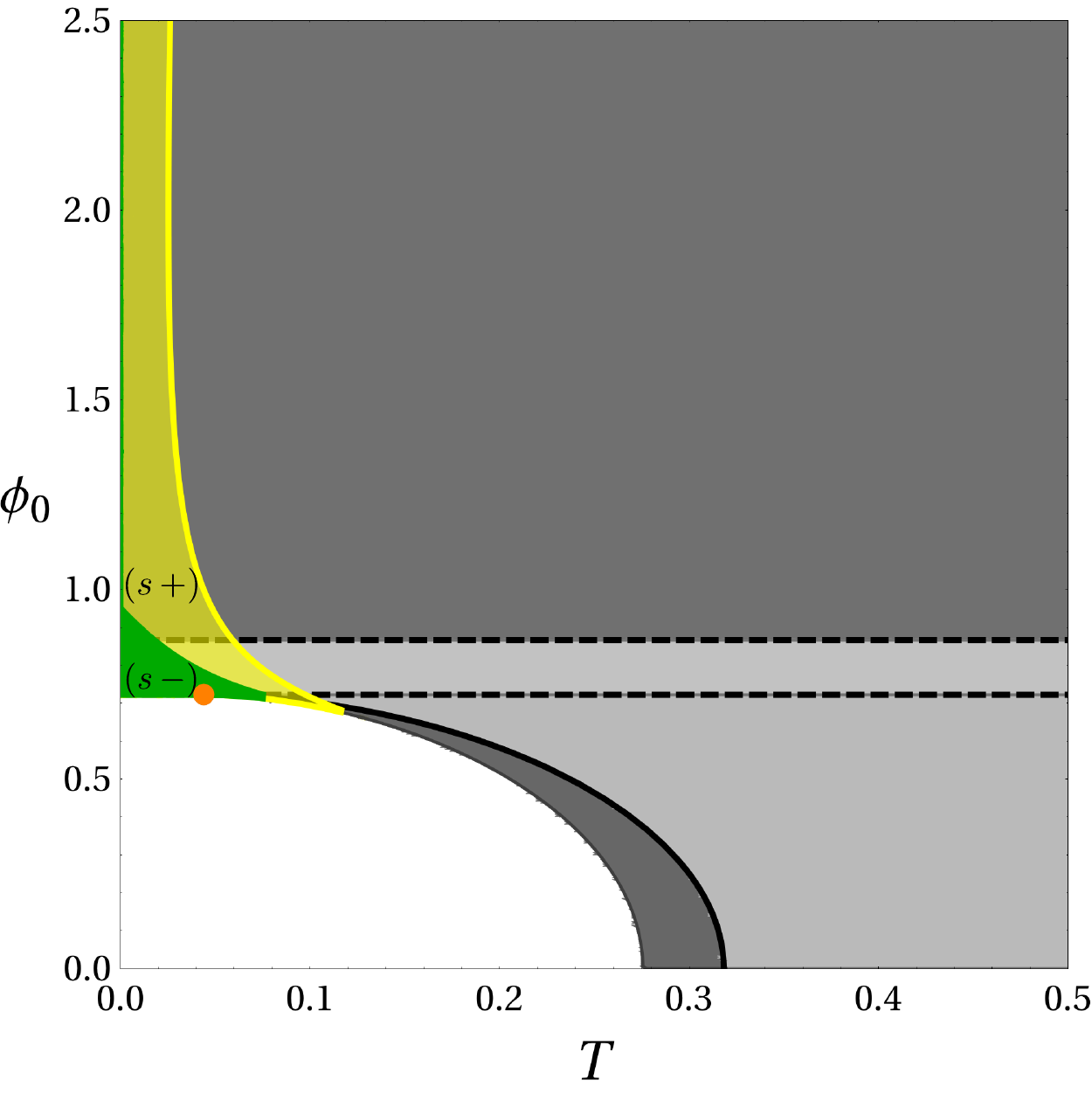}
\hspace{1cm}
\includegraphics[width=0.43\textwidth]{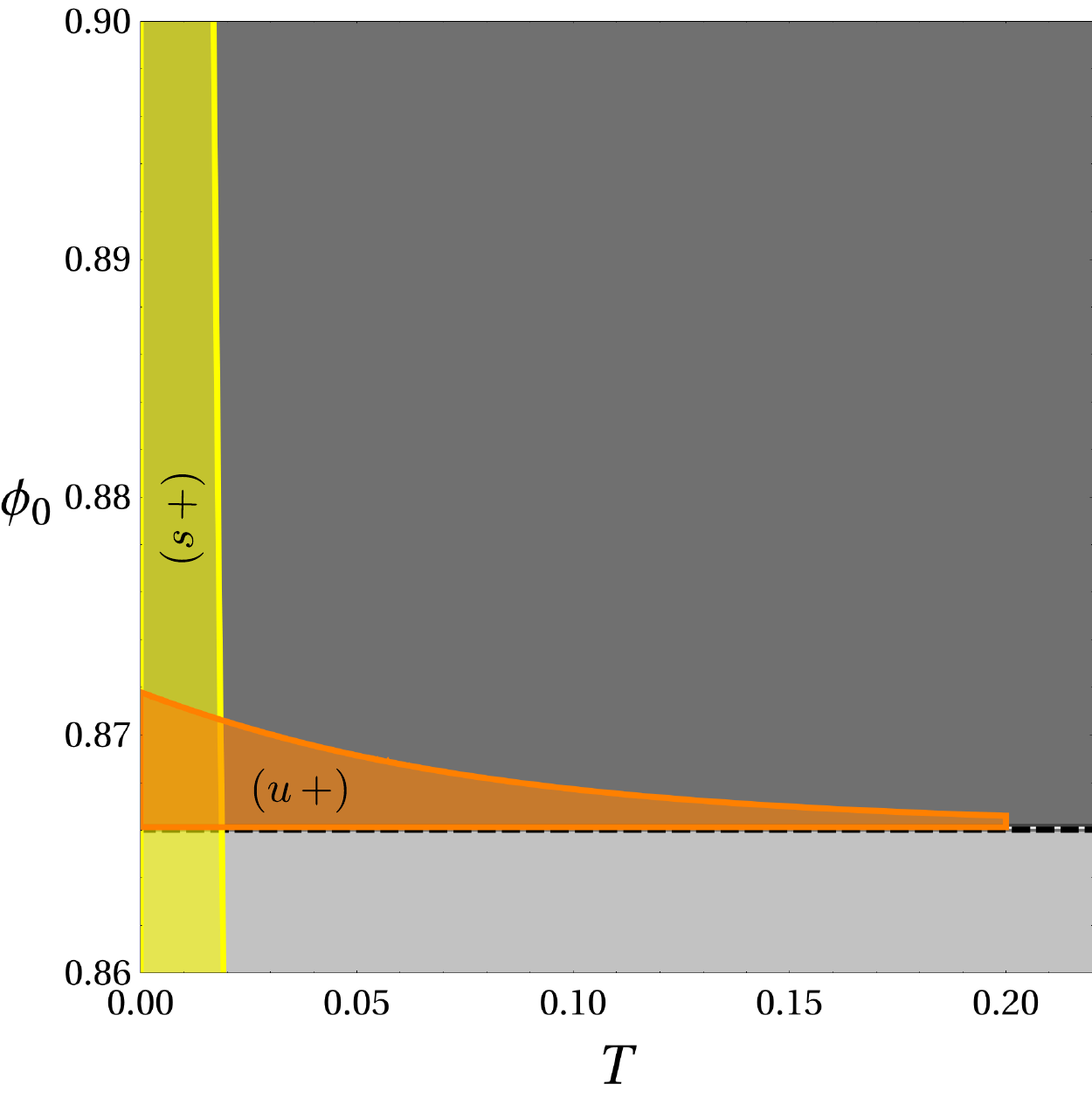}
}
\caption{\textbf{Left}: Probe bound states with $\phi_1/\,\phi_0=-0.6$, around the stable background. Note that the $(s\,\pm)$ regions dip below the dotted line. \textbf{Right}: Zoom of the $(u\,+)$ bound states above the dotted line with $\phi_1/\,\phi_0=0.6$.\label{fig:existdiag06gy}}

\end{figure}

\begin{figure}
\centering{
\includegraphics[width=0.43\textwidth]{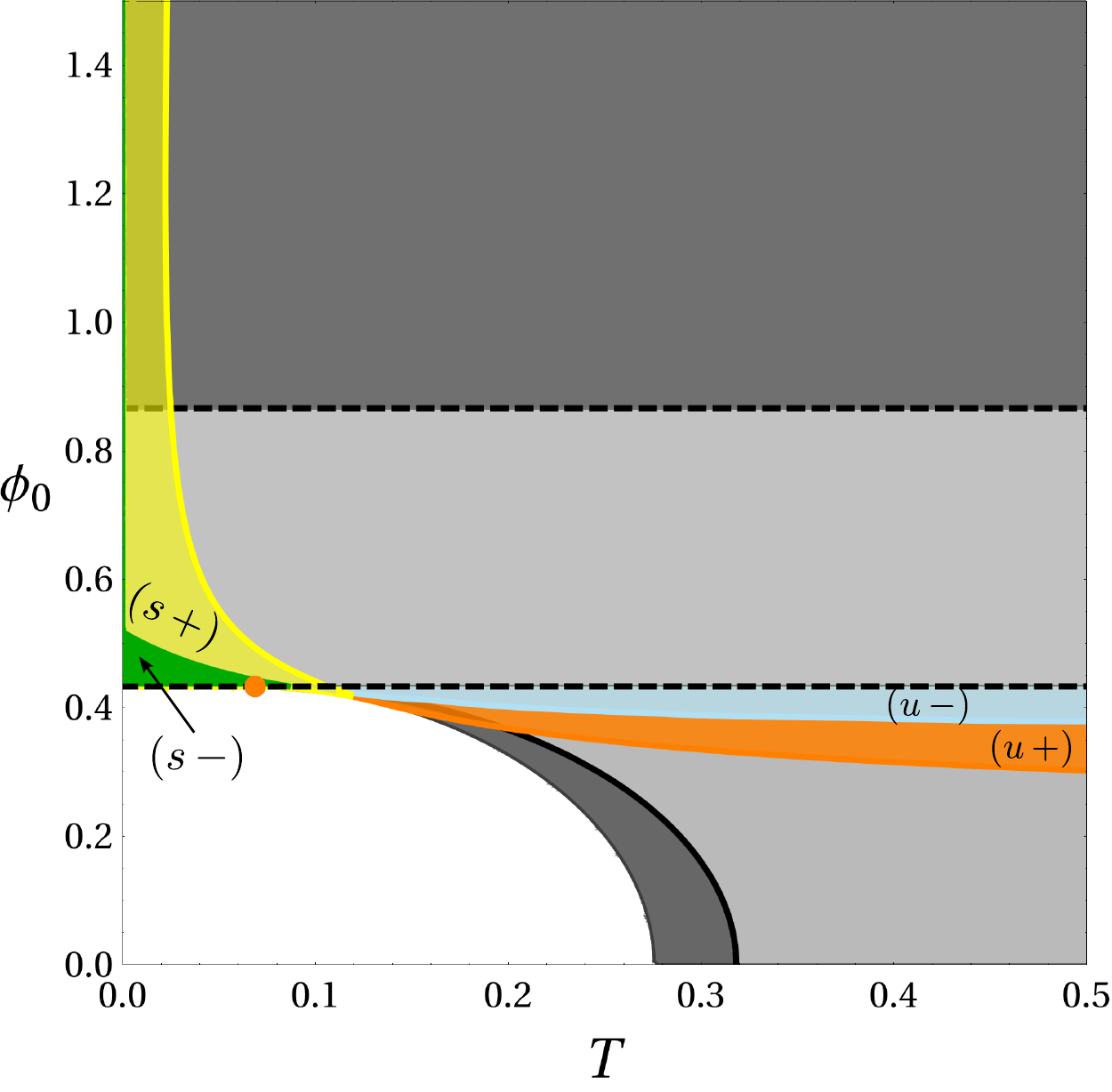}\
\hspace{1cm}
\includegraphics[width=0.43\textwidth]{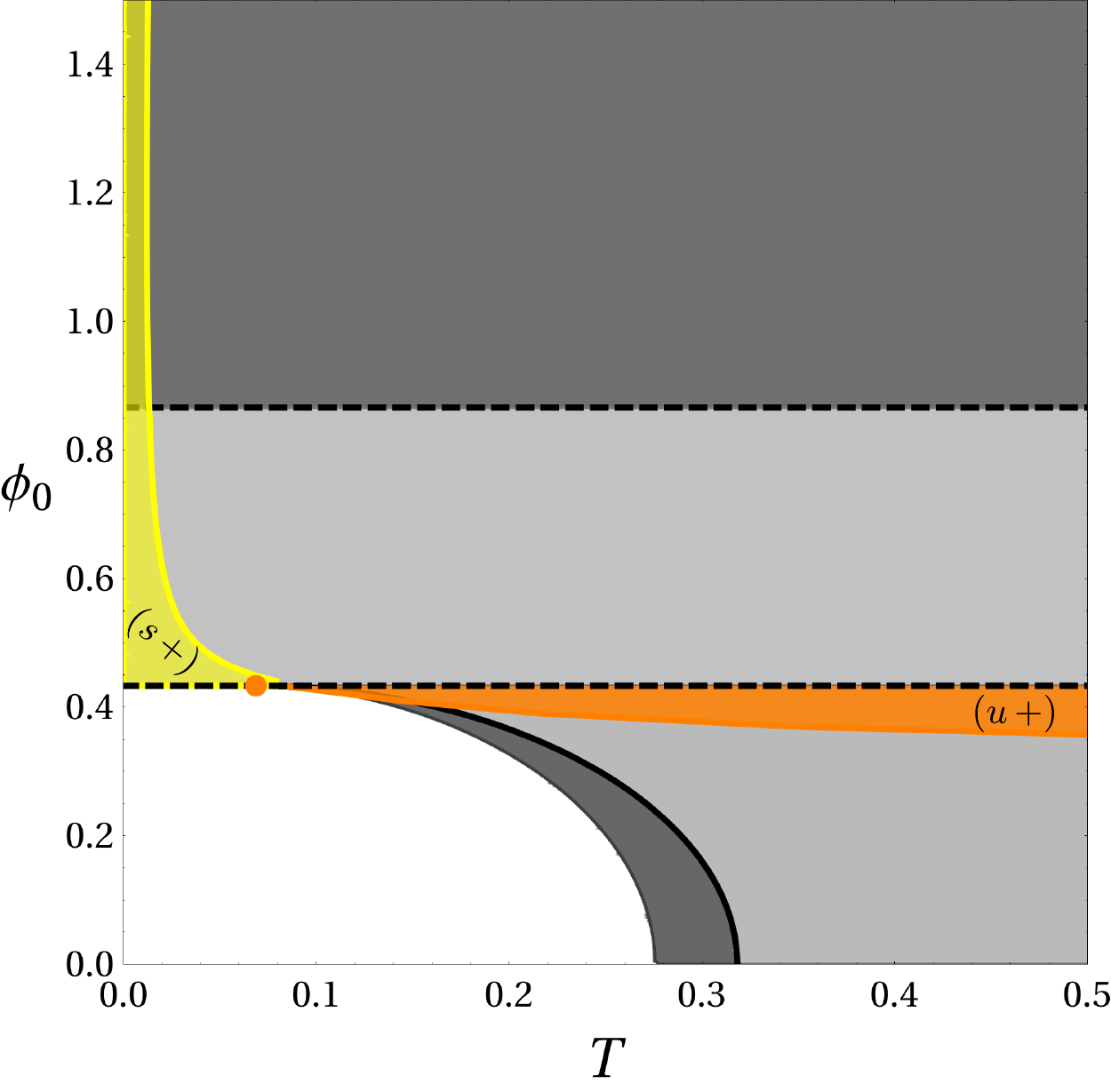}
}
\caption{Existence regions for probe bound states with $\phi_1/\,\phi_0=\pm1$, with the negative sign holding in the left hand column.  \label{fig:existdiag1}}

\end{figure}

We display the existence regions of probe bound states in figures~\ref{fig:existdiag049}-\ref{fig:existdiag1} which correspond to slices of phase space where the background potentials satisfy $\phi_1/\phi_0=\pm0.49\, , \pm0.6\,$ and $\pm1$. The bound state existence regions have many common features which we describe below.

\begin{enumerate}
\item Bound states around the stable black hole background ---the black hole with lowest free energy--- are represented by the green and yellow regions with labels $(s\,\pm)$ in figures~\ref{fig:existdiag049}-\ref{fig:existdiag1}. The green $(s\,-)$ regions demarcate where stable bound states exist, in the sense that these bound states have negative potential energy. Metastable bound states live in the yellow $(s\,+)$ regions. Bound states around the unstable black hole background are shown in the orange $(u\,+)$ and blue $(u\,-)$ regions of our diagrams, with the $(u\,-)$ regions representing stable bound states and $(u\,+)$ regions labelling metastable bound states.  

\item Recall that the probes are sensitive to the signs of $\phi_0$ and $\phi_1$. Figures~\ref{fig:existdiag049}, \ref{fig:existdiag06} and~\ref{fig:existdiag1} are slices of phase space where the potentials satisfy, respectively, $\phi_1/\phi_0=\pm0.49\, , \pm0.6\,$ and $\pm1$, with the minus sign holding in the left hand columns. When the potentials have opposite sign, there exist stable bound states between the probe and the black holes. As in~\cite{arXiv:1108.5821}, in a small region, there also exist stable (negative energy) bound states when the potentials have the same sign. In this case the $(u\,-)$ bound states lie in a thin sliver below the lower dotted line (where $\Delta_1$ changes sign). This happens for $\phi_1/\phi_0>1/2$ for arbitrarily high $T$.     

\item The $(u\,\pm)$ regions disappear as we cross the lower dotted line from below. This is expected since the background to which the probes are bound have diverging charge as we cross the dotted line from below and stop existing altogether above it. One caveat is shown in figure~\ref{fig:existdiag049zoom} where the $(u\,\pm)$ regions seep across the dotted line near the orange dot. These are probes bound to a black hole with negative free energy and are closer in nature to the bound state regions across the white gap than those across the dotted line. Naturally there are no bound states of type $(u\,\pm)$ above the dotted line when $\phi_1/\phi_0>1/2$ as we cross into a region where only one black hole exists.

\item While the $(u\,\pm)$ bound states generically disappear when crossing the lower dotted line from below, nothing analogous can be said for the $(s\,\pm)$ bound states above the dotted line as we cross it from above. Since nothing singular happens for the stable backgrounds as the lower dotted line is crossed, this matches with our expectations. A clear example of bound states dipping below the dotted line can be seen in the left hand column of figure~\ref{fig:existdiag06gy}.  

\item When $|\phi_1/\phi_0|=1$ there are no $(u\,\pm)$ regions above the dotted lines, even when the potentials have opposite signs. This should not be taken to mean that there are no bound states around the unstable black hole above the dotted line beyond a certain ratio of $\phi_1/\phi_0$. As in~\cite{arXiv:1108.5821}, the disappearance of bound states may indicate that the favored probes for forming bound states are not pure fluxed D6 branes in this region of parameter space. 

\item In all cases considered, the $(s\,+)$ regions open up at large $\phi_0$. By this we mean that bound states at large chemical potential exist for larger values of $T$. This is consistent with the existence of a large region of $(s\,+)$ bound states in the planar limit as shown in section~\ref{planarbs} below.
\end{enumerate}


\subsection{Probe bound states in the planar limit}\label{planarbs}

\begin{figure}
\centering{
\includegraphics[width=0.305\textwidth]{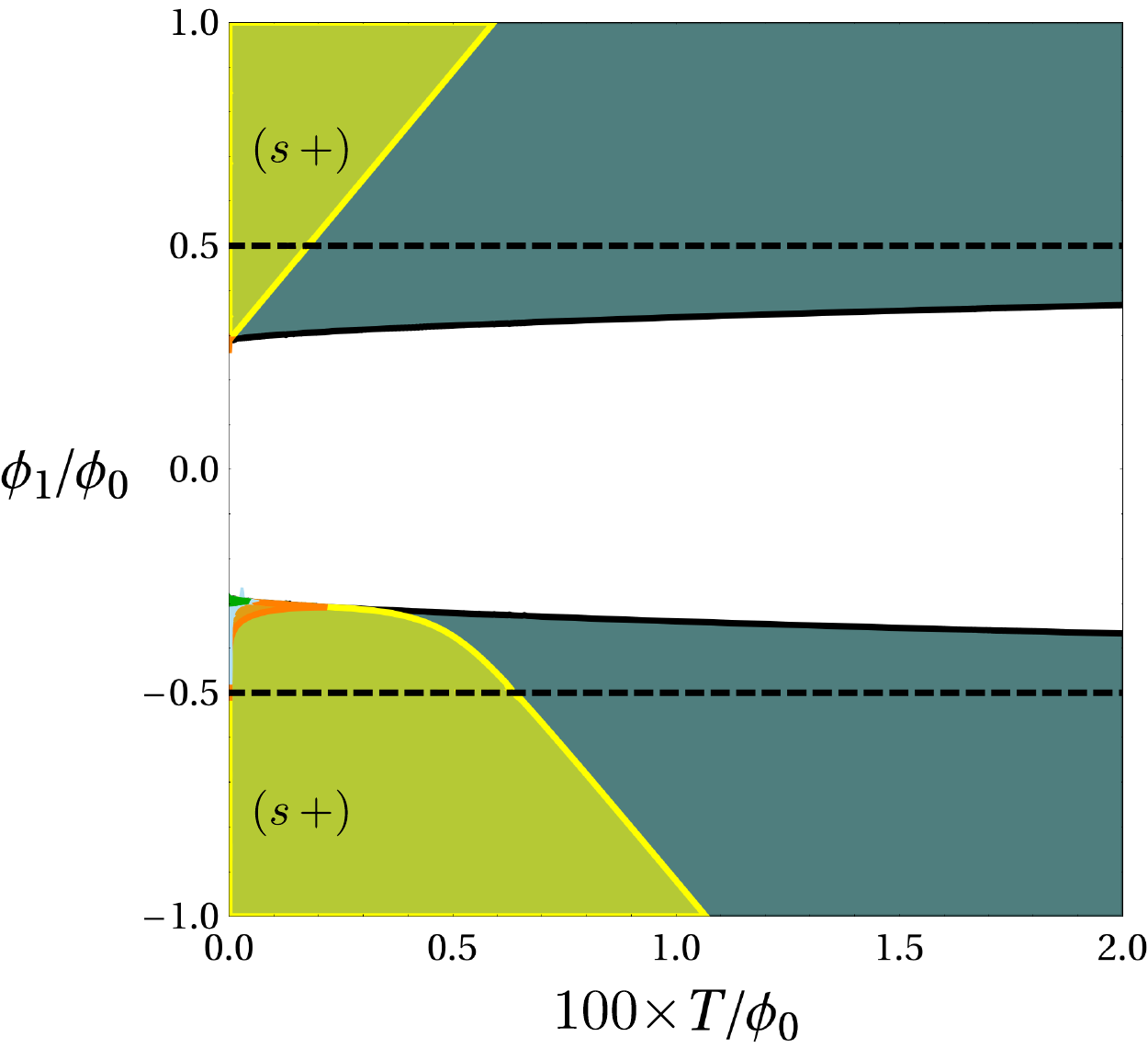}
\includegraphics[width=0.3205\textwidth]{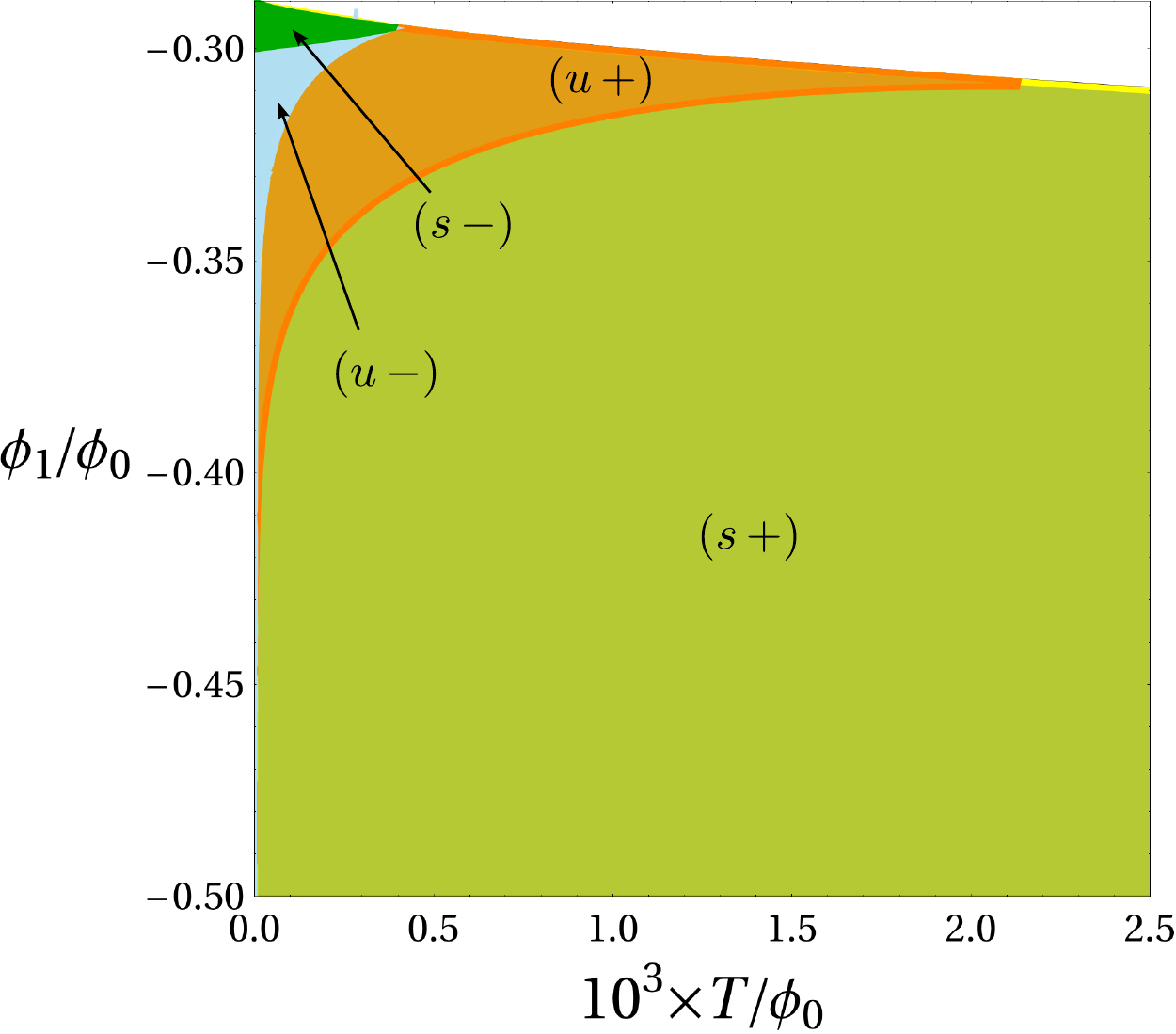}
\includegraphics[width=0.3405\textwidth]{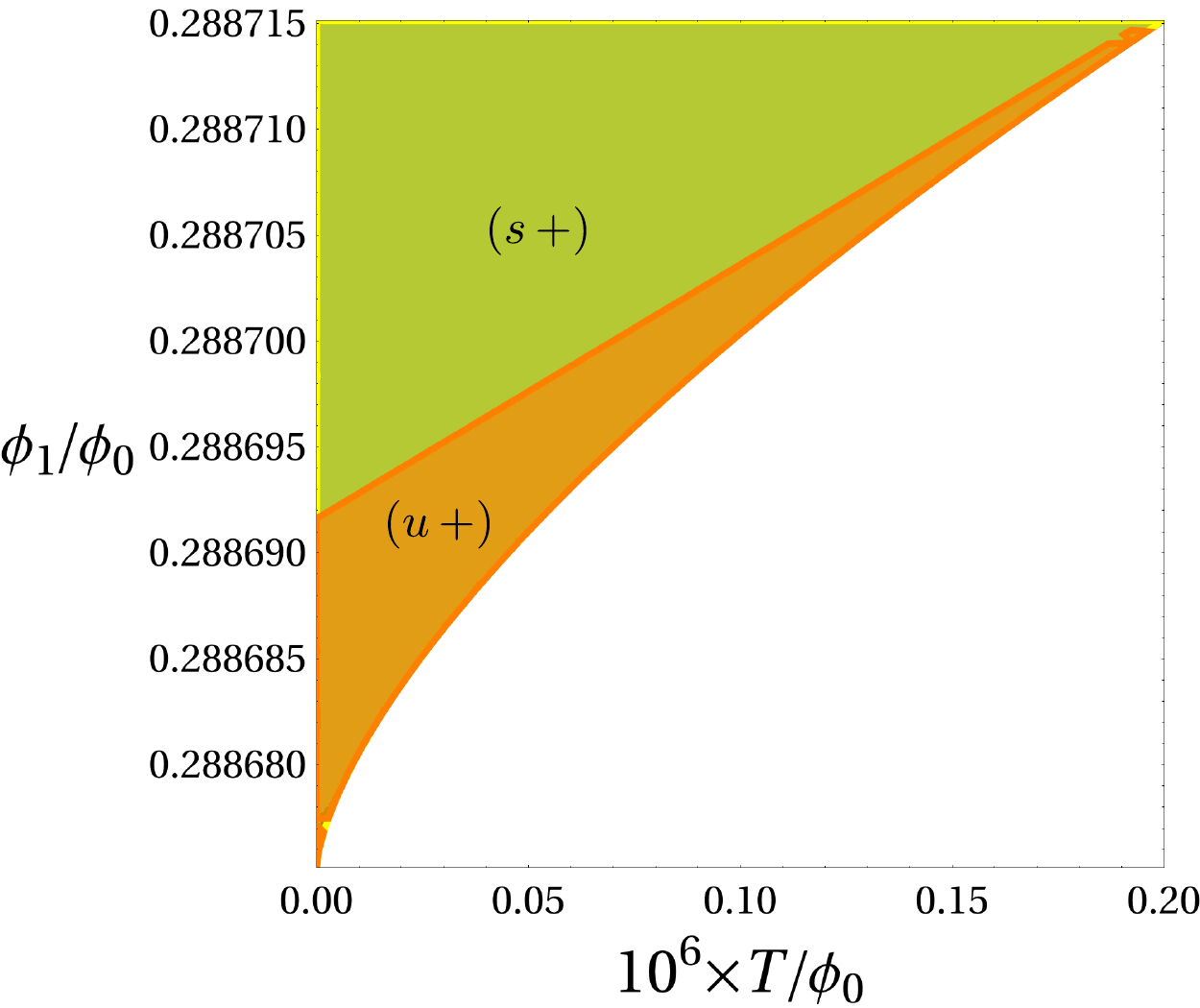}
}
\caption{Existence regions for planar bound states. The regions labeled $(s\,+)$ show bound states around the stable black brane with positive potential energy. If we zoom in closer to small $T/\phi_0$ near the boundary of the white gaps, we see more interesting features as shown in the two rightmost panels. 
\label{fig:existdiagplan}}
\end{figure}

One can obtain the probe potential in the planar limit (\ref{eq:planarscalings}) either directly from the probe particle action~(\ref{probeaction}) or by scaling the parameters in~(\ref{probepotdef}-\ref{probepotdefel}). In the latter case one must be careful to divide by an overall factor of $\lambda$ coming from the fact that we have scaled $dt = d\bar{t}/\lambda$ in the probe action. The planar black hole probe potential 
equals the spherical black hole potential except that the $\rho^2$ term under the square root disappears:
\begin{equation} \label{probepotdefplanar}
 { V}_{\rm grav} =  \frac{\sqrt{3}}{2} \sqrt{\left( 2{\rho}{r}_++ {f}_0 {f}_1^3-{u}_0 {u}_1^3 \right)\biggl[
  \biggl(\frac{{p}_1}{2 {f}_1}  + \frac{{q}_0}{{f}_0} \biggr)^2 + \frac{{f}_0}{{f}_1} \biggl(\frac{{p}_0}{6\,{f}_1}   -\frac{{q}_1}{{f}_0} \biggr)^2
 \biggr] } \, ~,
\end{equation}
and 
\begin{equation}\label{probepotdefelplanar}
{{V}}_{\rm em} =  - \frac{\phi_0 {q}_0 \rho}{f_0} - \frac{\phi_1 {p}^1 \rho}{f_1}   \, .
\end{equation}
Because of the extra scaling symmetry discussed in section \ref{sec:planarlimit}, we can scale out the appropriate powers of $\phi_0$ from the various quantities occurring in the expression for the potential, reducing its dependence on $\phi_0$ to an overall factor. Accordingly  all nontrivial dependence of the probe potential on the electric potentials and temperature will be in terms of scale invariant quantities e.g.\ the ratios $\phi_1/\phi_0$ and $T/\phi_0$.
The bound state existence regions are shown in figure~\ref{fig:existdiagplan}.

As expected from our spherical analysis, bound states with negative energy only exist when $\phi_1/\phi_0<0$. Bound states about the unstable black hole only live in a very thin sliver of parameter space for $\phi_1/\phi_0>0$.

\subsection{Analytic results for $T=0$} \label{sec:anal}

\begin{figure}
\centering{
\includegraphics[width=0.42\textwidth]{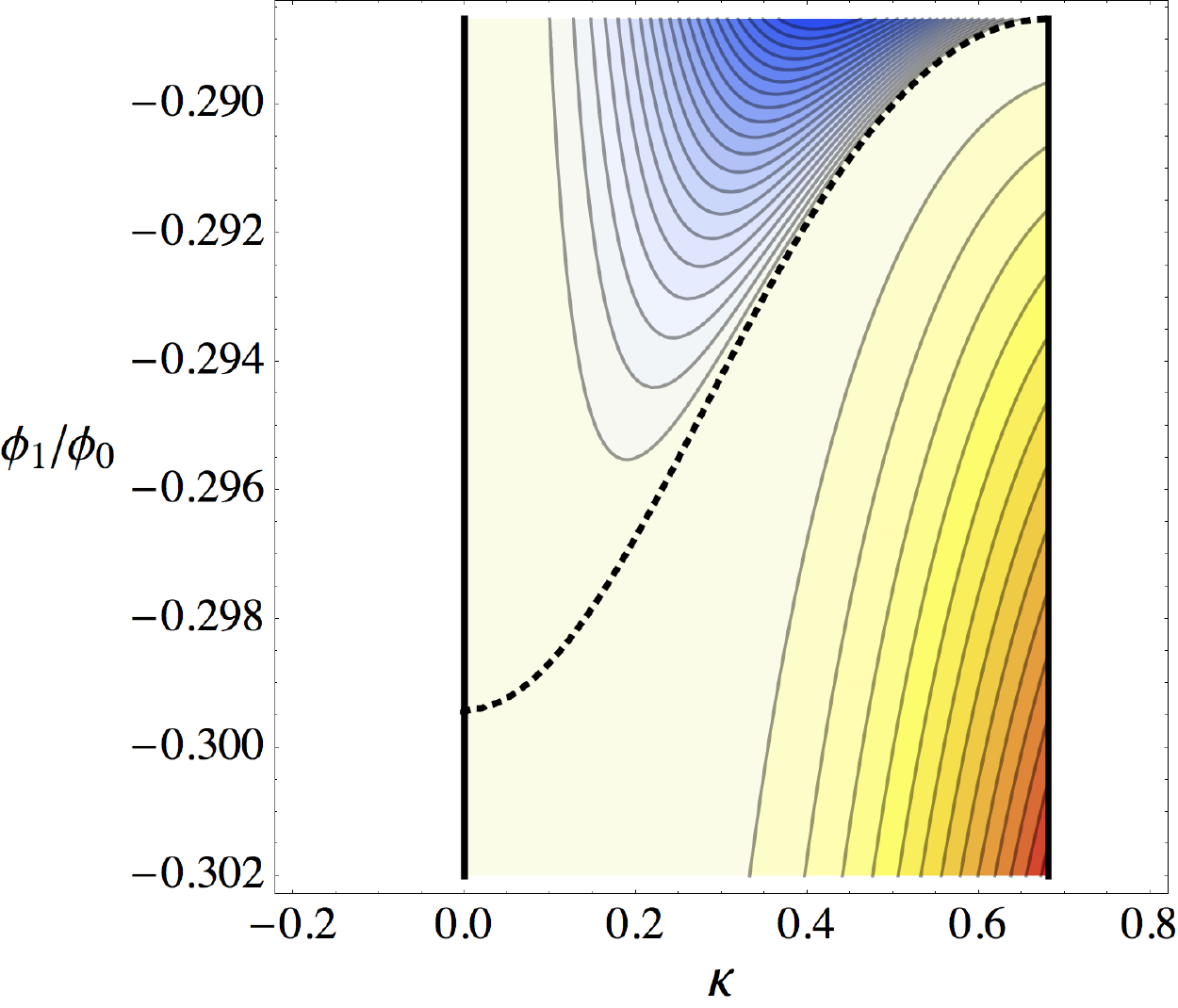}
\hspace{1cm}
\includegraphics[width=0.42\textwidth]{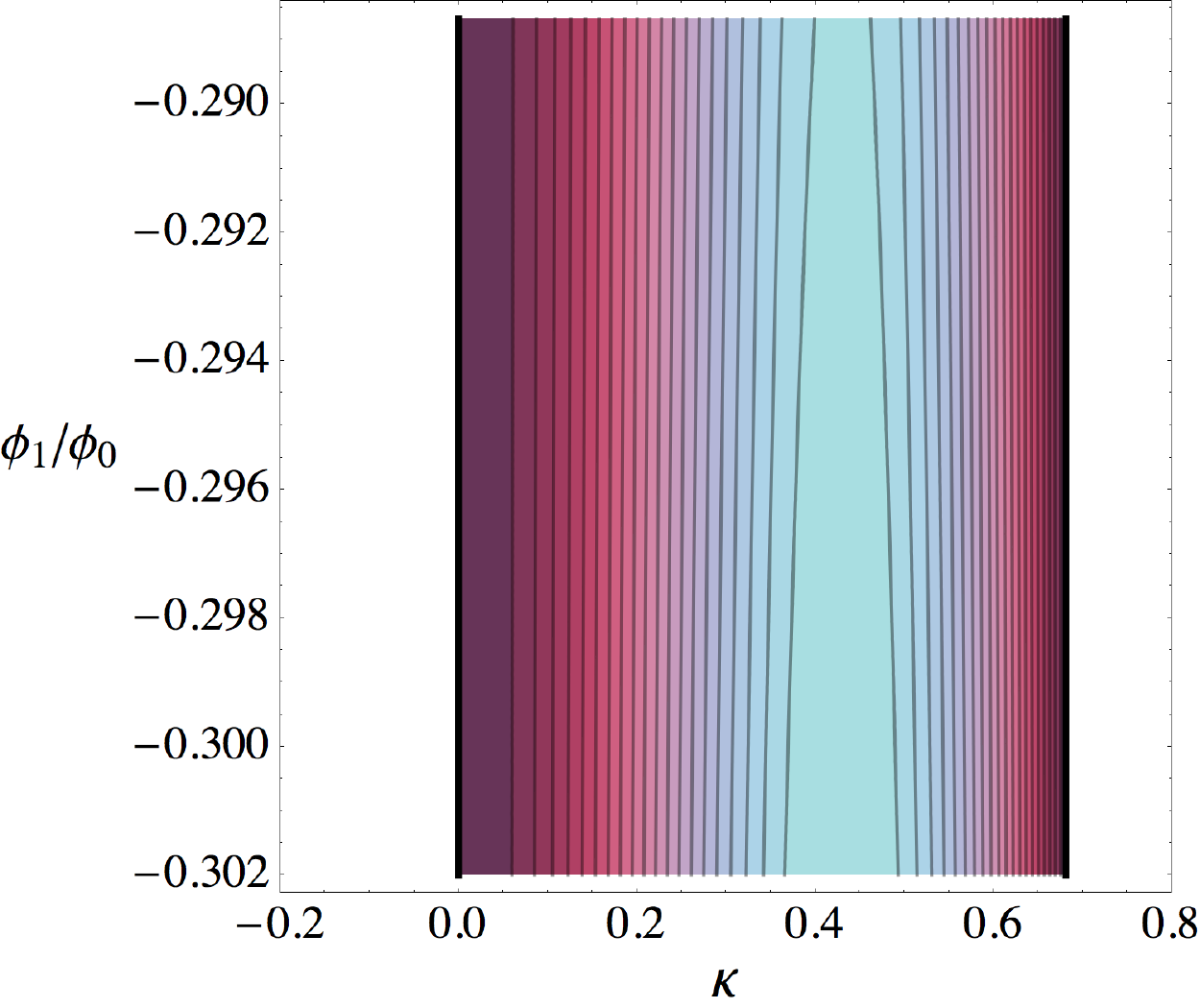}
}
\caption{{\bf Left}: Minimal values of the potential as a function of probe charge parameter $\kappa$ and background parameter $\phi_1/\phi_0$, computed using the expansion of $V_p$ to second order in $\tau$. The minima are negative above the dotted line, positive below. The lowest minimum attained for a given value of $\phi_1$ is $V_{\rm min} \approx -4 \times 10^{-4} \phi_1$. {\bf Right}: Separation $\rho_{\rm eq}=u_1 \tau_{\rm eq}$ of the minimum from the horizon, where $u_1=4|\phi_1|/\sqrt{3}$. Lighter is further away. The rescaled separation $\tau_{\rm eq}$ only depends on $\kappa$, not on the potentials. The maximal separation is given by $\tau_{\rm max} \approx 0.025$; at the edge values of $\kappa$ the separation drops to zero. 
\label{fig:minima}}
\end{figure}

In simple limits, it is straightforward to confirm our numerical results analytically. At zero temperature, the thermodynamically preferred planar solution is the $u_0=0$ solution discussed in section \ref{sec:hsvl}. In this limit the explicit probe potential for the charges (\ref{fluxD6}) becomes quite simple:
\begin{equation}\label{simpleplan}
 V_p = \frac{\phi_0 \kappa^3}{6} -   \frac{\phi_1 \, \kappa \, \tau}{1+\tau}
 +\frac{|\phi_1|}{3} \sqrt{(3+3 \tau + \tau^2)\bigl(\kappa^2+\frac{\tau}{1+\tau}\bigr)^3} ~ , \qquad
 \tau \equiv \rho/u_1 \, ,
\end{equation}
with $u_1=4 |\phi_1|/\sqrt{3}$. 
Expanded to first order at small $\tau$, this becomes, say for $\phi_1 > 0$:
\begin{equation}
 \frac{V_p}{\phi_1} = \frac{|\kappa|^3}{6} \biggl({\rm sgn}\, \kappa \cdot \frac{\phi_0}{\phi_1} + \sqrt{12} \biggr) + \frac{|\kappa|}{\sqrt{12}} \biggl( |\kappa|^2 + 3  - \sqrt{12} \, {\rm sgn}\, \kappa \biggr) \, \tau + {\cal O}(\tau^2) \, .
\end{equation}
Since we need $|\phi_1/\phi_0| \geq 1/\sqrt{12}$ to have a black hole solution, the zeroth order term is always nonnegative.\footnote{The fact that this is nonzero is an artifact of the degenerate limit $u_0 \to 0$. At any finite $u_0$, the potential will drop to zero for $\rho \ll u_0$.} The first order term is negative if  $0 < \kappa < \sqrt{\sqrt{12} - 3} \approx 0.68125$. In this case a bound state exists, which may have negative energy if $\phi_1/\phi_0$ is  sufficiently close to $-1/\sqrt{12}$, 
This is illustrated in figure \ref{fig:minima}. These observations are consistent with the numerical results of figure \ref{fig:existdiagplan}.

\begin{figure}
\centering{
\includegraphics[width=0.4\textwidth]{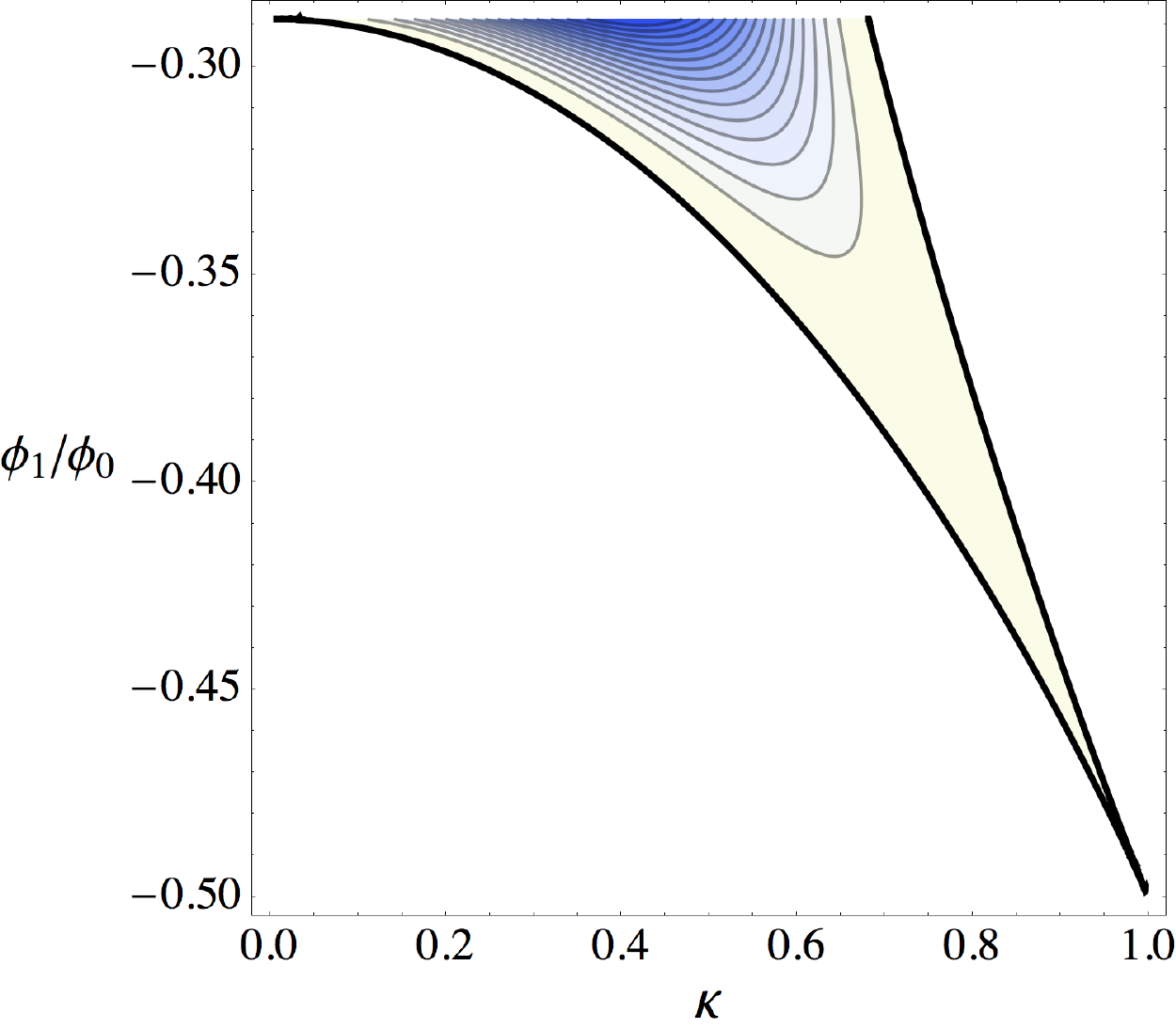}
\hspace{1cm}
\includegraphics[width=0.4\textwidth]{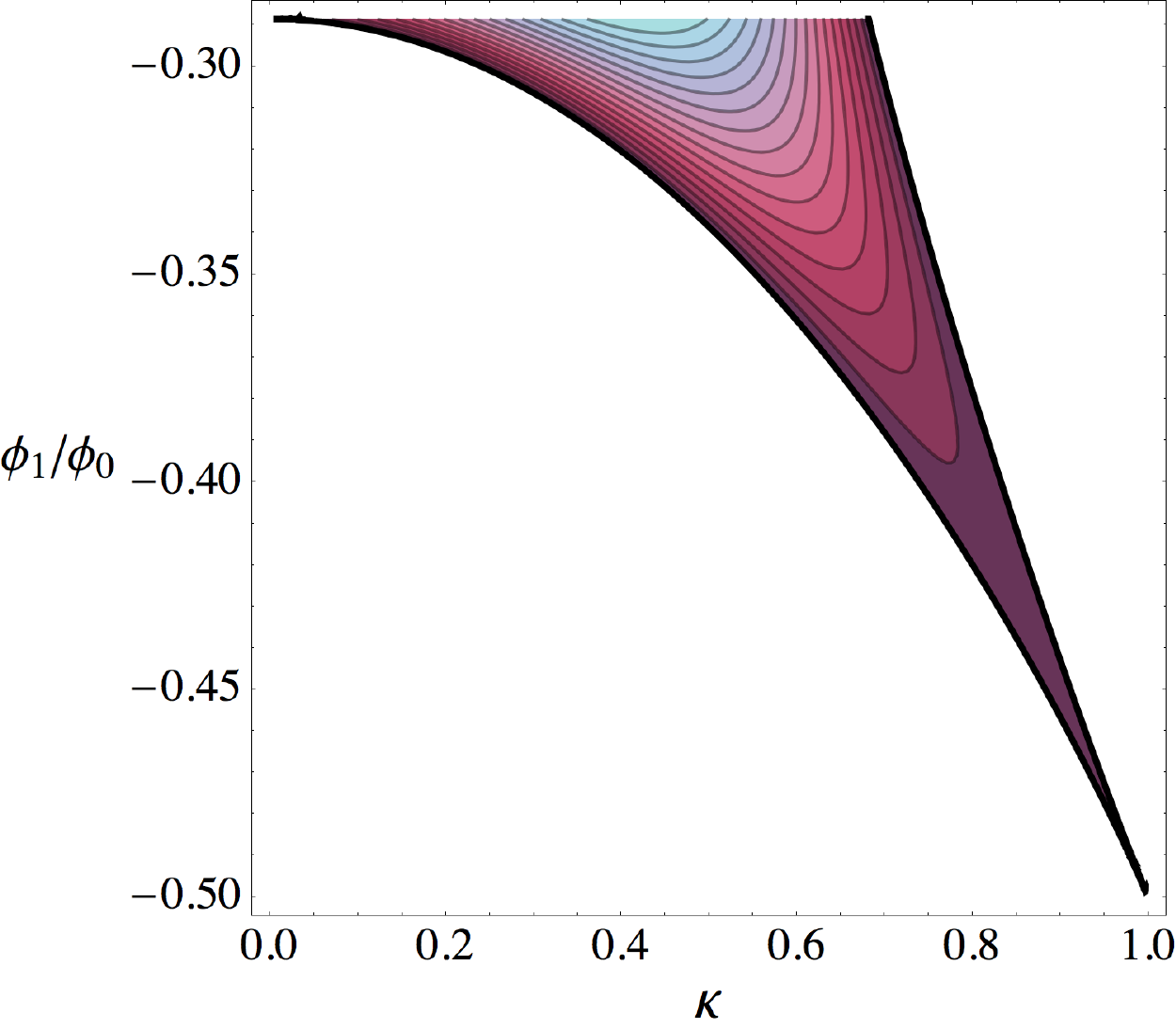}
}
\caption{{\bf Left}: Minimal values of the potential as a function of probe charge parameter $\kappa$ and background parameter $\phi_1/\phi_0$,  using the expansion of $V_p$ to second order in $\tau$. Minima exist (and are negative) within the black curve. {\bf Right}: Separation $\rho_{\rm eq}=u_0\tau_{\rm eq}$ of the minimum from the horizon. Lighter is further away, along the black curve the distance drops to zero.
\label{fig:potderiv}}
\end{figure}

We can repeat this analysis for the thermodynamically disfavored planar solution, again at $T=0$. In this branch, $u_1=2|\phi_0|/3$ and $u_0=\tfrac{|\phi_0|}{3}(12\phi_1^2/\phi_0^2-1)$. In this limit, $V_p$ is slightly more complicated than~(\ref{simpleplan}), however expanded to first order in $\tau=\rho/u_0$ we find: 
\begin{equation}
\frac{V_p}{|\phi_0|}=\left(-\frac{\kappa^3}{6}-\frac{\kappa}{2}\frac{\phi_1}{\phi_0}\,\left(12\tfrac{\phi_1^2}{\phi_0^2}-1\right)+\frac{1}{24}\sqrt{\left(1+12\,\tfrac{\phi_1^2}{\phi_0^2}\right)\left(-1+2\,\kappa^2+12\,\tfrac{\phi_1^2}{\phi_0^2}\right)^3}\right)\tau+{\cal O}(\tau^2)~.
\end{equation}
If the coefficient of $\tau$ is negative in this expansion, then the potential admits a minimum with negative energy. It is straightforward to check that this only happens for a special range of values with $0<\kappa<1$ and $-1/2<\phi_1/\phi_0<-1/\sqrt{12}$ shown in figure~\ref{fig:potderiv}. These results are consistent with those presented in figure~\ref{fig:existdiagplan}.

\begin{figure}
\centering{
\includegraphics[width=0.49\textwidth]{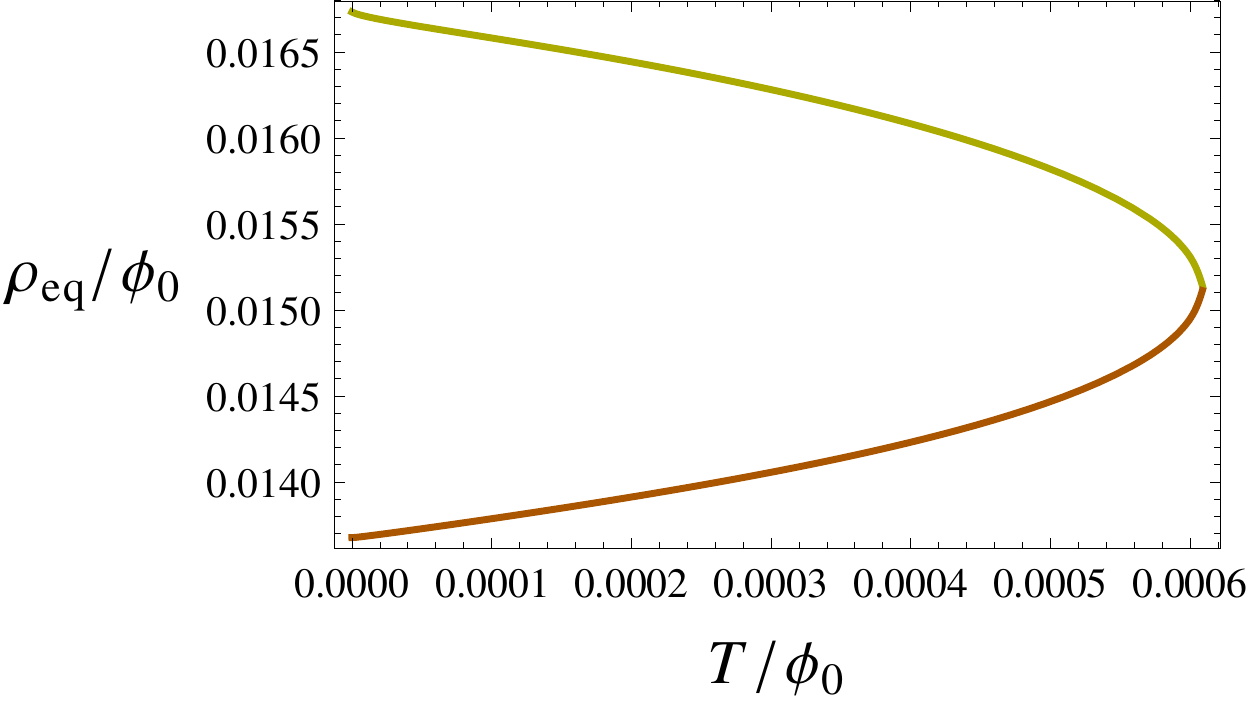} 
\includegraphics[width=0.49\textwidth]{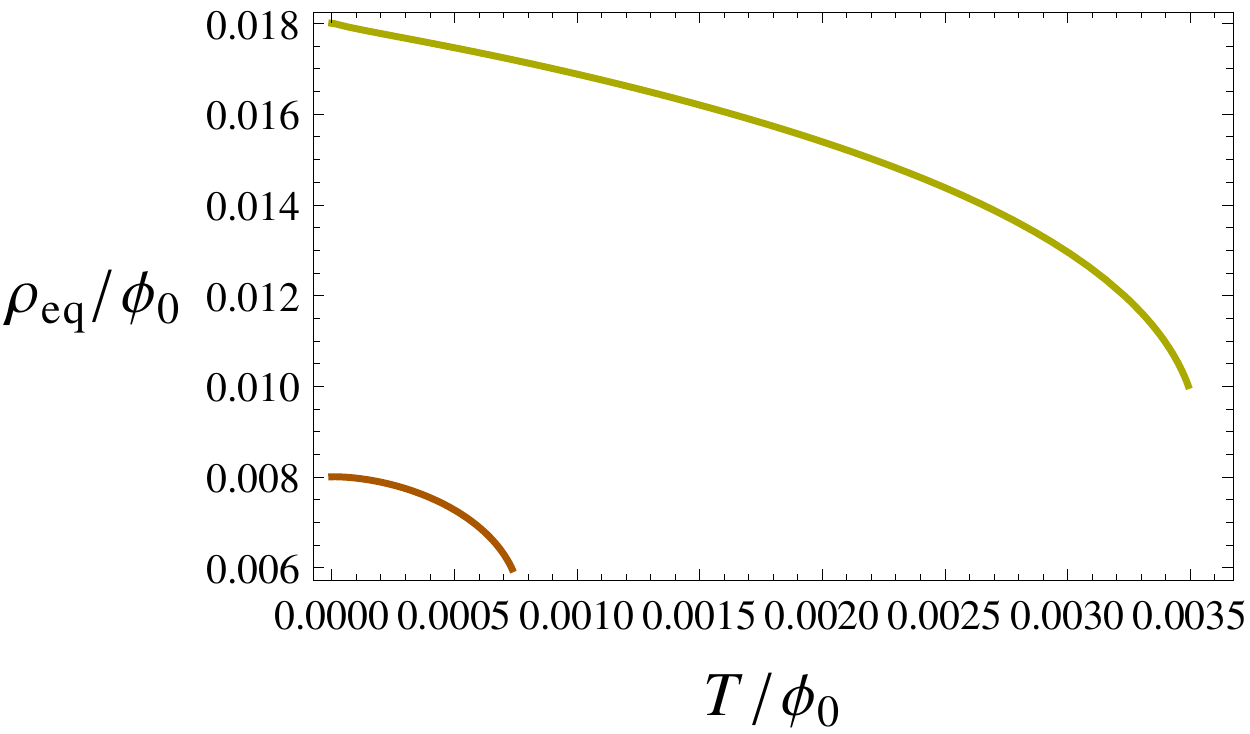}
}
\caption{{\bf Left}: Equilibrium distances $\rho_{\rm eq}/\phi_0$ for $\phi_1/\phi_0=-0.297$ with $\phi_0>0$. The upper curve shows bound state distances for probes bound to the stable black brane, the lower curve for probes bound to the unstable brane. The probe charge is chosen such that $V_p$ is lowest at its minimum. The two curves meet at the boundary of the white region where the solutions degenerate. {\bf Right}: Equilibrium distances $\rho_{\rm eq}/\phi_0$ for $\phi_1/\phi_0=-0.32$ and $\phi_0>0$. Again the upper curve shows bound state distances for probes bound to the stable black brane. Note that $\rho_{\rm eq}$ never vanishes within the existence regions for bound states. 
\label{fig:DistfofT}}
\end{figure}

The thick lines in figures~\ref{fig:minima} and~\ref{fig:potderiv} coincide with $\rho_{\rm eq}=0$ and represent the boundary of the allowed region of $\kappa$s admitting bound states for a given $\phi_1/\phi_0$ at $T=0$.  Naturally, one might wonder if $\rho_{\rm eq}=0$ identically at the edges of the various $(s/u\,\pm)$ regions in figure~\ref{fig:existdiagplan}. The answer is no. To show this, in figure~\ref{fig:DistfofT} we plot $\rho_{\rm eq}$ as a function of $T/\phi_0$ for fixed $\phi_1/\phi_0=-0.297$ and $-0.32$ for the numerically found probe charge such that $V_p$ is lowest at its minimum. Within the exitence region, $\rho_{\rm eq}$ never vanishes, remaining finite until the bound state disappears completely. 

\subsection{Small black holes, caged wall crossing and AdS-goop}\label{adsgoop}

\subsubsection{Small black hole limit}

Consider again the small black hole / asymptotically flat space limit discussed in section \ref{sec:flatlimit}, more specifically the flat space BPS limit, i.e.\ $\Delta_0 = \epsilon \delta_0$, $\Delta_1 = \epsilon \delta_1$, $\epsilon \to 0$, $\phi_0 \to \frac{\sqrt{3}}{2}$, $\phi_1 \to \frac{\sqrt{3}}{4}$. In this limit (\ref{Teq}) is solved on the small black hole branch by 
\begin{equation}
 u_1 = \frac{\epsilon}{4 \pi T} \sqrt{\delta_0 \delta_1} \, ,
\end{equation}
where $T$ is the temperature in AdS units, which can take any finite value. Furthermore $r_+ = \frac{\epsilon^2}{4 \pi T} \sqrt{\delta_0 \delta_1^3}$ and $u_0 = \frac{\epsilon}{4 \pi T} \sqrt{\frac{\delta_1^3}{\delta_0}}$, and if we restrict to values of $\rho$ of order $\epsilon$, the probe potential is given by
\begin{equation} \label{smallBHVp}
 {V}_p = \frac{\sqrt{3}}{2} \rho \biggl[ 
 \sqrt{
  \biggl(\frac{{p}_1}{2 f_1}  + \frac{{q}_0}{f_0} \biggr)^2 + \frac{f_0}{f_1} \biggl(\frac{{p}_0}{6\,f_1}   -\frac{{q}_1}{f_0} \biggr)^2
  } - \left(\frac{{p}_1}{2 f_1}  + \frac{{q}_0}{f_0} \right) 
\biggr] \, ,
\end{equation}
up to subleading terms at small $\epsilon \to 0$. This is minimized at ${V}_p=0$ when $\frac{{p}_0}{6\,f_1}   -\frac{{q}_1}{f_0}  = 0$, or equivalently at $r \approx \rho=\rho_{\rm eq}$ where
\begin{equation}
 {\rho}_{\rm eq} = \frac{\epsilon}{4 \pi T} \sqrt{\frac{\delta_1}{\delta_0}} \frac{{p}_0 \delta_1 - 6\, {q}_1 \delta_0}{6\, {q}_1 - {p}_0} = \sqrt{12} \; \frac{{p}^0 {Q}_0 - {q}_1 {P}^1}{6 \, {q}_1 - {p}_0} \, .
\end{equation}
Returning to the original, non-rescaled variables, this becomes 
\begin{equation} \label{rhoeq}
 \rho_{\rm eq} = \ell_p \frac{p^0 Q_0-q_1 P_1}{q_1 \sqrt{\frac{3}{\yz}}- p^0 \sqrt{\frac{\yz^3}{12}}} \, ,
\end{equation}
reproducing the well known BPS equilibrium separation formula \cite{Denef:2000nb}. Bound states of this kind exist if $1<\xi<\alpha$ or $\alpha<\xi<1$, where $\alpha \equiv \frac{\Delta_1}{\Delta_0}=\frac{6 \, Q_0}{P_1 \yz^2}$ and $\xi \equiv \frac{6 \, q_1}{p_0 \yz^2}$ (restoring the original $\yz$ dependence here to make the dependence on the scalar manifest). When $\xi \to 1$, the expression for $\rho_{\rm eq}$ given in (\ref{rhoeq}) diverges. In the asymptotically flat case, this corresponds to decay at marginal stability, also know as wall crossing: the bound state disappears from the spectrum once $\xi$ has crossed the wall. In the present case however, the divergence merely signals we exit the regime of validity of the small $\rho$ approximation. Indeed, since AdS acts as an infinitely deep gravitational potential well, the true radius cannot diverge; instead when $\rho_{\rm eq}$ becomes of order $\ell$ the bound state will start feeling the confining effect of AdS. We return to this below. When $\xi \to \alpha$, the bound state radius vanishes and the two centers merge. When $\alpha=1$, the bound states around the small black hole disappear altogether. This is easy to understand: At this locus, the background solution reduces to the constant scalar Reissner-Nordstrom solution, and without running scalars, there cannot be a stable potential. We refer to \cite{arXiv:1108.5821} for further discussion. 

\subsubsection{Caged wall crossing} \label{sec:confwall}

When $\rho$ is no longer restricted to order $\epsilon$ values and is allowed to get larger, the potential given in (\ref{smallBHVp}) --- i.e.\ the probe potential in asymptotically flat space --- is no longer accurate. Instead of the factor $\rho$, the gravitational part of the potential gets a factor $\sqrt{\rho^2+\rho^4}$. Thus the proper potential is ${V}_p = {V}_p(\text{above}) + \delta {V}_p$, where the correction term is (still to leading order at small $\epsilon$): 
\begin{equation}
 \delta {V}_p = \left( \sqrt{1+\rho^2} - 1 \right) \cdot \frac{\sqrt{3}}{2} \, \rho \, \sqrt{
  \biggl(\frac{{p}_1}{2 f_1}  + \frac{{q}_0}{f_0} \biggr)^2 + \frac{f_0}{f_1} \biggl(\frac{{p}_0}{6\,f_1}   -\frac{{q}_1}{f_0} \biggr)^2
  } \, .
\end{equation}
When $\rho$ is of order $\epsilon$, this is a negligible correction. When $\rho \gg \epsilon$ on the other hand, we have $f_0 \approx f_1 \approx \rho$, and
\begin{equation}
 \delta {V}_p \approx \left( \sqrt{1+\rho^2} - 1 \right) \frac{\sqrt{3}}{2} \, \sqrt{
  \bigl(\frac{{p}_1}{2}  + {q}_0 \bigr)^2 + \bigl(\frac{{p}_0}{6}   - {q}_1 \bigr)^2
  }  \, .
\end{equation}
The quantity multiplying the $\rho$-dependent factor is nothing but the (rescaled) mass of the probe in the vacuum; that is, $\delta {V}_p  \approx {m}_\gamma ( \sqrt{1+\rho^2} - 1 )$.
As alluded to earlier, the presence of this confining potential term is that no actual decay will happen when crossing the analog of a wall of marginal stability, i.e.\ when varying parameters such that we pass through $\xi \equiv \frac{6 \, q_1}{p_0 \yz^2}=1$ (from above or below depending on the ratio $\frac{\Delta_1}{\Delta_0}$). However, something nontrivial does happen when $\xi$ approaches 1. As long as $\xi$ is bounded away from 1, the minimum of the potential $\rho_{\rm eq}$ will be of order $\epsilon$. When $\xi$ approaches 1, this will rapidly increase to a much large radius, and roughly stabilise there. At the same time, the local minimum will get lifted well above its near-BPS value, thus becoming metastable for decay back into the global minimum at $\rho=0$. Eventually the local minimum may disappear altogether.   

To get some intuition, let us use the following toy model for the potential:
\begin{equation}
 V(\rho) = \left(\frac{\epsilon}{\rho} + \theta \right)^2  + \rho^2 \, .
\end{equation}
The first term represents the flat space potential, the second term the AdS correction.  This captures the typical behavior of the probe potential of interest quite well as long as $\rho$ is well below $1$ but not much smaller than $\epsilon$. Now, as long as $\theta \ll -\epsilon$, there will be a local minimum near $\rho = -\epsilon/\theta$ (obtained by minimizing the first term at zero), with energy $V \sim \epsilon^2/\theta^2$ (from the correction). This corresponds to bound state of size $\epsilon$, very close to its flat space BPS analog. When $\theta$ becomes positive, the flat space state disappears. In contrast, the full potential in AdS still has a local minimum, at $\rho \approx \theta^{1/3} \epsilon^{1/3}$, with an energy $V \sim \theta^2$ (for $\theta \gg \sqrt{\epsilon}$). These scalings with $\epsilon$ are consistent with numerical observations. Note however that this is entirely due to the gravitational trapping effect of AdS, the additional inter-particle interaction being now repulsive over the entire range of distances.

\subsubsection{AdS supergoop}

A natural question is how to generalize the two-particle black hole - probe picture developed so far to a system of $n>2$ interacting dyonic particles in AdS. In asymptotically flat space with unbroken ${\cal N}=2$ supersymmetry, at low energies and for well-separated dyons (which can be black holes, solitons or D-particles), a universal description is provided by a particular $\CN=4$ supersymmetric ``quiver'' quantum mechanics \cite{Denef:2002ru} (see also \cite{Avery:2007xf,Diaconescu:1997ut,D'Hoker:1985et,D'Hoker:1986uh, D'Hoker:1985kb,Feher:1989xw,Feher:1988th,Bloore:1992fv,Coles:1990hr,Horvathy:2006hx,Ivanov:2002pc,armenian,Anninos:2012gk,Manschot:2013sya,Manschot:2012rx,Manschot:2011xc,Manschot:2010qz}). The supersymmetry completely fixes the static potential and magnetic interactions up to a set of integers $\kappa_{ij}$ equal to the symplectic product of the electromagnetic charges of particle pairs $(i,j)$, $i,j = 1,\ldots,n$, and a set of real numbers $\theta_i$ determined by the charges and by vacuum moduli. In turn this completely determines the degeneracies of BPS bound states (which tends to be large due to the large Landau level degeneracies induced by the simultaneous presence of magnetic and electric monopole charges). Explicitly in flat space the $n$-particle static potential is of the form
\begin{equation}
 V^{(n)}_{\text{flat}} = \sum_{i=1}^n \frac{1}{2 m_i} \biggl( \sum_{j=1}^n \frac{\kappa_{ij}}{2|{\bf x}_i-{\bf x}_j| } + \theta_i \biggr)^2.
\end{equation}
The magnetic interaction is of Dirac monopole form and completely determined by the $\kappa_{ij}$; we refer to \cite{Denef:2002ru} for details. 

In AdS we do not have the same bulk supersymmetry structure, and hence it is not obvious what the appropriate generalization should be. However the considerations made in section \ref{sec:confwall}, as well as more elementary considerations regarding the effective Newtonian description of nonrelativistic particles confined to global AdS, suggest the following simple modification of the static potential:
\begin{equation}
 V^{(n)}_{\text{AdS}} = V^{(n)}_{\text{flat}} + \sum_{i=1}^n \frac{1}{2} \frac{m_i {\mathbf{x}}_i^2}{\ell^2}  \, ,
\end{equation}
where $\ell$ is the AdS length and ${\mathbf{x}}_i$ is the position of the $i$-th particle in isotropic coordinates. Indeed this is the effective Newtonian potential one gets for a nonrelativistic probe particle moving in global AdS$_4$, when expanding the metric in isotropic coordinates,
\begin{equation}\label{eq:isotropicads}
ds^2=\frac{-\bigl(1+\frac{\mathbf{x}^2}{4\ell^2}\bigr)^2 dt^2+d\mathbf{x}^2}{\bigl(1-\frac{\mathbf{x}^2}{4\ell^2}\bigr)^2}~,
\end{equation}
at small velocities and small potential energies. Isotropic coordinates are appropriate here, as they allow us to keep the translationally invariant flat space expressions for the static and magnetic interaction potentials. 

It would be interesting to study dynamical aspects of this system, along the lines of the analogous flat space study of \cite{Anninos:2012gk}. Due to the magnetic interactions, the dynamics has rather peculiar properties, with magnetic trapping, dynamical rigidity and precession drift being some of the more striking features. A key differences with the flat space system is that supersymmetry is broken. At the classical level one expects the high-dimensional moduli space to get lifted; at the quantum level one expects similarly the lowest Landau level to split up.

\section{Relaxation dynamics} \label{sec:relaxationdynamics}

In this section we initiate a study of the relaxation dynamics of metastable probe clouds. We will see that even when ignoring interactions between the probes, the system exhibits ``aging'' behavior typical for glasses. We begin by outlining the general ideas, and then apply this to our setup. 

\subsection{Slow relaxation and aging: general idea}

As observed in section \ref{sec:thermointerpr}, to leading order in the probe approximation, the probe potential can be identified with the system's free energy relative to the probe being inside the black hole. In particular, bound states with $V_{\rm min}<0$ are thermodynamically preferred and thus can be expected to be populated over time, while bound states with $V_{\rm min}>0$ are metastable. 

However, transitions of probes in and out of the black hole will generically be exponentially slow at large $\N$. A transition induced by thermal activation will have a rate suppressed by $e^{-\Delta F/T}$, where $\Delta F$ is the free energy barrier, while a transition mediated by quantum tunneling will have a rate suppressed by $e^{-I}$, where $I=\int p dq$ is the tunneling action for a trajectory crossing the barrier. Both exponents scale linearly with $\N$, hence transition rates will be exponentially suppressed at large $\N$. The coefficients $c$ in the transition rates $\Gamma \sim e^{-c \N}$ depends on the charges of the probe and on the background parameters. As we will see, at large $\N$ and $v$, there is a parametrically large number of probe charges that form bound states, leading to a broad, quasi-continuous distribution of values for $c$, and hence to a broad distribution of exponential time scales.

\subsubsection{Aging}

\def\CO{{\cal O}}

On general grounds, in such a situation, one may expect ``aging'' phenomena to occur, i.e.\ the system exhibits age-dependent relaxation behavior which breaks time translation invariance but exhibits approximate scale invariance. More concretely this means the following. Consider a system ``born'' at a time $t=0$, and say we are interested in some observable $\CO$. The system could for example be a glass sample produced by a rapid cooling quench at $t=0$, and the observable $\CO$ its dielectric constant. For ordinary, non-aging systems, equilibrium will be reached on microscopic time-scales, after which $\CO$ will be constant, up to small fluctuations decaying exponentially on some characteristic microscopic time scale $\tau$, independent of its age. In contrast, for an aging system, $\CO(t)$ will forever evolve, and in addition obeys 
\begin{equation}
 \CO(t_2)-\CO(t_1) \sim f(t_2/t_1)
\end{equation} 
for some function $f$. Thus, there is no time translation invariance, but instead we have scale invariance: the relaxation behavior depends on the age of the system, with all relevant time scales growing in proportion to age.

\subsubsection{Relation to metastability}

Let us sketch the basic idea of how aging can emerge from the presence of a very large number of exponentially long relaxation time scales with broadly and densely distributed exponents  \cite{amirshortreview,amirreview,amirtalk,mezard-aging,BouchaudPhysRep1990,Palmer1984}. Below, we will see in more detail how this is concretely realized in our setup. For now, let us just assume that at $t=0$ we quench the system of interest in some state that is not its equilibrium state, and that after this time it relaxes towards equilibrium along many different decay channels, characterized by exponentially large time scales $\tau(c) = \tau_0 e^{c N}$, with the set of values of $c$ smoothly distributed over some finite range. Suitable observables $\CO(t)$ will evolve in time accordingly, picking up contributions from a broad range of the metastable, decaying modes. Assuming the set of relaxation modes can be viewed as a continuum\footnote{we will give a discrete version of the argument below}, we can write
\begin{equation} \label{contdecay}
 \CO(t) = \int dc \, g(c) \, e^{-t/\tau(c)}\, , \qquad \tau(c) = \tau_0 \, e^{c N} \, .
\end{equation}
Here $g(c)$ is determined by the number density of relaxation modes with decay coefficient $c$, by the dependence of the observable on these modes, and by the initial occupation numbers of the modes, set by the quench at $t=0$. Let us assume all of these factors depend in a smooth, $N$-independent way on $c$, so that $g(c)$ can be taken to be a smoothly varying, $N$-independent function. Differentiating with respect to time and changing integration variables from $c$ to $\tau$, we get: 
\begin{equation} \label{ratechange}
  \frac{\partial}{\partial t} \CO(t)  = - \frac{1}{N} \int \frac{d\tau}{\tau^2} \, g(c_\tau) \, e^{-t/\tau} \approx - \frac{g(c_t)}{N} \, \frac{1}{t} \, , \qquad c_t = \frac{\log(t/\tau_0)}{N} \, .
\end{equation}
In the last step we made use of the assumption that $g$ is a slowly varying, $N$-independent function, implying that at large $N$, the integral only receives significant contributions from values of $\tau$ of the same order of magnitude as $t$, i.e.\ $c_\tau \approx c_t$. More explicitly, by expanding $g(c)$ around $c=c_t$, we get a $1/N$-expansion $\frac{\partial}{\partial t} \CO(t) = -\frac{1}{N} \frac{1}{t} \left( g(c_t) + \frac{\gamma}{N} g'(c_t) + \cdots \right)$. 

Since the dependence on $t$ of $c_t$ is logarithmic and $1/N$-suppressed, we can take $c_t$ to be approximately constant over many orders of magnitude.\footnote{For example if say $N=250$ and the microscopic time scale is $\tau_0 = 10^{-22} \, {\rm s}$, then for the range of time scales between $t \sim 1$ sec and $t \sim$ 1 day, $c_t$ ranges from $0.5$ to about $0.6$. For $t \sim$ 10 years, we get to $c_t \sim 0.7$.} Thus, if $t_*$ is the rough time scale at which we are doing measurements, (\ref{ratechange}) integrates simply to
\begin{equation}
 \CO(t_2) - \CO(t_1) \approx - \frac{g(c_{t_*})}{N} \, \log \frac{t_2}{t_1}.
\end{equation}
In other words, the system exhibits the (approximate) scale invariant aging behavior discussed above, with an aging rate set by $c_{t_*}$ and by the nature of the quench and other details of the system (which determines $g$). The essential feature leading to this conclusion is an approximately scale invariant distribution of relevant time scales, $dn \sim \frac{d\tau}{\tau}$. 

This kind of logarithmic aging behavior is observed in a huge variety of glassy materials, ranging from the length of wires carrying weights to the conductivity of electron glasses \cite{weber1835,ovadyahu2000,amirshortreview}. 

\subsubsection{Discrete case} \label{discretecase}

For our black hole system, we will not really have a continuum of time scales, but rather a discrete set, corresponding to the probe charges  allowing bound states. Nevertheless, a sufficiently finely spaced set of charges is sufficient to get the logarithmic aging behavior described above. To see this, start from the discrete version of (\ref{contdecay}), that is $\CO(t)  = \sum_i a_i \, e^{-t/\tau_i}$, where $\tau_i =  e^{c_i N}$. The rate of change of $\CO$ at time $t$ is then given by
$$
 \frac{\partial}{\partial t} \CO(t)  = -\frac{1}{t} \sum_i a_i \, \frac{t}{\tau_i} \, e^{-t/\tau_i} \, .  
$$
The idea is again that under suitable circumstances, this sum is dominated by terms with time scales $\tau_i$ of order $t$, that is by values of $c_i \approx c_t = \frac{\log (t/\tau_0)}{N}$. To see in more detail what suitable means, write $c_i = c_t + \delta_i$, so $\frac{\partial}{\partial t} \CO(t)  = -\frac{1}{t} \sum_i a_i \, e^{-\delta_i N - \exp(-N \delta_i)}$. The exponential factor is of order 1 when $|\delta_i| \lesssim 1/N$. It becomes exponentially small when $\delta_i \gg 1/N$, and double-exponentially small when $\delta_i \ll -1/N$. Hence, provided the spacing of $c_i$ values is finer than $1/N$ and there are no sharp peaks or gaps in the values of $a_i$, we have $\frac{\partial}{\partial t} \CO(t) \approx -\frac{1}{t} \sum_{i:|\delta_i| \lesssim 1/N} a_i$, hence for times $t_1$, $t_2$ roughly of order $t_*$: 
\begin{equation}
 \CO(t_2) - \CO(t_1) \approx -a_{t_*} \, \log\frac{t_2}{t_1} \, , \qquad  a_{t_*} =\sum_{i:|c_i-c_{t_*}| \lesssim \frac{1}{N}} a_i \, .
\end{equation}
In other words, we get smooth logarithmic aging provided the values of $\log \tau_i$ are roughly uniformly distributed with spacings of order 1 or less.

\subsection{Application to metastable clouds of probe charges} \label{sec:meta-aging}

\begin{figure}
\centering{
\includegraphics[width=\textwidth]{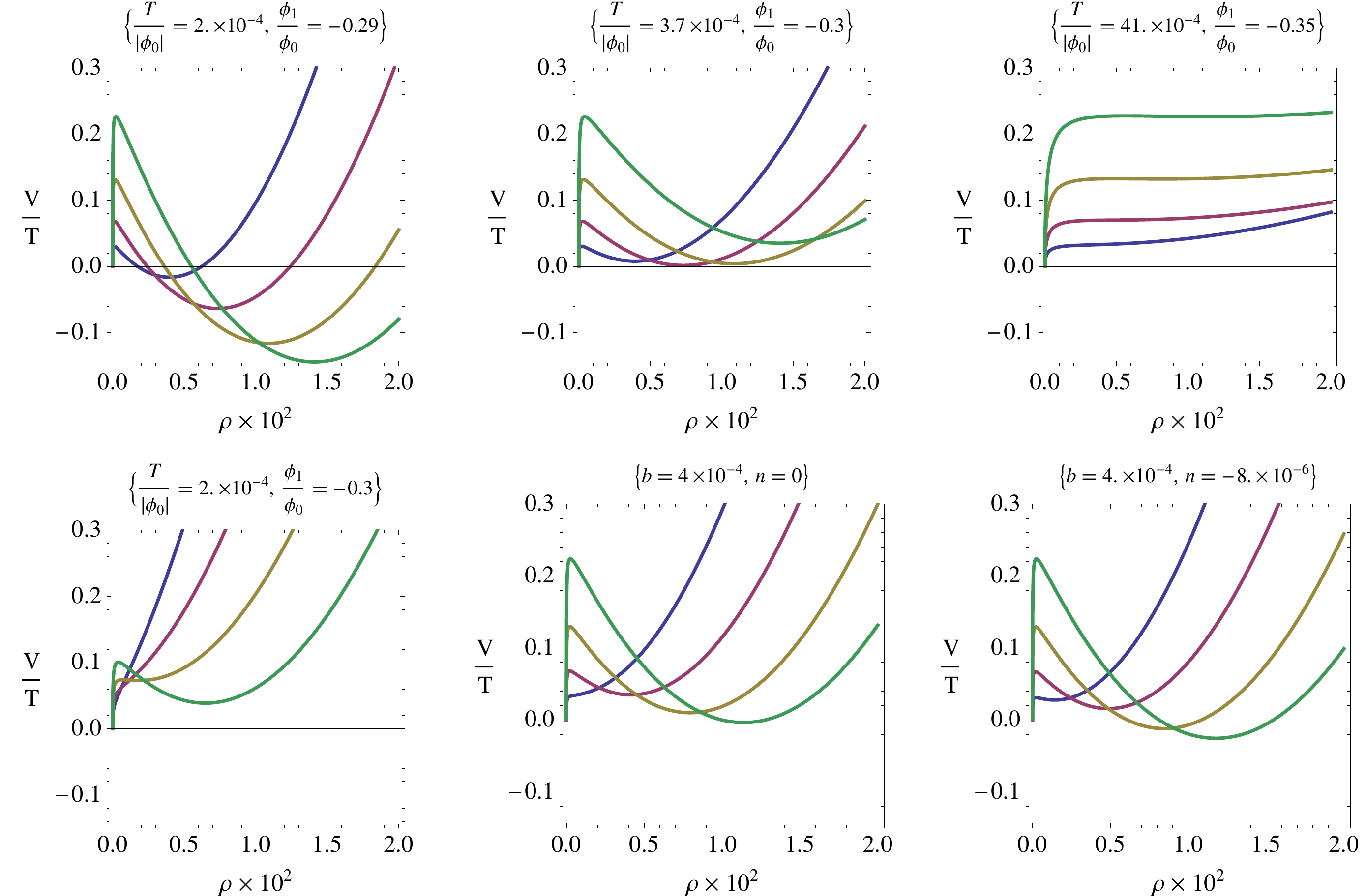}
}
\caption{Examples of probe potentials (divided by $T$) for a range of parameters. In each plot, we show the potential for $\kappa=0.15, 0.2, 0.25$ and $0.3$, respectively in blue, red, yellow, green. (Recall the probe charge parametrization by $(\kappa,b,n)$ given in (\ref{bnparametrization}).) The temperature and chemical potential are indicated above the plots. We label the panels by (row, column). In panel (1,1) we have $P^1=1$, $Q_0=-10^{-3}$ and $\delta m \equiv (M-M_0)/M_0 = 3 \times 10^{-7}$, where $M_0$ is the energy at zero temperature. In panel (1,2) we increased the energy to $\delta m = 10^{-6}$, and in $(1,3)$ to $\delta m = 10^{-4}$. In panel (2,1) on the other hand we kept $\delta m$ as in (1,1), but changed the D0-charge to $Q_0=-10^{-2}$. Finally panels (2,2) and (2,3) have the same background values as (1,1), but we changed the probe charge parameters $b$ and $n$ as indicated. 
\label{fig:examplepotentials}}
\end{figure}

Consider first a black brane in a region of parameter space where only metastable bound states exist, that is to say $V_{\gamma}^{\rm min} >0$ for all probe charges $\gamma$. In this case the equilibrium density (\ref{mgammadensity2}), $n_{\gamma}^{\rm eq} \sim e^{-V_{\gamma}^{\rm min}/T}$, will be exponentially small at large $N$ for all probe charges, since the probe potential $V_{\gamma}$ scales linearly with $N$. Imagine however that by a suitable quench procedure,\footnote{\label{fnquench} We will not try to explicitly describe such a procedure here but assume it can be done and only study the subsequent relaxation dynamics. Possibilities could include the injection of a hot gas of many particles in a cold black hole background, the collision of two black holes at very high energies, creating a plasma ball which subsequently decays into many charged particles (the analog of a collision of heavy ions hadronizing into jets of baryons), or a rapid change in the asymptotic parameters of the solution which may thermalize in part into a gas of charged particles of which a fraction will get trapped in potential wells. Since thermal relaxation proceeds on non-exponential time scales, the cloud thus formed will cool down relatively rapidly, after which it will follow the slow relaxation dynamics described here.} we populate the metastable states such that for a large set of charges $\gamma$, the densities $n_\gamma$ at time $t=0$ are \emph{not} exponentially small. The sizable, metastable charge cloud we have thus created will then slowly decay back into the black hole, with a broad distribution of many exponentially large time scales. Observables depending significantly on the amount of electric or magnetic charges in the cloud may therefore be expected to exhibit aging behavior. We will now argue in more detail that this is indeed the case.

\subsubsection{Time evolution of cloud particle densities}

In a classical stochastic picture, ignoring interactions between the probes, the time evolution of the probe number densities $n_\gamma(t)$ is given by
\begin{equation}
 \frac{dn_\gamma}{dt} = \Gamma_{\gamma}^{\rm out} - \Gamma_{\gamma}^{\rm in} \, n_\gamma \, , 
\end{equation}
where $\Gamma^{\rm out}_\gamma$ is the transition rate of probes out of the black hole into the metastable minimum and $\Gamma^{\rm in}_\gamma$ the reverse (absorption) rate. This is solved in general by
\begin{equation} \label{generalstochasticsol}
 n_\gamma(t) = n_{\gamma}^{\rm eq} + \bigl( n_\gamma(0) - n_{\gamma}^{\rm eq} \bigr) e^{-\Gamma^{\rm in}_\gamma \, t} \, , \qquad n_{\gamma}^{\rm eq} \equiv \frac{\Gamma^{\rm out}_\gamma}{\Gamma^{\rm in}_\gamma}\, ,
\end{equation}
where $n_{\gamma}^{\rm eq}$ is the equilibrium density. For simplicity we will ignore the possibility of quantum tunneling here and only consider classical thermal activation processes. We will also ignore non-exponential prefactors. In this case the transition rates are
\begin{equation}
 \Gamma^+_\gamma = e^{- E_\gamma^{\rm in}/T}, \qquad
 \Gamma^-_\gamma = e^{- E_\gamma^{\rm out}/T}
\end{equation} 
where $ E_\gamma^{\rm in}$ and $ E_\gamma^{\rm out}$ are the potential barrier heights in and out of the black hole, respectively, that is, $ E_\gamma^{\rm in} = V_{\gamma}^{\rm max} - V_{\gamma}^{\rm min}$ and $ E_\gamma^{\rm out} = V_{\gamma}^{\rm max} - V_{\gamma}^{\rm hor} = V_{\gamma}^{\rm max}$. This leads to the correct detailed balance equilibrium densities $n_{\rm \gamma}^{\rm eq} \sim e^{-V^{\rm min}_\gamma/T}$. The relaxation dynamics will thus be entirely determined by the barrier properties of the probe potentials. To get an idea of how these depend on the background and probe parameters, we display some examples of potentials in fig.\ \ref{fig:examplepotentials}.

\subsubsection{Distribution of relaxation time scales and condition for aging}

\begin{figure}
\centering{
\includegraphics[width= \textwidth]{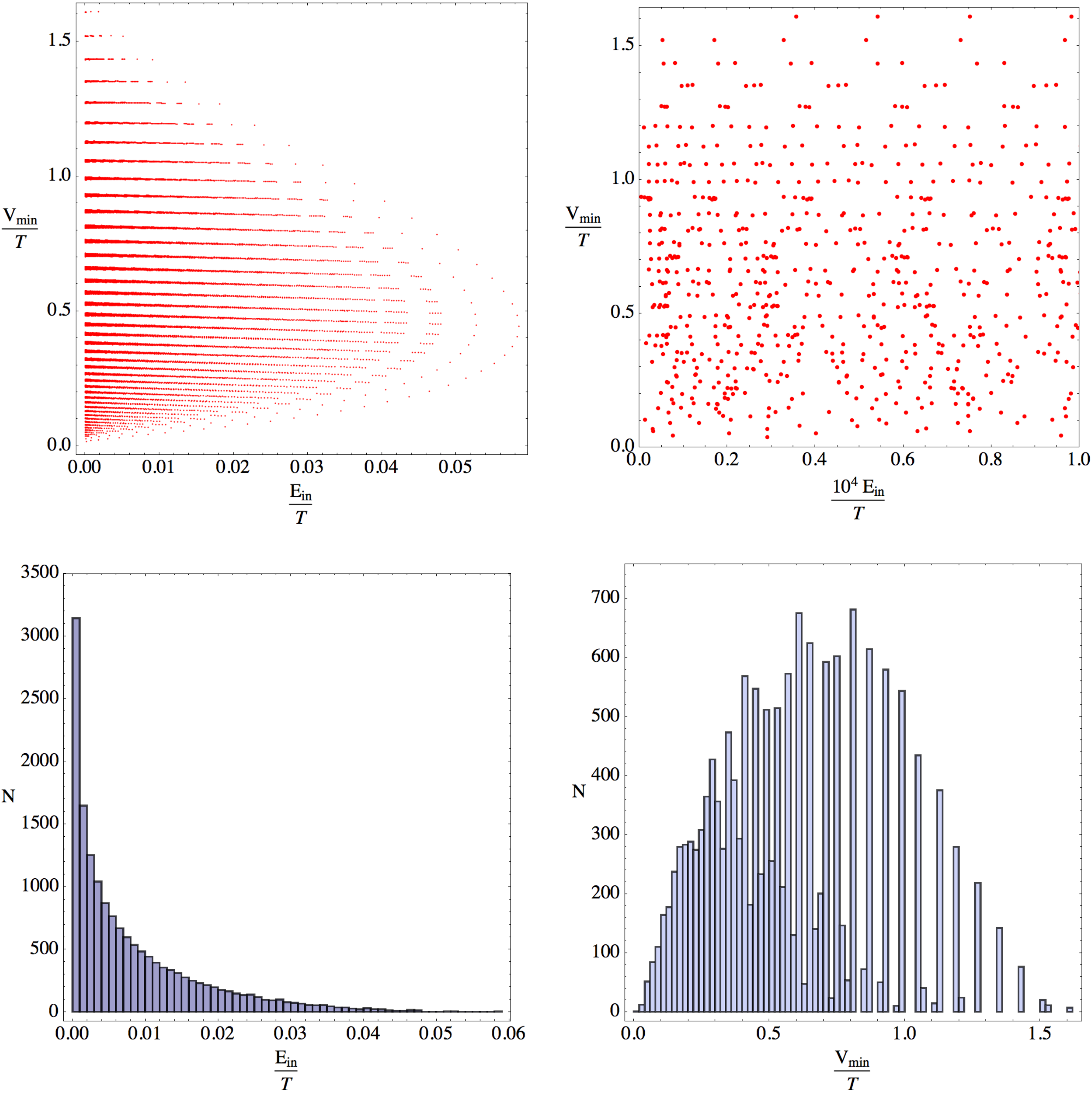}
}
\caption{Distributions of absorption barriers and potential minima for $v=100$, $Q_0=-10^{-3}$, $P^1=1$, $\delta m=10^{-5}$ ($T=1.3 \times 10^{-3}$, $\phi_0=-1.1$, $\phi_1=0.33$). In this case there are 15,862 charges with $p^0=1$ leading to bound states, displayed as dots in the upper left panel. When the system ages, states are decimated at a logarithmic pace from the left to the right. The upper right panel shows a close-up for very small values of the absorption barrier heights. The lower panels show histograms of respectively absorption barriers and potential minima. For $p^0=2$, there would be $2^3=8$ times as many points, distributed over a region scaled up by a factor of 2, and similarly for higher values of $p^0$. The peak in the density of states at small absorption barriers means that there will be a deviation of perfect logarithmic aging behavior over many orders of magnitude of time; more specifically aging will be faster when the system is young.
\label{fig:barrierdistr}}
\end{figure}

In the case under study (only metastable bound states), we can set $n_\gamma^{\rm eq} = 0$ for all practical purposes, so (\ref{generalstochasticsol}) reduces to a simple exponential decay of the densities $n_\gamma$, with time scale
\begin{equation}
 \tau_\gamma = e^{c_\gamma N} \, , \qquad c_\gamma = \frac{1}{N} \frac{ E^{\rm in}_\gamma}{T} = p^0 v \, \widehat{c}_\gamma \, , \qquad \widehat{c}_\gamma =\frac{\widehat{E^{\rm in}_\gamma}}{\widetilde{T}} \, .
\end{equation}
Here we explicitly reinstated the scaling with $N$ and $v=\sqrt{\frac{N}{k}}$, using $V \propto \frac{N g}{v}$ and $g = p^0 v^2$. Observables depending on the cloud charge densities (for example the conductivity and other transport coefficients, which will pick up a contribution from the cloud) can be expected to evolve in time with this spectrum of exponential decay time scales, matching the assumptions of our general discussion in section \ref{discretecase}. Thus observables of this kind can generically be expected to exhibit smooth logarithmic aging behavior, provided at least the spacing $\Delta c_\gamma$ of $c_\gamma$ values is much smaller than $1/N$. 

To see under which conditions this is true, we need to take into account charge quantization. Besides $p^0$, the integrally quantized probe charges are then $p^1 = p^0 v \hat{p}^1$, $q_1 = p^0 v^2 \hat{q}_1$ and $q_0 = p^0 v^3 \hat{q}_0$, where we recall that $\hat{p}^1=\kappa$, $\hat{q}_1=\frac{\kappa^2}{2}-b$, $\hat{q}_0=-\frac{\kappa^3}{6}+b \kappa + n$. This implies that the quantized values of the rescaled charge variables are quantized with the following spacings:
\begin{equation}
 \Delta \kappa = \Delta \hat{p}^1 = \frac{1}{p^0 v} \, , \quad  
 \Delta b = \Delta \hat{q}_1 = \frac{1}{p^0 v^2} \, , \quad
 \Delta n = \Delta \hat{q}_0 = \frac{1}{p^0 v^3} \, .
\end{equation}
The region in $(\hat{p}^1,\hat{q}_1,\hat{q}_0)$-space allowing bound states has finite volume. At fixed $p^0$, the number of quantized charges in this volume scales as $1/(\Delta \hat{p}^1 \, \Delta \hat{q}_1 \, \Delta \hat{q}_0) = (p^0)^3 v^6$. Recall that the validity of the probe approximation requires $g \ll N$, that is $p^0 \ll N/v^2 = k$. If we allow values of $p^0$ up to $p^0_{\rm max} = \epsilon N/v^2$ for some fixed $\epsilon \ll 1$, we thus get
\begin{equation}
 \CN \equiv \mbox{Total number of charges forming bound states} \sim \frac{\epsilon^4 N^4}{v^2} \, .
\end{equation}
Over this range of charges, $c_\gamma$ takes values from 0 to an order $(p^0)_{\rm max} v$ upper bound. The average spacing of $c_\gamma$ values near a generic point can therefore be expected\footnote{This is somewhat naive, but borne out by explicit enumeration of barriers in examples, as exemplified in fig.\ \ref{fig:barrierdistr}} to scale as $\Delta c_\gamma \sim p^0_{\rm max} v / \CN =v/\epsilon^3 N^3$. Hence the time scale spacing condition discussed in section \ref{discretecase}, $|\Delta c_\gamma| \ll 1/N$, reduces to 
\begin{equation}
 \frac{N^2}{v} \gg \frac{1}{\epsilon^3} \, .
\end{equation}
Recalling (\ref{abjmlike}), we see that the left hand side equals $(\ell/\ell_p)^2$ (or the central charge of the CFT if there is a holographic dual), so this is nothing but the condition that the classical gravity description is reliable! 

We conclude that whenever we trust the gravitational description, metastable probe clouds quenched to order one densities at time $t=0$ will exhibit logarithmic aging behavior when relaxing back to the pure black hole state. 

\subsubsection{Explicit example: aging of cloud D0-charge}

\begin{figure}
\centering{
\includegraphics[width= \textwidth]{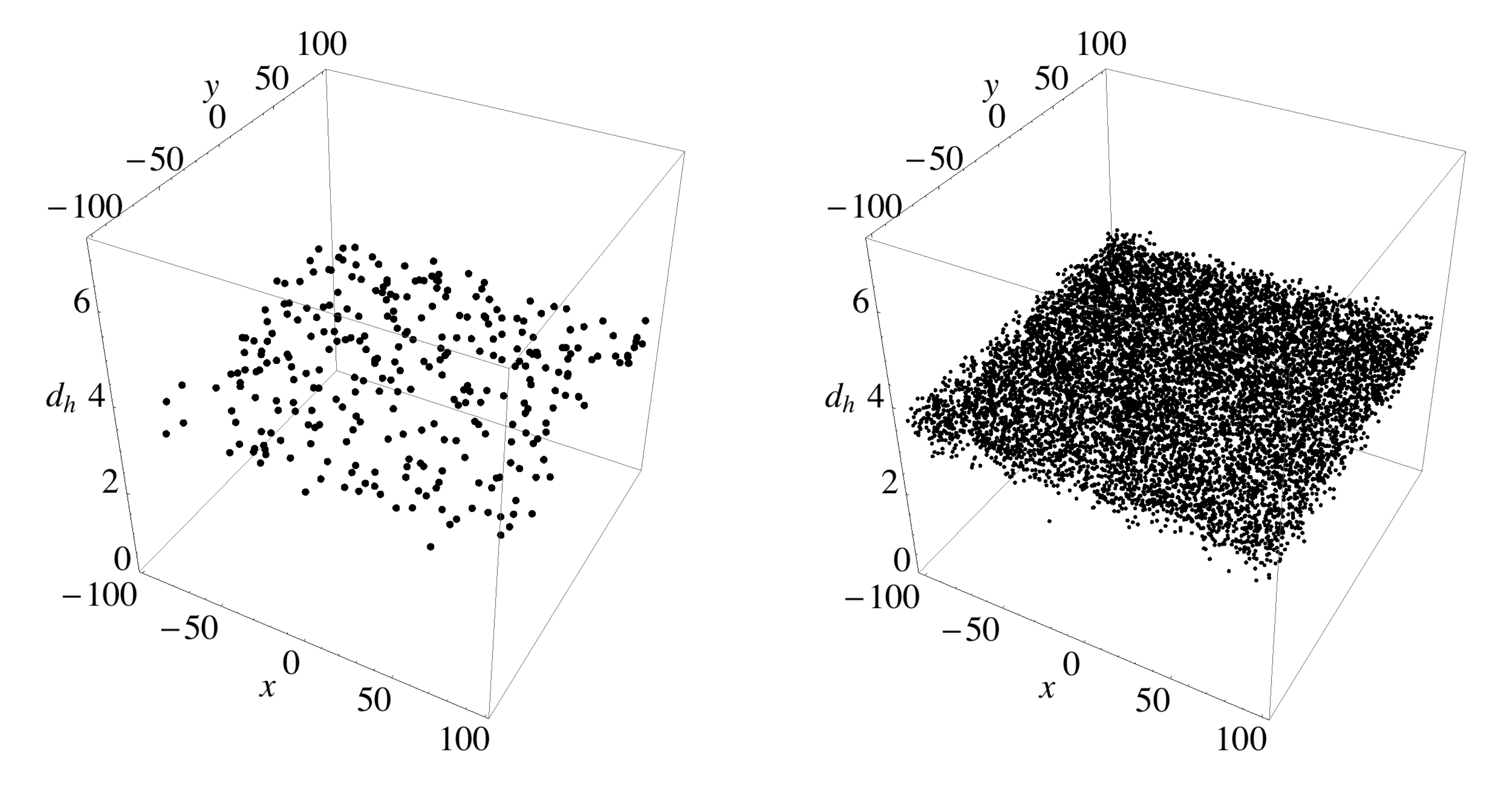}
}
\caption{Random samples of probe bound states for $Q_0=-10^{-3}$, $P^1=1$, $\delta m = 10^{-5}$, $v=100$, quenched at $n_\gamma(0) \propto e^{-m_\gamma/T_0}$ with $T_0 = v N/20$. The height $d_h=\upsilon_h-\upsilon_p$ is the ``optical distance'' between probe and  horizon, as defined in  (\ref{opticaldist}). The sample on the left has 300 probes in the cloud, i.e.\ a density $n_{\rm tot} = 0.75 \times 10^{-2}$.  The sample on the left has 10,000 probes, i.e.\ $n_{\rm tot} = 0.25$. 
\label{fig:holes}}
\end{figure}

\begin{figure}
\centering{
\includegraphics[width= \textwidth]{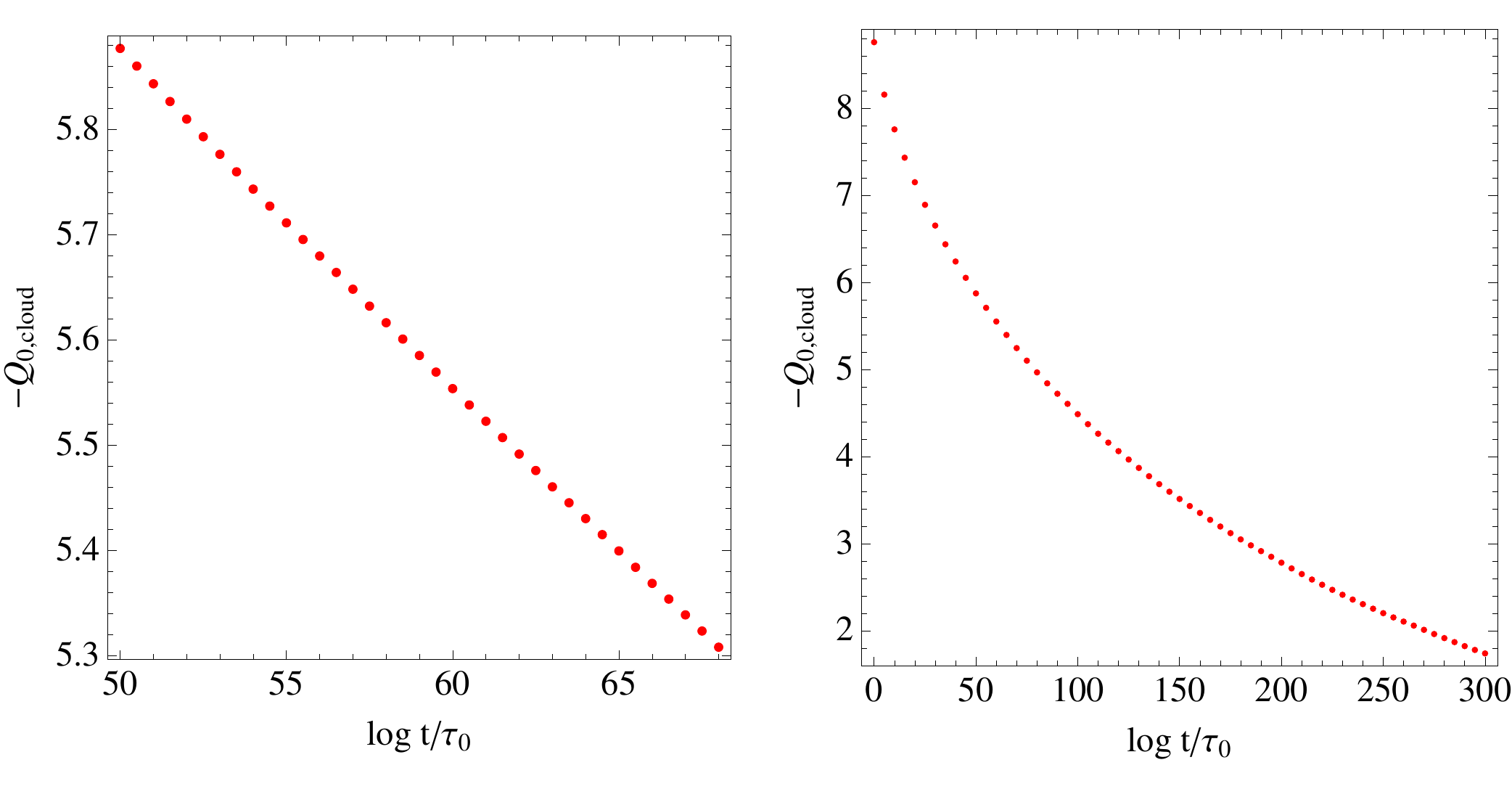}
}
\caption{Logarithmic aging of the cloud D0-charge. The background and quench parameters are the same as for figs.\ \ref{fig:barrierdistr} and \ref{fig:holes}, and we took $N=200$. If we imagine for concreteness the microscopic scale $\tau_0$ to be $\tau_0 = 10^{-22} \, {\rm s}$, then the plot on the left shows the aging behavior for $t$ ranging from 1 second to 1 year. The plot on the right then shows the aging behavior for $t$ ranging from femtoseconds to $10^{100}$ years. The deviation from exact logarithmic aging at this range of scales is due to the fact that the density of states  is higher for smaller barrier heights $E_{\rm in}/T$, as can be seen in fig.\ \ref{fig:barrierdistr}.
\label{fig:agingplots}}
\end{figure}

To check the above assertions, we consider an explicit example. Let the observable of interest be the total D0-charge in the cloud:
\begin{equation}
 \CO(t) = Q_{0}^{\rm cloud} = \sum_\gamma \gamma_0 \, N_\gamma(0) \, e^{-t/\tau_\gamma}  \, .
\end{equation}
Here $\gamma_0$ is the D0-charge component of the charge vector $\gamma$, and $N_\gamma(0) = \int n_\gamma(0)$ is the total charge in the region of interest, at time $t=0$. To be concrete, we will assume we quench the initial particle densities to be proportional to what their abundance would be in flat space at a high temperature $T_0=v N/20$, i.e.\ $n_\gamma(0) \propto e^{-m_\gamma/T_0}$. (At this temperature, the distribution of charges is dominated by charges with $p^0=\pm1, \pm2$.) We show the exact results for a specific choice of parameters in fig.\ \ref{fig:agingplots}. Clearly the results are in excellent agreement with the general discussion above. 

A natural question is whether there is a correlation between the distance of a bound probe to the horizon and the life expectancy of the bound state. Translated to the dual CFT, this becomes the question whether there is a correlation between the size of the inhomogeneity corresponding to the probe bound state and its life expectancy, with larger size mapping to shorter distances to the horizon. In section \ref{sec:pbhbopt} we will see that the relevant notion of distance here is the ``optical distance'', defined in (\ref{opticaldist}). The optical distance to the black hole horizon is proportional to the size of the charge inhomogeneity in the CFT. A similar question can be asked about correlations between free energy and distance/size. 

\begin{figure}
\centering{
\includegraphics[width= \textwidth]{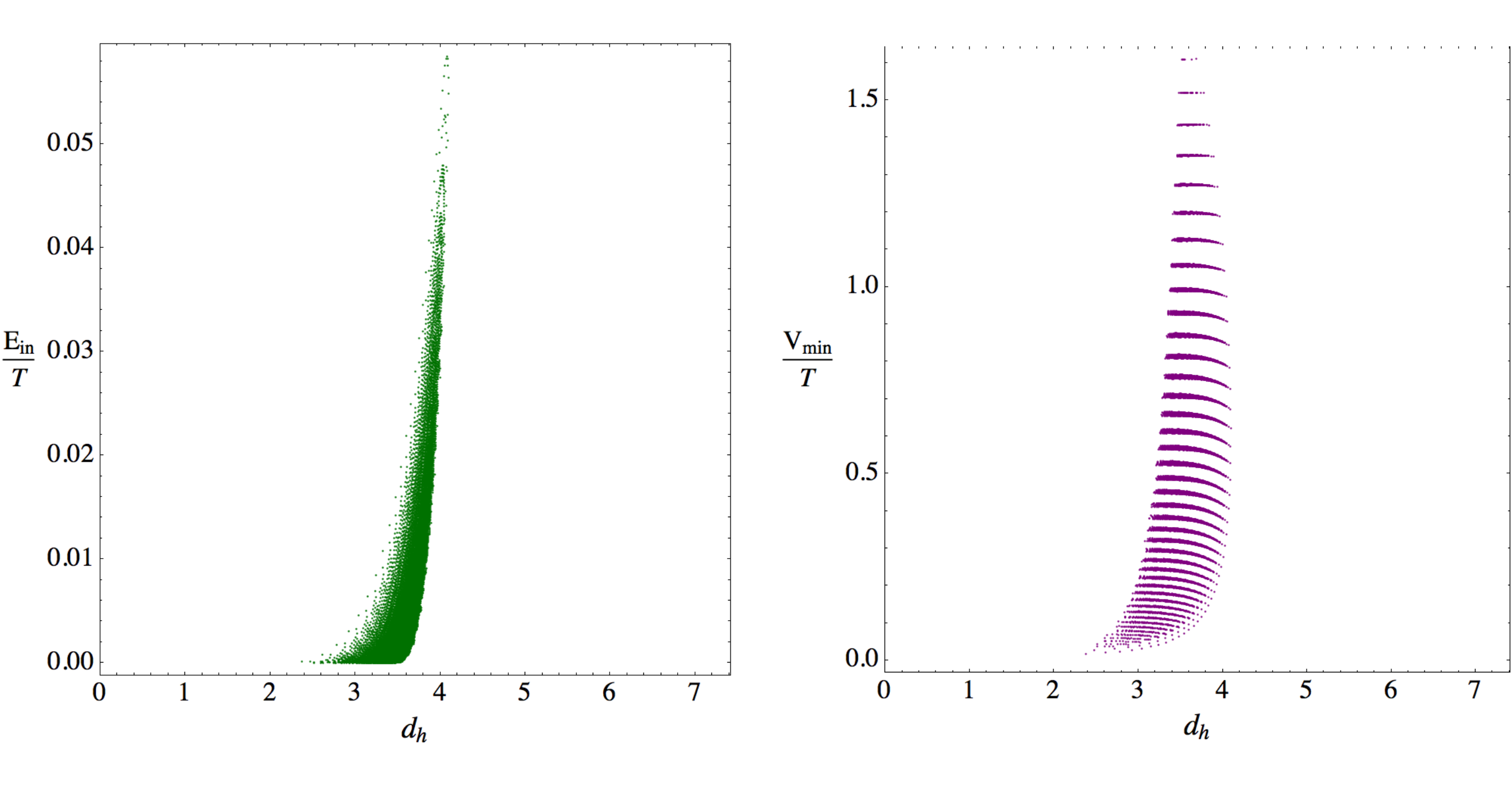}
}
\caption{Correlation between optical distances $d_h$ of bound probes from the horizon and their absorption barrier heights (left) and free energies (right). $d_h=0$ is the horizon, $d_h=7.4$ the boundary of AdS.
\label{fig:distplotsmeta}}
\end{figure}

To explore these questions, we plot in \ref{fig:distplotsmeta} the optical distance versus barrier height and free energy for all 15,862 charges with $p^0=1$ forming bound states. A first thing to note is that all bound states are localized within a fairly narrow band of distances, roughly in-between the horizon and the boundary of AdS. This is also evident from fig.\ \ref{fig:holes}. Thus, in the CFT, the inhomogeneities will have a fairly narrow range of characteristic scales. There is a mild correlation between distances and barrier heights, with the probes closest to the black hole having relatively low absorption barriers, and therefore relatively short lifetimes, and the probes most far away having the longest lifetimes. In the CFT, this translates to the smallest structures being the most stable. The bound states closest to the black hole also all have relatively low free energies, although those the most far away are not the ones with the highest free energies.

\subsection{Relaxation when stable bound states exist} \label{sec:aging-stab}

\begin{figure}
\centering{
\includegraphics[width= \textwidth]{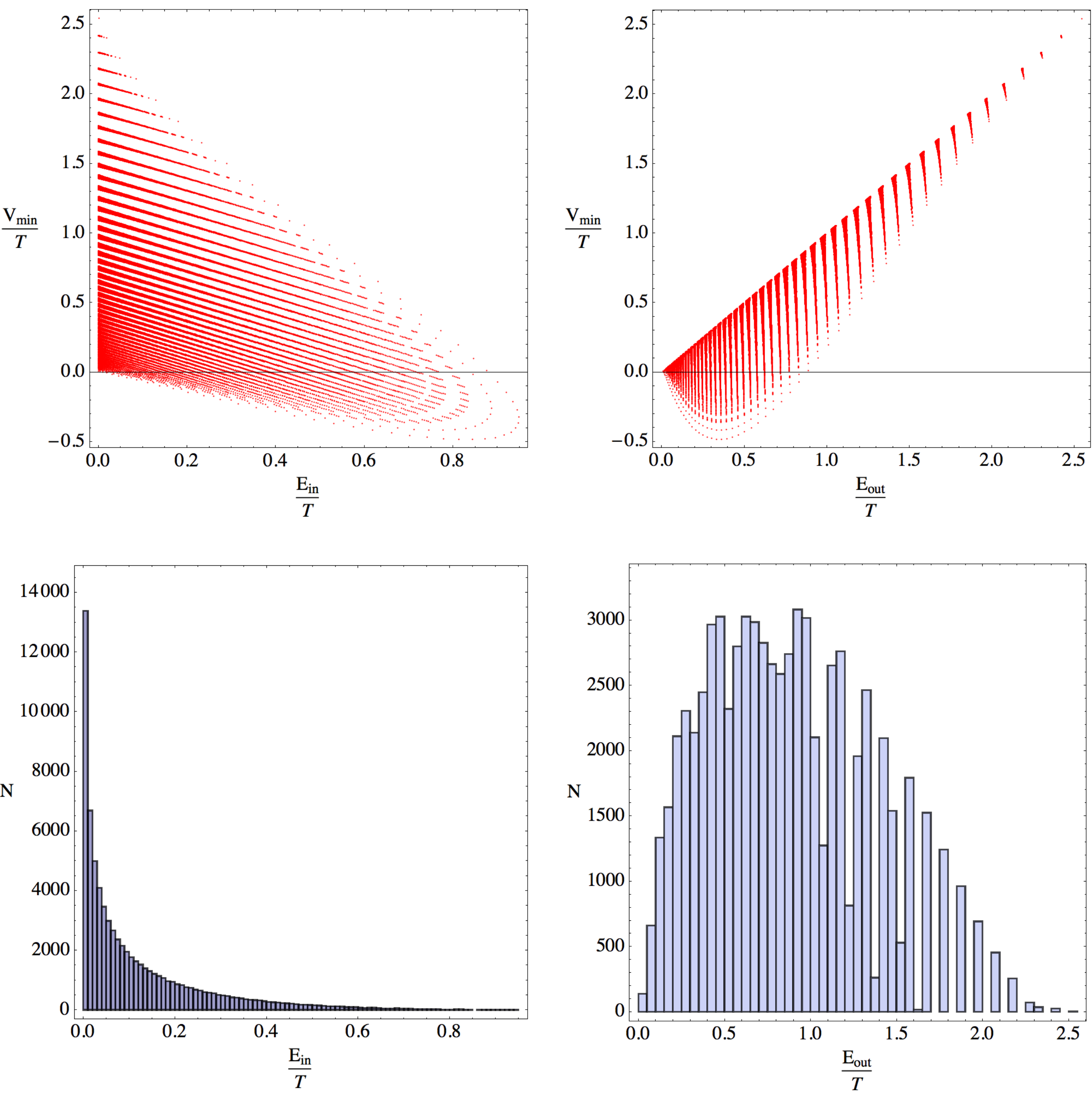}
}
\caption{Distributions of absorption and emission barriers for $v=100$, $Q_0=-10^{-3}$, $P^1=1$, $\delta m=10^{-7}$ ($T=1.3 \times 10^{-4}$, $\phi_0 = -1.1$, $\phi_1 = 0.33$). In this case there are 72,240 charges with $p^0=1$ leading to bound states. Occupied metastable ($V_{\rm min}>0$) states  are decimated at a logarithmic pace from the left to the right in the upper left panel. Unoccupied stable ($V_{\rm min}<0$) states are populated from the left to the right in the upper right panel. Notice that the lower free energy states take exponentially longer to populate than the higher ones, conceivably causing massive failure to properly equilibrate. 
\label{fig:barrierdistr10m7}}
\end{figure}

\begin{figure}
\centering{
\includegraphics[width= \textwidth]{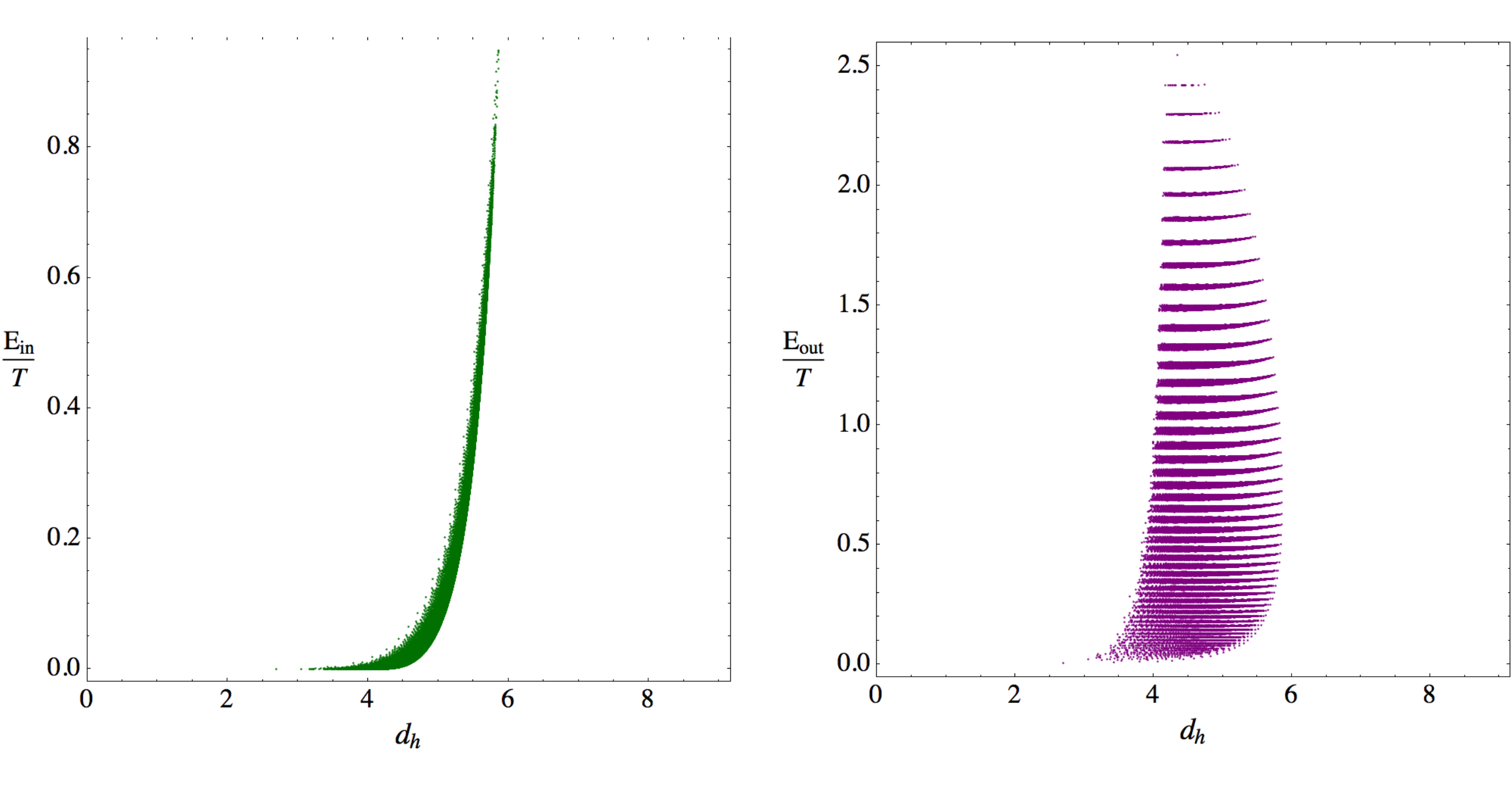}
}
\caption{Correlation between distances and barriers, analogous to fig.\ \ref{fig:distplotsmeta}, except that the plot on the right now has $E_{\rm out}/T$ on the vertical axis (as this is what sets the (naive) time scales for population of the stable minima). A sharper correlation is noticeable between absorption barrier heights and distances. A mild correlation exists between distances and emission barrier heights; the stable bound states closest to the horizon will form at the earliest times. A striking feature is again that bound states have a characteristic scale of order half the optical distance between horizon and boundary, despite the considerably lower temperature.
\label{fig:distplotsmeta10m7}}
\end{figure}

When stable bound states exist, that is states with $V_{\rm min}<0$, the relaxation dynamics changes considerably. In fact in the presence of such minima, the system quickly runs out of the regime where we have the simplifying control of the probe approximation, since now the formal equilibrium densities are exponentially large rather than exponentially small. Of course what this really means is that the probe approximation breaks down after some finite time. At least three effects may be expected to play a key role in the subsequent dynamics. First, due to charge depletion, the background will change by a non-negligible amount. Second, the presence of a large density of probes in a potential well may favor or disfavor the arrival of other probes, depending on their charges. And third, probes may begin to cluster and clump together, thus either forming larger black holes, or multi-centered bound states amongst each other. Over time, the system may thus be expected to sink irreversibly deeper into complicated bound states involving large numbers of centers of which the backreaction can no longer be neglected. 

Although this is conceivably the truly glassy regime, it clearly falls outside of the scope of the probe approximation. Nevertheless, some qualitative features about the onset of this phase can be made already with the results we have in hands. Let us assume we start with a black brane \emph{without} any bound probes, for example by a cooling quench of the temperature or the chemical potentials, starting from a high temperature black brane. From this point in time on, the negative potential minima will start to get populated, at a rate given by $\Gamma_\gamma^{\rm out} \sim e^{-E_{\rm out}/T}$. Naively, one might think the minima with the lowest $V_{\rm min}$ will get populated first, but interestingly, this is not so. In fact quite the opposite is happening, as is evident from fig. \ref{fig:barrierdistr10m7}: The lowest minima are shielded by the highest barriers, and they get populated at time scales many orders of magnitude larger than the shallower minima. In fact, the system will be far out of the probe regime long before the lowest free energy minima have even started to accumulate any noticeable charge. This is an explicit example of how for glassy systems, the relaxation path followed over time does not need to coincide at all with the steepest descent path towards the formal static equilibrium point.

\section{Holographic interpretation} \label{sec:holointerpret}

In the previous sections we have demonstrated the existence of black hole bound states in the probe approximation. The total number of different probe charges allowing bound states is proportional to $N^2/v \gg 1$, and besides the constraints related to the validity of the probe approximation, there is no limit in principle on the size or number of bound black holes. In the large $\N$ limit, each of these configurations corresponds to a (meta)stable macroscopic thermodynamic state, with individual black holes representing ``pockets'' of thermalized degrees of freedom existing at different positions and scales. They survive out of equilibrium for exponentially long times, and as we have seen in section \ref{sec:relaxationdynamics}, their relaxation dynamics naturally gives rise to logarithmic aging, eternally long in the large $N$ limit. These features are typical for glassy / amorphous systems, and thus we are lead to the hypothesis that these black hole bound states are in fact holographic descriptions of glassy phases of CFTs with a gravity dual.


As a first step to make this idea more precise, we now turn to a number of  observations relevant to the holographic interpretation of our results, assuming a dual CFT exists. For simplicity, and because it has the most straightforward thermodynamical interpretation, we will again focus on the planar limit. In this section we will make the distinction between rescaled variables introduced in section \ref{scalingsymm} and the original variables explicit again.

\subsection{Holographic dictionary for background} \label{sec:scales}

In this section we review the standard AdS-CFT dictionary for thermodynamic states dual to plain black holes or black branes. We will use the opportunity to fix  some normalizations conventions. 

The putative dual CFT has central charge proportional to 
\begin{equation} 
 C_{\rm CFT} \equiv \frac{\ell^2}{\ell_p^2} = \frac{\N^2}{v} \, .
\end{equation}
Spherical black holes are dual to thermal states of the CFT on a 2-sphere of radius $R$. Bulk energies in units of $1/\ell$ are identified with CFT energies in units of $1/R$; for example $\ell \, T = R \, T_{\rm CFT}$, $\ell \, \phi = R \, \phi_{\rm CFT}$, $\ell \, M =R \, E_{\rm CFT}$, and so on. Planar black holes are dual to thermal states on the infinite 2-dimensional plane. They are obtained by zooming in on a small solid angle of the 2-sphere and taking the radius $R$ of the 2-sphere to infinity while keeping the intensive variables fixed in the CFT. Indeed, defining $\lambda \equiv R/\ell \to \infty$, thermodynamic quantities will scale with $\lambda$ exactly as in the planar limit discussed in section \ref{sec:planarlimit}. With this identification, the barred intensive thermodynamic variables introduced there are directly identified with their CFT counterparts: $T_{\rm CFT} = \frac{\ell}{R} T = \bar{T}$, $\phi_{\rm CFT} = \bar{\phi}$. The barred extensive quantities on the other hand get identified with planar \emph{densities} of the CFT, upon multiplication by a factor $1/4 \pi \ell^2$; for example the entropy density of the CFT, defined as the entropy per unit coordinate volume, is $s \equiv s_{\rm CFT} \equiv \frac{S}{4 \pi R^2} = \frac{\bar{S}}{4 \pi \ell^2}$, and the energy density is $e \equiv e_{\rm CFT} = \frac{\bar{M}}{4 \pi \ell^2}$. 

The CFT interpretation of bulk electromagnetic response properties depends on the duality frame chosen in the bulk \cite{Witten:2003ya}, for the following reason. In the standard AdS-CFT dictionary, changes of the asymptotically constant mode of a bulk vector potential correspond to changes of external sources from the point of view of the CFT. In particular the bulk path integral is to be performed with these boundary values held fixed. The vector potential modes falling off as $1/r$ on the other hand are interpreted as currents; they are the response to the sources. More precisely, in gauge invariant terms, the component of the electric field normal to the boundary is identified with the charge density, $j^0 \sim \vec n \cdot \vec E$, and components of the magnetic field parallel to the boundary are identified with the current density: $\vec j \sim \vec n \times \vec B$. On the other hand the components of the electric field parallel to the boundary and the magnetic field component normal to the boundary are identified with external sources. Electromagnetic duality exchanges electric and magnetic fields, and from the above it is clear that this symmetry does not commute with the dictionary. In fact, 4-dimensional S-duality acts as a symmetry relating different 3-dimensional CFTs \cite{Witten:2003ya}.


In the explicit bulk Lagrangian (\ref{modelLag}), we assumed a duality frame in which $Q_0$ is electric and $P^1$ is magnetic. On the other hand, in our discussion of the thermodynamics of the background, we have been working in a grand canonical ensemble with fixed potentials $\phi_0$ and $\phi_1$, which are more naturally interpreted in a duality frame in which both $Q_0$ and $P^1$ are considered to be electric. So let us begin by assuming we are working in the latter frame. The bulk D0- and D4-charges are then identified with two global $U(1)$ charges in the CFT. Denoting the associated CFT charge densities by $J^t_0$ and $J^t_1$, we have (for homogeneous planar solutions) the identifications $J^t_0 = \frac{\bar{Q}_0}{4 \pi \ell^2}$, $J^t_1 = \frac{\bar{P}_1}{4 \pi \ell^2}$.

To summarize, CFT quantities are related as follows to the dimensionless, scaling invariant tilde-variables of section \ref{scalingsymm} (which we used for example in all the phase diagrams of the preceding sections):
\begin{equation}
 T_{\rm CFT} = \frac{1}{\ell} \, \widetilde{T} \, , \qquad
 \phi_{0,\rm CFT} = \frac{N}{\yz^2 \ell} \, \widetilde{\phi}_0 \, , \qquad
 \phi_{1,\rm CFT} = \frac{N}{\ell} \, \widetilde{\phi}_1 \, , 
\end{equation}
and
\begin{equation}
 s = \frac{\N^2}{\yz} \frac{\widetilde{{S}}}{4 \pi \ell^2} \, , \qquad
 e = \frac{\N^2}{\yz \ell} \frac{\widetilde{{M}}}{4 \pi \ell^2} \, , \qquad
 J^t_0 = \N \yz \frac{\widetilde{{Q}}_0}{4 \pi \ell^2} \, , \qquad
 J^t_1 = \frac{N}{\yz} \frac{\widetilde{{P}}^1}{4 \pi \ell^2} \, , \label{bkgdens}
\end{equation}
where for example $\widetilde{{S}} = \pi \sqrt{\widetilde{{u}}_0 \widetilde{{u}}_1^3}$. Transport coefficients are easily obtained by making use of the general formulae of e.g.\ \cite{Iqbal:2008by}. The D0-charge DC conductivity $\sigma_0$, susceptibility $\Xi_0$ and diffusion coefficient $D_0$ are:
\begin{equation}
 \sigma_0 = \frac{y_{\rm hor}^3}{12 \pi} = \frac{v^3}{12 \pi} \frac{\widetilde{u}_0^{3/2}}{\widetilde{u}_1^{3/2}} \, , \qquad
 \Xi_0 = \frac{J^t_0}{\phi_{0,\rm CFT}} = \frac{v^3}{12 \pi \ell} \, \widetilde{{u}}_0 \, , \qquad
 D_0 = \ell \, \frac{\widetilde{u}_0^{1/2}}{\widetilde{u}_1^{3/2}} \, .
\end{equation}
Here we made use of the explicit expressions given in section \ref{sec:planarlimit}.
 Similarly the D4-charge transport coefficients are
\begin{equation}
 \sigma_1 = \frac{1}{\pi y_{\rm hor}} = \frac{1}{\pi v} \frac{\widetilde{u}_1^{1/2}}{\widetilde{u}_0^{1/2}} \, , \qquad
 \Xi_1 = \frac{J^t_1}{\phi_{1,\rm CFT}} = \frac{1}{\pi v \ell} \, \widetilde{{u}}_1 \, , \qquad
 D_1 = \ell \, \frac{1}{\widetilde{u}_0^{1/2} \widetilde{u}_1^{1/2}} \, .
\end{equation}
The charge transport coefficients satisfy the Einstein relation $\sigma = \Xi D$, as they should \cite{Iqbal:2008by}. As always (in single black hole setups at finite temperature), the viscosity is given by $\eta=s/4 \pi$. The expressions given above imply various relations between CFT quantities which are specific to the system under study, for instance $s = C_{\rm CFT}/4 \sqrt{D_0 D_1^3}$.


Finally, we briefly return to the issue of the choice of duality frame. If we had chosen a frame in which the D4 is magnetic and the D2 electric, then our background would have been interpreted in the CFT (different from the original CFT) as having zero charge density $J_1^{'t}$ and chemical potential $\phi'_{1,\rm CFT}$, but a nonzero magnetization field and a constant magnetization density. At this level, this can perhaps be viewed as merely a different use of words, but it becomes important when we want to deduce the effect of the presence of probes bound to the black hole, to which we turn next.

\subsection{Holographic dictionary for probes}

We now turn to the holographic interpretation of the black hole bound states.  Consider first the case of pure Maxwell electrodynamics with Lagrangian ${\cal L}=-\frac{1}{4g^2} F_{\mu\nu} F^{\mu\nu}$, and a particle with $q$ units of electric charge at rest in a fixed planar empty AdS background, i.e.\ in a metric $ds^2=\ell^2 \frac{-d t^2 + d z^2 + d x^2 + d y^2}{{z}^2}$ (where ${z} \equiv \ell^2/{\rho}$). Without loss of generality we can assume the particle to be at $(x,y,z)=(0,0,z_p)$. Since the metric is conformally flat and Maxwell's equations are conformally invariant, the electromagnetic field is identical to the field produced by a particle at rest in flat space.  The electrostatic potential satisfies Dirichlet boundary conditions at the plane $z=0$, that is $A_t = 0$ and hence $F_{tx}=F_{ty}=0$ (or $E_{\|}=0$) at $z=0$. This is nothing but the classic textbook problem of a charge in the presence of an infinite perfect conductor at $z=0$, solved most elegantly by the method of image charges. The potential is thus, with our charge conventions (compare to (\ref{modelLagsimple}), (\ref{probeaction}) and (\ref{ABexpr})):
\begin{equation} \label{electricpot}
 A_t = \frac{g^2 q}{4\pi} \biggl( \frac{1}{\sqrt{(z-z_p)^2 + x^2 + y^2}} -  \frac{1}{\sqrt{(z+z_p)^2 + x^2 + y^2}}  \biggr) \, .
\end{equation}
The expectation value of the charge density in the dual CFT is given by the electric field strength at the boundary \cite{Witten:1998qj} (as is the induced charge density on the conducting plate in the classic electrostatics problem): 
\begin{equation} \label{AdS4jt}
 j^t = \frac{1}{g^2} \, F_{zt}|_{z=0} 
 = \frac{q}{2\pi} \frac{z_p}{(z_p^2 + x^2 + y^2)^{3/2}}  \, .
\end{equation}
We fixed the normalization by requiring the density to integrate to the total charge $q$. The radius of the charge density peak is $R \sim z_p = \ell^2/\rho_p$. 



\begin{figure}[t]
\centering{
\includegraphics[width=0.49\textwidth]{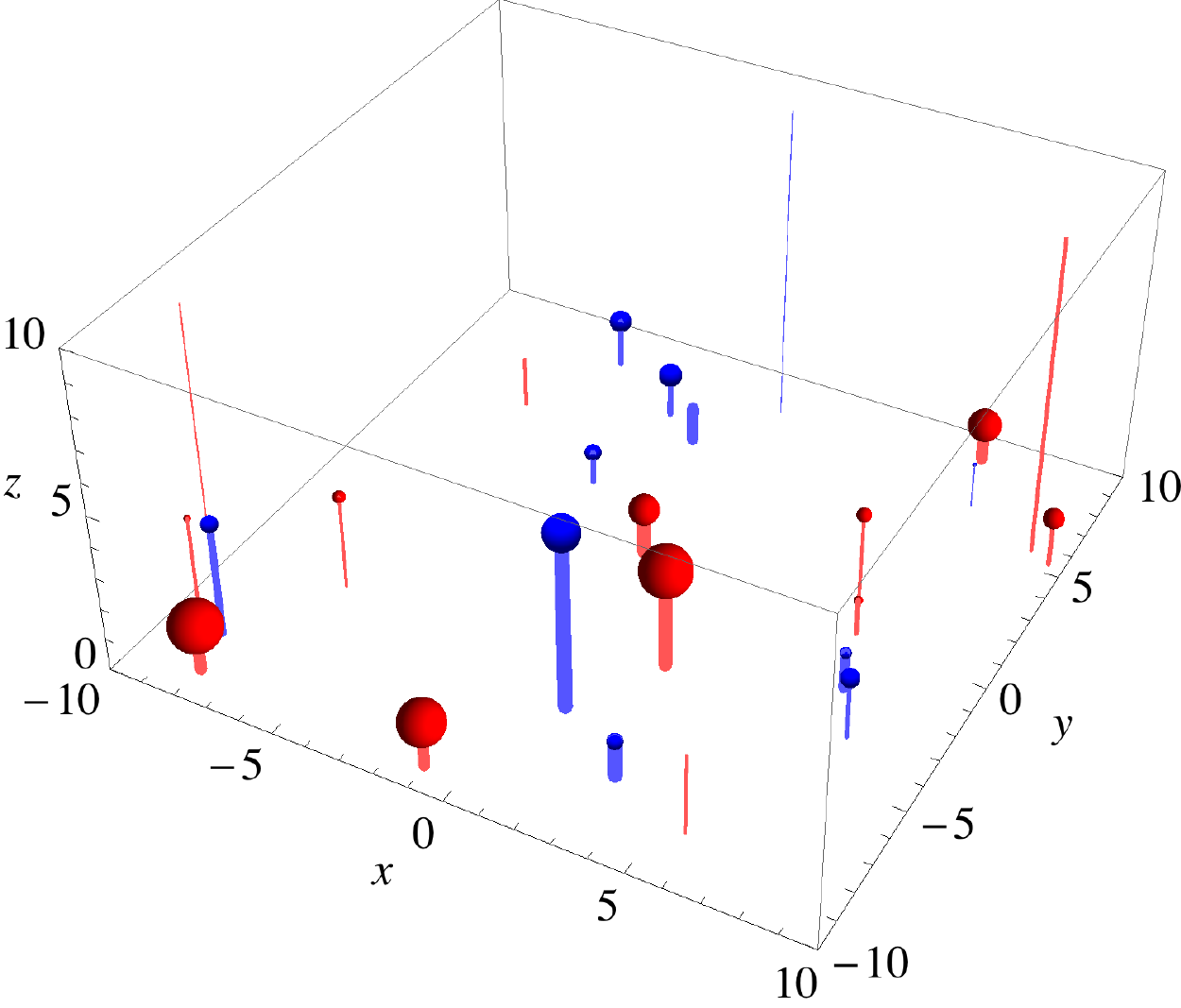}
\includegraphics[width=0.49\textwidth]{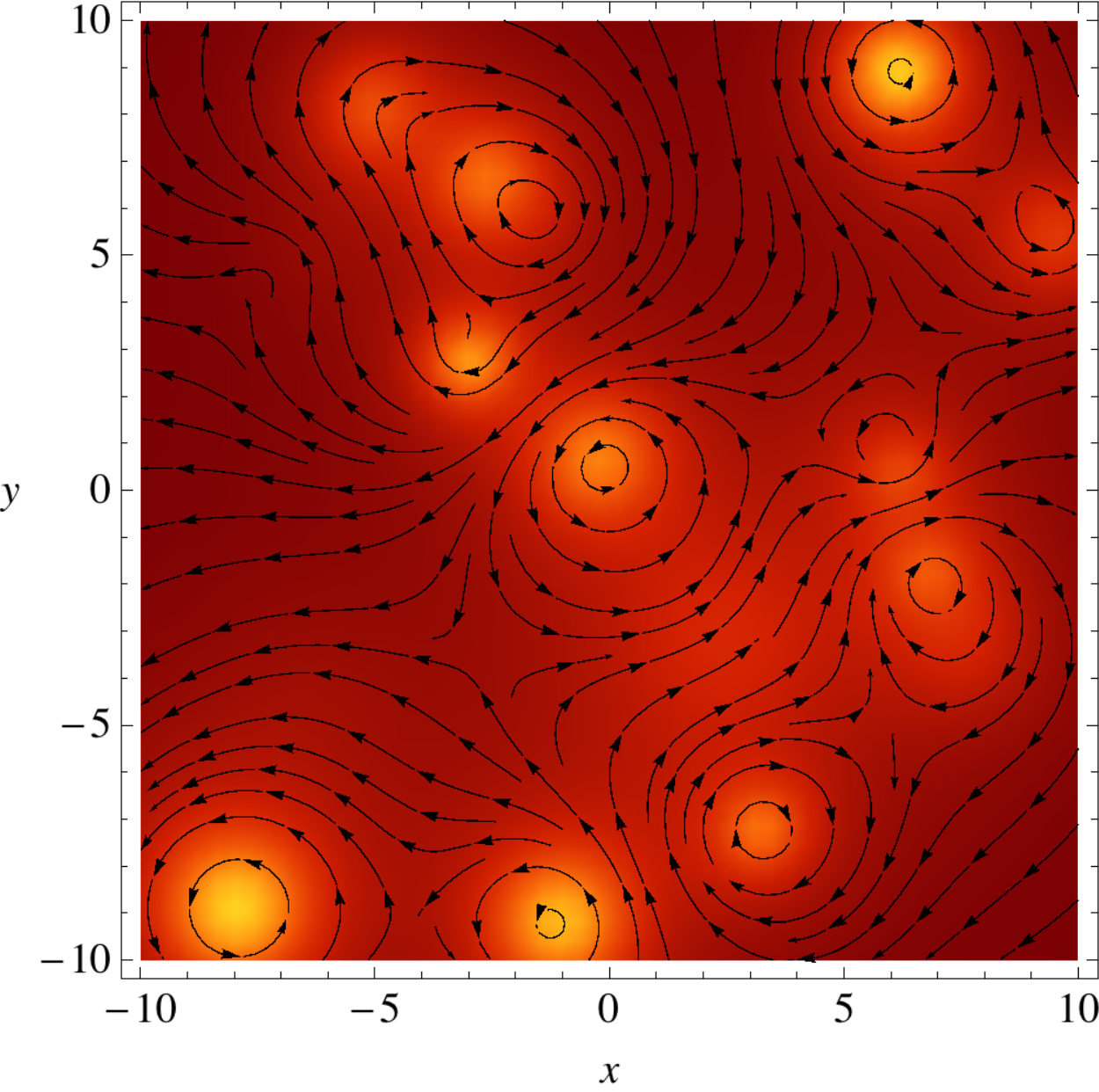}
}
\caption{{\bf Left}: A random collection of probe black holes (artificially made up, unrelated to any of our actual examples), represented by spheres. The size of each sphere is proportional to the D0-charge, while the thickness of the line projecting the probe onto the boundary $z=0$ is proportional to the D6-charge. Red/blue = positive/negative D6-charge. Notice that because of the symmetry (\ref{D6D2symmetry}), one expects positive and negative D6-charge probes to be present in equal abundance. {\bf Right}: Corresponding 3-currents in the CFT. Brighter means higher charge density $j_0^t$, flow lines indicate the direction of the current $\vec j_0$. D0-charge determines charge density, D6-charge determines current density. Smaller values of $z_p$ lead to smaller structures. Positive and negative D6-charges produce oppositely circulating currents. \label{fig:vortices}}
\end{figure}

Let us now consider instead a magnetically charged particle. Dirichlet boundary conditions on the vector potential imply $B_{\perp} \equiv F_{xy}=\partial_x A_y - \partial_y A_x = 0$; they forbid magnetic flux through the $z=0$ boundary surface. The boundary conditions thus break electromagnetic duality: The magnetic field sourced by a magnetic charge, subject to the boundary conditions at hand, is not obtained by dualizing the electrostatic field \ref{electricpot}, as this would give a magnetic field with $B_{\|}=0$ instead of $B_\perp =0$. Rather it is obtained by dualizing the electrostatic field of a point charge with boundary conditions $E_\perp=0$. This can again be constructed by the method of image charges, but this time with an image charge $+q$ instead of $-q$. The nonvanishing components of the electromagnetic field strength at $z=0$ are then $(F_{xt},F_{yt})=\frac{g^2 q}{2 \pi s^3}(x,y)$. This dualizes to the magnetostatic fields
$(F_{zx},F_{zy}) = \frac{p}{s^3}(-y,x)$ where $p$ is now the magnetic charge. In the CFT dual, this corresponds to a  medium with zero net charge density but with a nontrivial stationary vortex current, 
\begin{equation}
 (j^x,j^y) = \frac{p}{g^2 (z_p^2 + x^2 + y^2)^{3/2}} (-y,x) \, .
\end{equation}
This can also be viewed as a ``magnetization'' current $j =  \nabla \times m = (\partial_y m,-\partial_x m)$ where the magnetization density is 
\begin{equation}
 m = \frac{p}{g^2 (z_p^2 + x^2 + y^2)^{1/2}} \ .
\end{equation}
In the context of  two dimensional incompressible fluid dynamics (see e.g.\ \cite{wayne}), $m$ is called the stream function, and $\omega \equiv \nabla \times j = -\nabla^2 m$ is called the vorticity. The total current through a line from the origin to infinity is given by $m(\infty) - m(0) =\frac{p}{2 \pi z_p}$.  

Putting things together, we see that a general dyonic particle with charge $(q,p)$ at $(x,y,z)=(0,0,z_p)$ will correspond to a charge density $j^t = \frac{q  z_p}{2 \pi (z_p^2 + x^2 + y^2)^{3/2}}$ and a magnetization density $m=\frac{p}{g^2 (z_p^2 + x^2 + y^2)^{1/2}}$.

Applying this to our model in the duality frame where the D0 and D4 charges are considered to be electric charges (and the scalar kept fixed), we see from (\ref{modelLagsimple}) that we have $g^2_0 = g^2_{D0} = \frac{3}{2 \pi \yz^3}$ and $g^2_1 = g^2_{D4} = \frac{\yz}{4\pi}$. Hence for an arbitrary probe charge $(p^0,p^1,q_1,q_0)$, we get, in the notation (\ref{probehats}) with $\hat{p}^0 \equiv 1$, the following D0 and D4 charge and magnetization densities:
\begin{align} 
 j^t_0&= \frac{p^0 v^3}{2 \pi} \, \frac{\hat q_0 z_p}{(z_p^2 + x^2 + y^2)^{3/2}} \, , 
 &j^t_1&=\frac{p^0 v}{2 \pi } \, \frac{\hat{p}^1 z_p}{(z_p^2 + x^2 + y^2)^{3/2}} \, , \\
 m_0 &=  \frac{p^0 v^3}{12 \pi} \,  \frac{1}{(z_p^2 + x^2 + y^2)^{1/2}}
 &m_1 &=  \frac{p^0 v}{\pi}  \, \frac{\hat{q}_1}{(z_p^2 + x^2 + y^2)^{1/2}} \, .
\end{align}
Note that under the symmetry (\ref{D6D2symmetry}), the magnetizations flip sign, while the charge densities remain invariant. In a duality frame with D4-charge considered to be magnetic, the roles of $\hat{p}^1$ and $\hat{q}_1$ would be exchanged, with the former giving rise to a magnetization density and the latter to a charge density.

For values of $\tilde{\rho}_p = \rho_p/\ell$ of order 1, $z_p$ is of order $\ell$, causing the current density to be concentrated in a region of order $\ell$. The  charge density due to the probe will generically be much smaller than the background charge density (\ref{bkgdens}) provided $p^0 v^2 \ll N$, which, not surprisingly, was the condition for the probe approximation to be valid. However since the background magnetic field vanishes, the magnetizations and corresponding spatial currents are entirely due to the probe. 

Probes located at different positions will produce these currents appropriately translated in the $(x,y)$-plane, and multiple probes will produce currents which are superpositions of single probe currents. An example is shown in figure \ref{fig:vortices}.

More generally, the probes will also source the scalar and the metric, which in the CFT corresponds to fluctuations in the expectation value of some scalar operator and in the energy-momentum tensor. This can be studied in a similar way but we will not do this here.

For global AdS, a similar analysis can be done, although we can no longer make use of the simple map to flat space electromagnetism, so the gauge field propagator is somewhat more involved. We give the relevant expressions in appendix \ref{vector}. To get a solution involving magnetic charges which is also consistent with Dirichlet boundary conditions on the vector potential associated to our choice of duality frame, the total magnetic charge must be zero. In a dynamic setup, where we start off with a purely electrically charged black hole, this will be guaranteed by charge conservation.

\subsection{Probes in a black brane background} \label{sec:pbhbopt}

\begin{figure}
\centering{
\includegraphics[width= \textwidth]{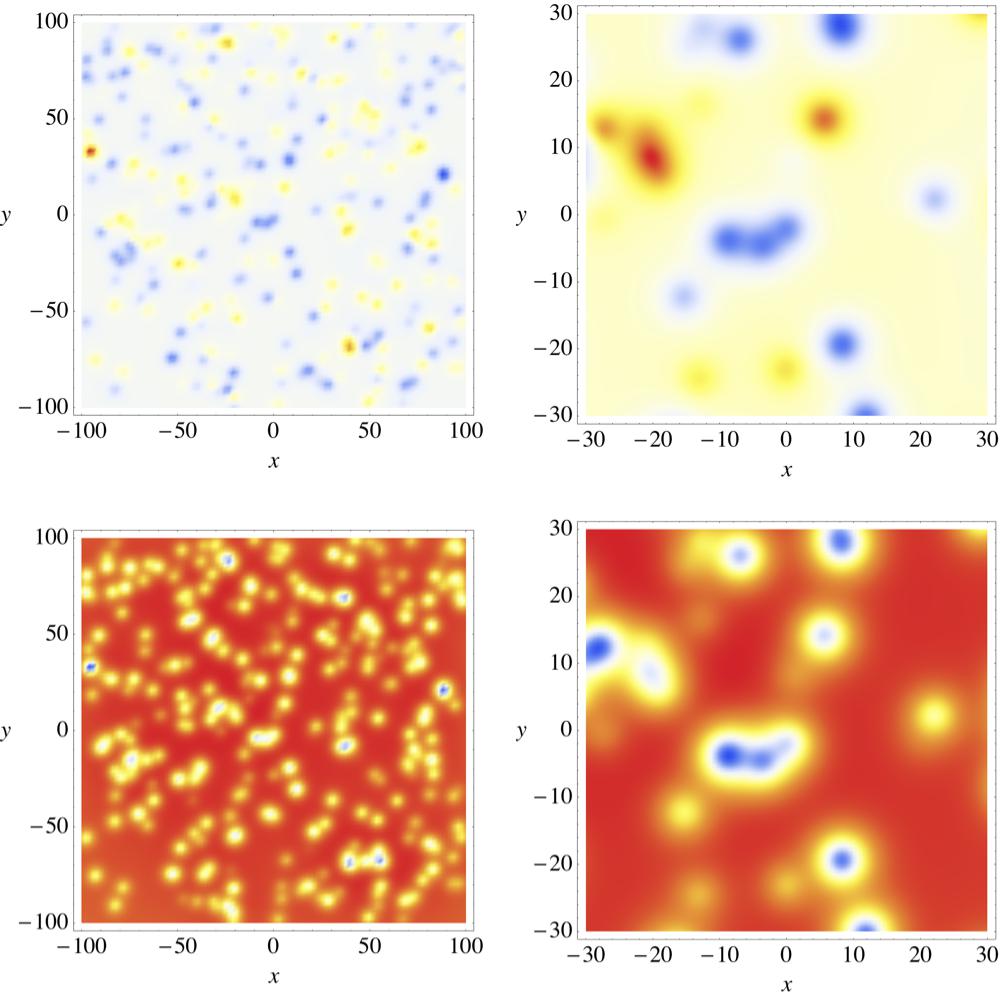}
}
\caption{Holographic projection of D2 (top) and D0 (bottom) charge density fluctuations due to the cloud shown in fig.\ \ref{fig:holes} on the left, for the choice of duality frame in which the D0 and D2 charges are electric. Red = positive, blue = negative. Notice the lumps always have negative D0 charge (same as background), while the D2 charge can have either sign. At this fairly low density, the individual lumps are still clearly discernible. In a duality frame in which the D0 or D2 charge is considered magnetic, we get smilar looking magnetization or vorticity densities instead, i.e.\ circulating vortex currents analogous to fig.\ \ref{fig:vortices}. \label{fig:densplots}}
\end{figure}

We now consider the system of actual interest, a probe charge in a general planar black hole background with a radially varying coupling constant. The background metric has the general form 
\begin{equation}
ds^2 = - g_{tt}(\rho) \, dt^2 + g_{\rho \rho}(\rho) \, d\rho^2 + g_{xx}(\rho) \left( dx^2 + dy^2 \right)~,
\end{equation}
with AdS$_4$ asymptotics at $\rho=\infty$, and the Maxwell Lagrangian takes the form ${\cal L} = \frac{1}{4 g(\rho)^2} F^{\mu \nu} F_{\mu \nu}$. In our setup, we have $g_{tt}=V(\rho)$, $g_{\rho\rho}=1/V(\rho)$, $g_{xx}=W(\rho)$ and e.g.\ $\frac{1}{g_{D0}^2}=\frac{2\pi v^3}{3}\, y(\rho)^3$, with the relevant explicit expressions given in (\ref{planarmetricansatz})-(\ref{planarwarp}).

In this case, it is no longer possible to find analytic solutions. In appendix \ref{app:particlesinBHBG} we obtain a general approximate solution based on a WKB analysis. The final result is given in (\ref{jseries}), to which (\ref{approxjt}) is a good enough approximation for our purposes:
\begin{equation} \label{jteqwkb}
 j^t(r) = \frac{q}{2\pi} \, \eta_p \left( 
 \frac{\upsilon_p}{(x^2+y^2+\upsilon_p^2)^{3/2}}
 - \frac{2 \upsilon_h-\upsilon_p}{(x^2+y^2+(2 \upsilon_h-\upsilon_p)^2)^{3/2}} \right) \, .
\end{equation}
Here 
\begin{equation}
\eta_p \equiv \frac{g(\rho_p)}{g(\infty)} \frac{g_{tt}^{1/4}(\rho_p)}{g_{xx}^{1/4}(\rho_p)} \, , 
\end{equation}
and $\upsilon_h$, $\upsilon_p$ are the ``optical distances''\footnote{Notice that in terms of the $\upsilon$ coordinate defined by $d\upsilon=\sqrt{g_{\rho\rho}/g_{xx}} \, d\rho$, de metric becomes spatially isotropic: $ds^2=-g_{tt} dt^2 + g_{xx}(d\upsilon^2+dx^2+dy^2)$.} from the boundary $\rho=\infty$ of AdS to the horizon $\rho=0$ and to the probe  $\rho=\rho_p$, that is
\begin{equation} \label{opticaldist}
 \upsilon_h \equiv \int_0^{\infty} d\rho \, \sqrt{\frac{g_{\rho\rho}}{g_{xx}} } \, , \qquad 
 \upsilon_p \equiv \int_{\rho_p}^{\infty} d\rho \, \sqrt{\frac{g_{\rho\rho}}{g_{xx}} } \,  .
\end{equation}
The first term in (\ref{jteqwkb}) is similar to the empty AdS solution (\ref{AdS4jt}), and the second term can be interpreted as due to an image charge behind the black hole horizon. Higher order corrections to this formula can similarly be interpreted as due to more image charges, obtained by subsequent mirroring over the horizon and boundary planes; see the appendix for more details. 

\begin{figure}
\centering{
\includegraphics[width= \textwidth]{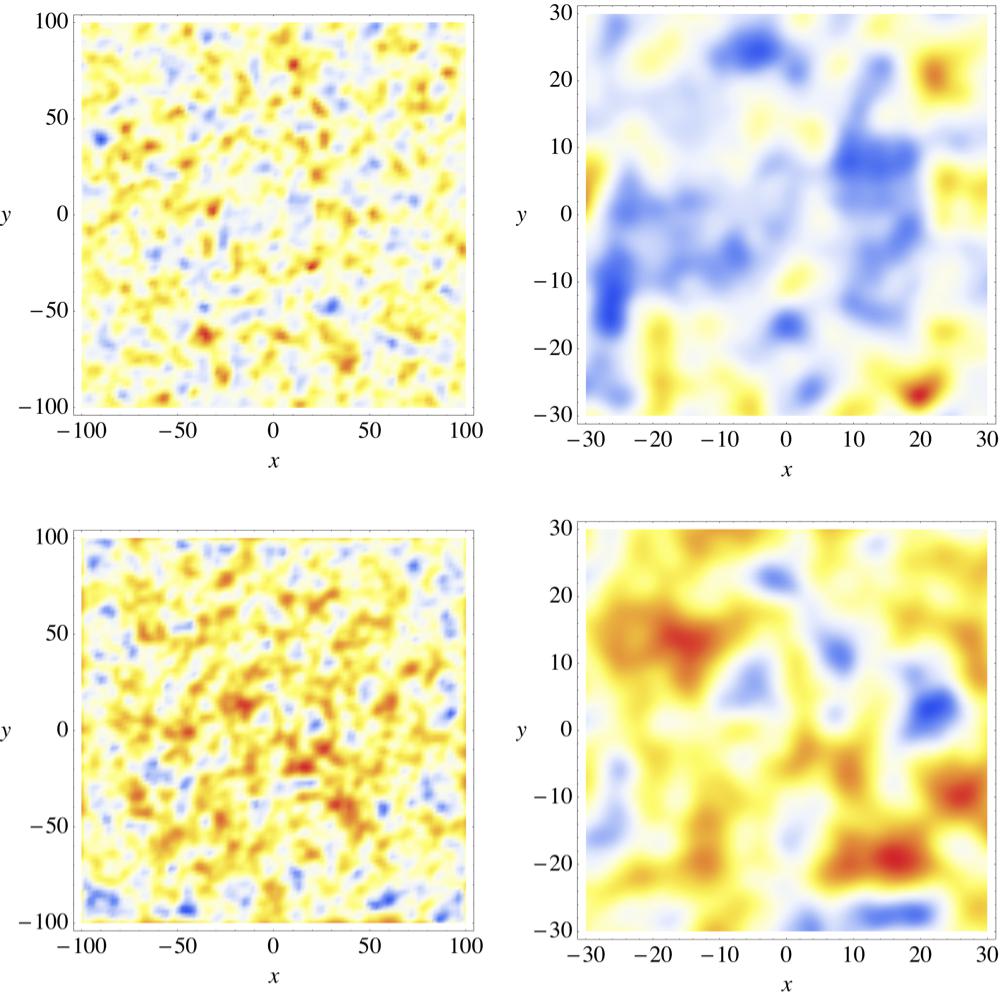}
}
\caption{Same as in in fig.\ \ref{fig:densplots} but now showing the profiles corresponding to the high density case, \ref{fig:holes} on the right. Although the individual charge disks are no longer discernible, their characteristic size would show up in a spectral analysis of the plot.
\label{fig:densplots2}}
\end{figure}

Using this, we can now compute the CFT charge density profiles corresponding to any cloud of bound probe particles. Actual examples are shown in figs.\ \ref{fig:densplots} and \ref{fig:densplots2}. At low density (fig.\ \ref{fig:densplots}), we can still discern the charge disks associated to individual probes, at high density  (fig.\ \ref{fig:densplots2}) this is no longer the case. (In this regime we would also expect the probe approximation to break down and interaction effects to become important.) From the boundary CFT point of view, the aging process described in section \ref{sec:meta-aging} corresponds to a gradual, exponentially slow ``melting'' of the charged/magnetized lumps into the homogeneous background. On the other hand, the decay of the homogeneous state into  stable bound states outlined in section \ref{sec:aging-stab} corresponds to a gradual increase in inhomogeneity, up to and beyond the situation shown in fig.\ \ref{fig:densplots2}.\footnote{If this transition must proceed through thermal activation, the process will be exponentially slow again. If it can proceed through some classical dynamical instability --- which we did not analyze --- the process may be fast. }

\subsection{Probe dynamics and transport} \label{sec:transport}

Recall that the bound probe charges are all magnetically charged with respect to the background black hole, which means that they will be magnetically trapped by the background --- classically, when kicked, they get stuck on circular orbits, quantum mechanically they form localized Landau droplets. Another way of thinking about this is that separated electric and magnetic charges come with intrinsic angular momentum stored in the electromagnetic field, so conservation of angular momentum will tend to obstruct free motion of the probes. Similar rigidifying magnetic interactions occur between the different probes, as well as through backreaction polarization effects of the probes on the black hole horizon. Thus we get a significant obstruction to spontaneous clumping or ordering effects one might naively have expected in the cloud, and various transport coefficients such as conductivity and viscosity may be strongly affected. 

At a more basic level, due to thermal activation, we may expect some of the probes to wobble around in their magnetic traps while other lay dormant. Furthermore, clusters of cloud particles may rearrange themselves and relax in a hierarchical cascade due to mutual interactions. In the CFT, this will show up as \emph{dynamical} inhomogeneities. Even at high density, in contrast to the charge densities themselves, these may still be expected to show up as distinguishable, locally active regions, since such regions will be relatively sparse, and their kinetic energy will only slowly be dissipated to other regions. In this way, inhomogeneities caused by probes will be different from inhomogeneities caused by a disordered horizon. The emergence of dynamical inhomogeneities appears to be a characteristic though not fully understood features of supercooled liquids near the glass transition \cite{kineticreviews,glassperspectives,Keys2011}. The effect we just described may be a holographic incarnation of this. 

The presence of a sufficiently dense probe cloud may also lead to a dramatic increase in viscosity, characteristic of approach to the glass transition. The shear viscosity $\eta$ can be viewed as momentum conductivity, more precisely e.g.\ conductivity of $p_y$-momentum in the $x$-direction. When black branes are stirred in the $y$-direction, the induced momentum does not propagate far along $x$ on the horizon --- rather it falls quickly into the black hole. As a result, the viscosity of a black brane is very low, leading to the famously low $\eta/s = 1/4\pi$. However, this changes completely in the presence of a cloud with  rigidifying magnetic interactions, as these interaction may have a strong drag effect, possibly leading to an enormous increase in momentum transport, i.e.\ an enormous increase in shear viscosity. Moreover, magnetic charges in the cloud may be expected to dramatically reduce charge transport efficiency by the black brane, by the ``eddy current brake'' mechanism \cite{eddy_current_brake}. Such a dramatic drop in (global) charge (i.e.\ matter) transport efficiency is another feature characteristic of the approach to the glass transition.

We leave exploration of these intriguing ideas to future work.

\subsection{Strings as an obstruction to string theory realizations} \label{sec:strings}

It would be desirable to have an explicit dual CFT realization of our setup. The model we have studied can be characterized as the bosonic sector of an $\CN=2$ Fayet-Iliopoulos gauged supergravity with cubic prepotential. The two massless $U(1)$s we have are sourced by charges which are parametrically heavier than the AdS scale --- they can be thought of as wrapped D0, D2, D4 and D6 branes in type IIA.

In the flat space case, this model is a universal subsector of any type IIA Calabi-Yau compactification, providing a consistent truncation of the corresponding four dimensional effective theories. The model we studied is basically the simplest possible uplift of this to AdS. It would therefore seem logical that it should be equally easy to embed this model in string theory. In particular, flux compactifications, such as type IIA on ${\mathbb{CP}}^3$ with $N$ units of RR 6-form flux and $k$ units of RR 2-form flux through the ${\mathbb{CP}}^3$ (this one specifically being dual to the ABJM quiver Chern-Simons CFT \cite{Aharony:2008ug}), or related compactifications \cite{Cassani:2012pj,Cassani:2009ck,Franco:2008um,Franco:2009sp,Tomasiello:2010zz,Tomasiello:2007eq,Martelli:2008rt,Gabella:2012rc,Martelli:2004wu,Gauntlett:2004hh,Aharony:2008gk}, would appear to be natural candidates.

However there is a general obstruction to this idea. Any AdS$_4 \times M_6$  compactification of type IIA string theory which is supported by fluxes will have the property that some linear combination of the $U(1)s$ obtained by naively reducing the RR potentials coupling to wrapped D-branes is in fact Higgsed and thus massive. The mechanism for this was exhibited explicitly for $\mathbb{CP}^3$ e.g.\ in \cite{Aharony:2008ug}. A general diagnostic for a $U(1)$ being Higgsed is that magnetic monopole charges necessarily come with confining strings attached; they are magnetic flux lines squeezed together by the Meissner effect. Now, if some compact $p$-cycle is threaded by $n$ units of Ramond-Ramond magnetic $p$-form flux, a D$p$-brane wrapped around this cycle will necessarily come with $n$ fundamental strings attached. This follows directly from Gauss' law for the D-brane worldvolume gauge theory: the flux creates a background charge for this gauge field, which due to the compactness of the brane must be canceled by the charge carried by open string endpoints. Thus, at least one of the RR U(1)s present in the original Calabi-Yau compactification must be Higgsed by turning on fluxes. This can also be seen  more directly in flux compactifications of supergravity, with the Higgs scalar emerging from the reduction of the dilation-axion. More specifically, for compactifications on ${\mathbb{CP}}^3$ with $N$ units of RR 6-form flux and $k$ units of RR 2-form flux, wrapped branes carrying D6- and D2-charge will generically come with strings attached, except for the specific ratio (proportional to $N/k$) of the charges for which the two flux tadpoles exactly cancel each other. Only the corresponding combination of the D0- and D4- $U(1)s$ survives as a gauge symmetry, the other one becomes massive. 

Thus, if we try to embed our model in string theory in this way, we would have to accommodate these features. Generic probes would have to come with fundamental strings attached (stretched from the horizon to the probe or between the probes), and the massive photon would decay exponentially rather than polynomially. It can be checked that generically these ``stringy'' effects scale in exactly the same way with $N$ and $v$ as the other forces we considered. For example the probe potential (\ref{VpVp}) for a D6 scales as $V_p \sim \N \yz/\ell$. On the other hand, making the identification of the string length as in \cite{Aharony:2008ug}, $\ell_s = \ell/\sqrt{\yz}$, a string stretched over a coordinate distance $\Delta \rho$ would have an energy of order $E_s \sim \Delta \rho/\ell_s^2 \sim \yz/\ell $ (times something of order 1, assuming $\Delta \rho/\ell$ is of other 1). But by the above arguments, a single D6 comes not with one, but with $N$ strings attached. Hence $E_s \propto N \yz/\ell$, the same scaling as the potential. This turns out to be the case for various other similar comparisons of scales. We conclude that in these models, the features we have exhibited are not obviously obliterated, nor are they obviously preserved. 

There are of course compactifications which can consistently be truncated to the model we consider. The simplest case in perhaps M-theory on AdS$_4 \times S^7$, which corresponds to the case $k=1$ of the IIA $\mathbb{CP}^3$ compactification considered above. The problem with these is that they have very light charged matter, with masses of the order of the AdS scale, which will tend to condense and form superconducting condensates \cite{Gubser:2008px,Hartnoll:2008vx}. This would again qualitatively affect our discussion. To physically trust our model, we need all charged matter to be parametrically heavy, which in at least the simpler examples means charges should be wrapped D-branes in a type II picture; in the usual Freund-Rubin compactifications, towers of charged KK modes tend to have masses going all the way down to the AdS scale \cite{Duff:1986hr}. This is not to say that in such cases glassy models are excluded. It is quite possible that analogous considerations can be made in the presence of light charged matter. But it would alter the analysis of this paper.

Borrowing language originating from the study of AdS$_5$ - CFT$_4$ pairs \cite{Witten:1998xy,Gubser:1998fp,Berenstein:2002ke}, we might call the heavy charges we have been assuming ``baryonic'' charges. Indeed since it takes as many quarks as there are colors to make a baryon, states with nonzero baryon number in the CFT are guaranteed to be heavy at large $N$. From the bulk dual point of view, baryons are heavy because they correspond to internally wrapped branes. Similar considerations hold for the AdS$_4 \times Y_7$ - CFT$_3$ analogs \cite{Klebanov:2010tj,Donos:2012sy}. Examples are M-theory compactifications on Sasaki-Einstein manifolds with nonzero betti number, such as $Q^{111}=SU(2)^3/U(1)^2$, or quotients thereof \cite{Cassani:2012pj,Cassani:2009ck,Franco:2008um,Franco:2009sp,Klebanov:2010tj,Donos:2012sy,Martelli:2008rt,Gabella:2012rc,Martelli:2004wu,Gauntlett:2004hh}. Although this comes closer, our model is again not quite a consistent truncation of the low energy effective action of such models \cite{Cassani:2012pj,Cassani:2009ck,Gauntlett:2009zw}; there are additional light scalars involved, which again may be expected to qualitatively change the analysis. A rather different class of flux compactifications involving Calabi-Yau orientifolds was studied in \cite{DeWolfe:2005uu} and black brane solutions in this setup were constructed in \cite{Torroba:2013oua}. In this case the obstruction to the most obvious attempts at embedding the model appear to be that the orientifold projection eliminates the desired massless $U(1)$s, which is surely related to the above general considerations.

It would be very interesting to follow a more direct top-down approach and see if bound states of the type we have analyzed here persist in models with a UV completion in string theory. In particular, in view of its genericity, it would be of interest to investigate the effect of the Higgsed $U(1)$, specifically the impact of necessarily having strings attached to the probes.

\section{Conclusions and outlook} \label{outlook} 

We have accomplished the following in this work:
\begin{enumerate}
 \item We have mapped out the complete thermodynamic phase diagram of a general class of nonextremal charged AdS black hole solutions with running scalars, uncovering a rich phase structure.
 \item We have established the existence of finite temperature stationary bound states of these black holes with probe black holes. This implies the existence, in principle, of the corresponding nonlinear solutions to the coupled Einstein equations. Black branes can form stable and metastable bound states with arbitrary numbers of different charges, to a large extent trapped by magnetic forces. This leads to metastable, strongly disordered states with an extensive configurational entropy. In contrast to other studies of disordered holography \cite{Hartnoll:2008hs,Kachru:2009xf,Fujita:2008rs,Ryu:2011vq,Adams:2011rj,Adams:2012yi,Saremi:2012ji}, the disorder is not induced by sources (which remain uniform), but is spontaneously generated.
\item We have mapped out the regions in thermodynamic state space where such bound states can form, and have extensively studied their properties, including their dynamical and thermodynamical stability and the distributions of their radial sizes and barrier heights.  
\item To the extent allowed by the probe approximation, and neglecting mutual probe interactions, we have studied the relaxation dynamics of clouds of bound probes, and established they exhibit logarithmic aging behavior characteristic for many amorphous systems. The aging rate at a given time scale is set by the density of states with a given barrier height set by the logarithm of the time scale.  
\item We have determined the detailed holographic map from bulk bound probe configurations to charge and vorticity densities in the dual CFT. Typically,  quenched clouds of bound states map to structures of comparable characteristic sizes, but vastly different exponential lifetimes. At the static level, these bound state homogeneities may be hard to distinguish from horizon inhomogeneities. However they will have rather different dynamical signatures. In particular they may naturally lead to the striking dynamical heterogeneities and correlations that are observed in supercooled liquids. 
\end{enumerate}
The overall picture we propose is that whereas plain, smooth black branes are holographic duals of fluids, amorphously ``fragmented'' black branes represent more glassy phases of matter. Most of the analysis of this paper was done in a probe approximation, and therefore the kind of fragmented branes we had actual control over were black branes dressed with clouds of charged probe black holes. Nevertheless, already at this level we could demonstrate characteristic features of glassy relaxation, including logarithmic aging. If our overall picture is correct, such cloudy branes can be expected to interpolate between fluids and genuine glasses, and may exhibit the properties of supercooled liquids approaching the glass transition. If so, holography may provide important new insights into the still elusive nature of the glass transition, as it provides direct access to the thermodynamic state space, allows for efficient computation of transport coefficients (which are the prime diagnostic for the approach to the glass transition), and is particularly powerful exactly in regimes that are relevant to glassy physics. However, more work is needed towards this goal, including the following:
\begin{enumerate}
 \item As we have seen, the black hole horizon itself is thermodynamically unstable (perturbatively in some case, nonperturbatively in others) to formation of inhomogeneities, similar to the instabilities found in \cite{Gregory:1993vy,Gubser:2000ec,Gubser:2000mm,Caldarelli:2008mv,Nakamura:2009tf,Donos:2011bh,Donos:2011qt,Donos:2013gda,Iizuka:2013ag,Cremonini:2012ir,Donos:2013wia,Withers:2013kva,Rozali:2013ama,Maeda:2009vf,Albash:2009iq}. The relative importance of this compared to the inhomogeneities caused by bound particles needs to be assessed. For some but not all of the black branes forming bound states, there is indeed a perturbative thermodynamic instability in the grand canonical ensemble, which suggests instability towards inhomogeneities (phase mixtures) in the microcanonical ensemble. One may wonder if such phase mixtures may be exhibited by a purely thermodynamic analysis, similar to \cite{Horowitz:2007fe}. If we restrict to mixtures of the black brane phases described in this paper, the answer is no. For two phases to coexist, their temperature, chemical potentials and pressure (free energy density) must be equal. It can be checked that for our black brane phases, this implies the phases are identical. It seems therefore that a direct analysis of spatially inhomogeneous solutions, along the lines of \cite{Gregory:1993vy,Gubser:2000ec,Gubser:2000mm,Caldarelli:2008mv,Nakamura:2009tf,Donos:2011bh,Donos:2011qt,Donos:2013gda,Iizuka:2013ag,Cremonini:2012ir,Donos:2013wia,Withers:2013kva,Rozali:2013ama,Maeda:2009vf,Albash:2009iq}, will be necessary to investigate this question. 
  
  \item Supercooled liquids and glasses are produced by thermal quenches, i.e.\ fast cooling of the liquid phase. The properties of the resulting phase depend crucially on the cooling rate. We did not work out in any detail a concrete holographic realization of such a quench (although we did give some suggestions in footnote \ref{fnquench}). It would be interesting to do so. Since the result of a quench is effectively immediate, it is plausible that the classical instabilities  mentioned above may play an important role; for example one could imagine a Gregory-Laflamme type instability \cite{Gregory:1993vy} (augmented by quantum effects to allow horizons to split off) spitting out large droplets of black hole brane, along the lines of e.g.\ \cite{Caldarelli:2008mv}, getting subsequently trapped in metastable potential wells.
  
\item The setups we studied in which the probe approximation remains valid at all times, namely clouds of probe charges that are not too dense and that are metastable, do not describe genuine glasses, since they ultimately return to the liquid (bare black brane) state. To get a system that does not relax back to the liquid phase, we need to go to regimes where stable bound states exist. Here we did see indications of relaxation dynamics getting irreversibly lost in the free energy landscape, a characteristic feature in many theories of glasses. However in this case we exit the probe regime in finite time. Therefore to really probe the glass phase and the glass transition, it will be important to go beyond the probe approximation. Some of the features that can be expected to arise were outlined in section \ref{sec:aging-stab}. We should also point out that even within the probe approximation, neglecting probe-probe interactions on exponentially long time scales is physically not justified, in particular not if these interactions may lead to probe black holes merging into larger and therefore more stable black holes. Taking probe interactions into account at this level can be done without having to solve for the fully backreacted geometries. 
  
\item We restricted to relaxation through classical thermal activation. At sufficiently low temperatures, quantum tunneling will become the dominant channel. The amplitude for tunneling through a barrier is suppressed by an exponential factor $e^{-\int |p| dq}$, which can be computed directly from e.g.\ (\ref{Hamiltonian}).

\item It would be very interesting to compute holographic transport coefficients in the presence of  black hole bound states. The approach to the glass transition by supercooled liquids is characterized by dynamical arrest without any static structural changes, and associated to this a dramatic increase in shear viscosity and decrease in diffusion coefficients. Since we no longer have a single horizon, but rather a fragmented conglomerate of horizons, there is no reason for the universal results for viscosity and conductivity to remain valid. In fact, tracing the reasoning of \cite{Iqbal:2008by}, it is clear that the presence of matter in the bulk will significantly alter these universality results. As we mentioned in section \ref{sec:transport}, there are good reasons to believe even modest clouds could lead to dramatic increase in the shear viscosity.

\item As discussed in section \ref{sec:strings}, there are obstacles to finding an explicit holographic dual of our model as it stands. Although the theory we start from is very similar to the low energy effective theory dual to the ABJM CFT \cite{Aharony:2008ug}, it misses the Higgs scalar that renders one of the $U(1)s$ massive, and forces D6- and D2-charges to come with strings attached. We argued that this feature is in fact universal for any Freund-Rubin-like string flux compactification. It would therefore seem quite important to see what the impact is of adding this feature to the analysis. 

\item More generally, it would be useful to propose simplified models to capture the essential physics of glassy holography in a more transparent way. Our model was motivated primarily because it was the simplest uplift of asymptotically flat $\CN=2$ supergravity to AdS, making it likely a priori that black hole bound states would be found. But obviously, if we do not insist to this relation, much simpler models might be possible. Indeed if we extrapolate to the fullest extent the real-world observation that virtually all known liquids form glasses when cooled sufficiently fast, we should expect glassy states to appear in setups simpler than ours.

\item Finally, it will of course be extremely interesting to ultimately extract general lessons from the holographic picture for the general theory of the glass transition. The geometrization of scale hierarchies, the natural symbiosis of thermodynamic and kinetic aspects and the easy access to out-of-equilibrium physics that are offered by holography, all of crucial importance for any theory of glasses, make us think that there are indeed important lessons to be learned from the holographic approach. Conversely, one may hope that empirical knowledge of the properties of glasses will then lead to a better understanding of the fundamental landscape of quantum gravity, de Sitter space, and the universe itself.

\end{enumerate}

\section*{Acknowledgements}

It is a great pleasure to thank Curtis Asplund, Tatsuo Azeyanagi, George Coss, Alessandra Gnecchi, Monica Guica, Sean Hartnoll, Thomas Hertog, Diego Hofman, Laurens Kabir, Jorge Kurchan, Gim Seng Ng, Omid Saremi, Edgar Shaghoulian, Alessandro Tomasiello, Gonzalo Torroba, Toine Van Proeyen, Bert Vercnocke and Alberto Zaffaroni for valuable discussions. The authors would also like to thank the workshop ``Cosmology and Complexity" in Hydra where part of this work took place. Frederik Denef would like to thank the theory group at Milano Bicocca for warm hospitality. This work was supported in part by DOE grant DE-FG02-91ER40654, NSF grant no. 0756174 and by a grant of the John Templeton Foundation. The opinions expressed in this publication are those of the authors and do not necessarily reflect the views of the John Templeton Foundation. The final stages of this work were supported in part by the National Science Foundation under Grant No. PHYS-1066293 and the hospitality of the Aspen Center for Physics. Lucas Peeters was supported by a Fellowship of the Belgian American Educational Foundation.

\appendix


\section{Probe degeneracy of states and cloud densities} \label{app:degstates}

The energy of a probe of charge $q$ and position dependent rest mass $m$ in a background metric $ds^2 = -\nu^2 dt^2 + h_{ij} dx^i dx^j$ and electromagnetic field $A_\mu$ is 
\begin{equation} \label{Hamiltonian}
 H = \nu \sqrt{m^2+h^{ij} (p_i+q A_i)(p_j+q A_j)} + q A_0 \, .
\end{equation}
The semiclassical 1-particle density of states per unit volume and energy is 
\begin{equation}
 g(E,\vec x) = \frac{1}{(2 \pi \hbar)^3} \int d^3 p \, \delta(H-E) \, \Omega(q,m) \, ,
\end{equation}
where $\Omega(q,m)$ corresponds to the internal state degeneracy for the given charge and local rest mass. Integrating this over a large range of coordinates and energies gives the number of states available to the particle in this range. Doing the integral with the above expression for $H$ substituted and denoting the kinetic energy by $\epsilon \equiv E - q A_0 - m \nu \geq 0$, this becomes
\begin{equation}
 g(\epsilon,\vec x) = \frac{\Omega(q,m)}{(2 \pi)^3} \, \frac{4 \pi \sqrt{h}}{\nu^3}  \, \sqrt{\epsilon(2 m \nu + \epsilon)} \, (m\nu + \epsilon)  \, .
\end{equation}
Assuming a dilute gas so interactions are negligible and particle densities are exponentially small, the expected number density of particles of mass $m$ and charge $q$ at temperature $T$ and chemical potential $\mu$ is then
\begin{equation}
 \langle n_{q,m}(\vec x) \rangle = \int_0^\infty d \epsilon \, g(\epsilon,\vec x) \, e^{-(q(A_0-\mu)+m \nu + \epsilon)/T} \, .
\end{equation}
(The density is defined such that the total number in a region ${\cal R}$ is $\int_{\cal R} d^3 x \, n_{q,m}(\vec x)$.)
In the nonrelativistic regime, i.e.\
\begin{equation} \label{lowTnonrel}
 T \ll m \nu \, ,
\end{equation}
this is approximately 
\begin{equation} \label{mqdensity}
 \langle n_{q,m}(\vec x) \rangle \approx  \frac{\Omega(q,m)}{(2 \pi)^3} \, \frac{4 \pi \sqrt{h}}{\nu^3} \, \sqrt{\frac{\pi}{2}}
  \left(m \nu T\right)^{3/2} \, e^{-(q (A_0-\mu) + m \nu)/T} \, .
\end{equation}
To apply to our setup, note that we can identify 
\begin{equation}
 V_{em} = q A_0 \, , \quad V_{\rm grav} = m \nu \, \quad
 V_p = q A_0 + m \nu \, \quad \nu = \sqrt{V} \, \quad \sqrt{h} = \frac{W}{\sqrt{V}} \, ,
\end{equation}
with the various quantities appearing here defined in sections \ref{sec:solution} and \ref{hhads}. By comparing to the discussion in section \ref{sec:thermointerpr}, we see moreover that we should take $\mu=0$, if, as we do, we take $A_0$ to be zero at the black hole horizon. Equation (\ref{mqdensity}) then translates to
\begin{equation} \label{mgammadensity}
 \langle n_\gamma(\vec x) \rangle \approx  \frac{1}{(2 \pi)^3} \, \frac{4 \pi W}{V^2} \, \sqrt{\frac{\pi}{2}}
  \left(V_{\rm grav,\gamma} T\right)^{3/2} \, \Omega(\gamma) \, e^{-V_{p,\gamma}/T} \, ,
\end{equation}
where $\Omega(\gamma) = 1$ if the probe is a structureless particle and $\Omega(\gamma) = e^{S(\gamma)}$ if the probe is a black hole. The low temperature condition (\ref{lowTnonrel}) needed for the nonrelativistic approximation to be valid translates to $T \ll V_{\rm grav}$, which at finite separation from the horizon is satisfied under our assumptions, since $V_{\rm grav} \propto \N$. Of course at the horizon, the nonrelativistic approximation breaks down together with the rest of low energy field theory.

\section{Gauge Field Propagator in Global AdS$_4$}\label{vector}

The electric potential due to a stationary charge $q$ sitting at a point $\vec{x}_p$ in Minkowski space is given by
\begin{equation}\label{eq:potflatspacecharge}
A_t=\frac{q}{4\pi|\vec{x}-\vec{x}_p|}~.
\end{equation}
This seemingly simple expression gives us a lot of information about the electric field of a particle in flat space. Notably, we can discern that multipole moments of the electric field get washed out as we get farther away from the particle. This is an obvious sanity check, as a point charge sitting at $\vec{x}_p$ is no different than a point charge sitting at the origin when regarded by a far away observer. 


We wish to determine the exact form of $A_\mu$ in analogy with~(\ref{eq:potflatspacecharge}). That is, for a static particle sitting at an arbitrary point $\vec{x}_p$ in the bulk of AdS$_4$ with metric given by
\begin{equation}\label{eq:metricads4}
ds^2=-\left(1+\frac{r^2}{\ell^2}\right)dt^2+\frac{dr^2}{\left(1+\frac{r^2}{\ell^2}\right)}+r^2d\Omega^2~.
\end{equation}
We follow the derivation of~\cite{D'Hoker:1999jc}, which is formulated in Euclidean space. This amounts to taking $t\rightarrow i\tau$ in~(\ref{eq:metricads4}).

The action of a gauge field in Euclidean AdS$_4$ is given by
\begin{equation}
S_A=\int d^4x\sqrt{g}\left(\tfrac{1}{4}F^{\mu\nu}F_{\mu\nu}-A_\mu J^\mu\right)~,
\end{equation} 
and its response to an external current $J^\nu$ is
\begin{equation}\label{eq:gaugesat}
A_\mu(x)=\int d^4x'\sqrt{g}\,G_{\mu\nu'}\left(x,x'\right)J^{\nu'}\left(x'\right)~,
\end{equation}
where $G_{\mu\nu'}\left(x,x'\right)$ is the propagator. Maxwell's equations $\nabla_\mu F^{\mu\nu}=-J^\nu$ impose
\begin{equation}
\nabla^\mu\left(\partial_\mu G_{\nu\nu'}-\partial_\nu G_{\mu\nu'}\right)=-g_{\nu\nu'}\frac{\delta(x,x')}{\sqrt{g}}~.
\end{equation}

The expression for the gauge invariant part of $G_{\mu\nu'}\left(x,x'\right)$ can be given in a manifestly coordinate independent way. To do this we note that Euclidean AdS can be constructed by embedding the hyperboloid
\begin{equation}\label{eq:hyperboloid}
-X_0^2+X_E^2+X_1^2+X_2^2+X_3^2=-\ell^2
\end{equation}
in 5-dimensional minkowski space with metric
\begin{equation}
ds^2_{\text{5d}}=g_{\mu\nu}^{\text{5d}}dX^\mu dX^\nu=-dX_0^2+dX_E^2+dX_1^2+dX_2^2+dX_3^2~.
\end{equation}
We obtain the metric~(\ref{eq:metricads4}) by parametrizing the hyperboloid as
\begin{gather}
X_1=x=r\sin\theta\cos\phi~,\quad X_2=y=r\sin\theta\sin\phi~,\quad X_3=z=r\cos\theta \nonumber\\
X_0=\sqrt{\ell^2+r^2}\cosh\left(\tau/\ell\right)~,\quad X_E=\sqrt{\ell^2+r^2}\sinh\left(\tau/\ell\right)~.\label{eq:globalAdScoords}
\end{gather}
For two points corresponding to $\vec{X}$ and $\vec{X}'$ on the hyperboloid in~(\ref{eq:hyperboloid}), we define a bilinear 
\begin{equation}
u\left(X,X'\right)=-1-\frac{P\left(X,X'\right)}{\ell^2}
\end{equation}
where $P\left(X,X'\right)=g_{\mu\nu}X^\mu X'^\nu$ is the dot product in the ambient minkowski space. The quantity $P$ is related to the geodesic distance $D$ between points $\vec{X}$ and $\vec{X}'$ by $P=\cosh\,D/\ell$. In terms of $u$, the gauge invariant part of the propagator is given by
\begin{equation}
G_{\mu\nu'}\left(x,x'\right)=-\left(\partial_\mu\partial_{\nu'}u\right)F(u)~,
\end{equation}
where 
\begin{equation}
F(u)=\frac{1}{4\pi^2}\frac{1}{u(2+u)}~.
\end{equation}
In terms of the coordinates~(\ref{eq:globalAdScoords}), $u\left(X,X'\right)$ is given by
\begin{equation}
u=-1-\frac{\vec{x}\cdot\vec{x}'}{\ell^2}+\sqrt{1+\frac{r^2}{\ell^2}}\sqrt{1+\frac{r'^2}{\ell^2}}\cosh\left(\frac{\tau-\tau'}{\ell}\right)~,
\end{equation}
where $\vec{x}\cdot\vec{x}'$ the standard flat Euclidean dot product between the two vectors and $r^2=\vec{x}\cdot\vec{x}$.

We wish to evaluate~(\ref{eq:gaugesat}) for a point charge sitting motionless at $\vec{x}_p$, that is 
\begin{equation}
J^{\nu'}\left(\vec{x}'\right)=\left(q\frac{\delta\left(\vec{x}'-\vec{x}_p\right)}{\sqrt{g}},0,0,0\right)~.
\end{equation}
This boils down to computing
\begin{equation}
A_\mu=-\left.\frac{q}{4\pi^2}\int d\tau' \left(\partial_\mu\partial_{\tau'}u\right)\frac{1}{u(2+u)}\right\rvert_{\vec{x}'=\vec{x}_p}~.
\end{equation}
Because $F(u)$ is even in $\tau'$ and $\partial_{\tau'}u$ is odd, the integral vanishes for all components of $A_\mu$ except $A_\tau$. Computing the integral is straightforward and the final result is
\begin{equation}\label{eq:pointchargepotential}
A_\tau=\frac{q}{4\pi^2\ell}\left(\frac{2+w-v}{\sqrt{v\,(2+w)}}\arctan\biggl[\frac{\sqrt{v\,(2+w)}}{v}\biggl]+\frac{2+v-w}{\sqrt{w\,(2+v)}}\arctan\biggl[\frac{w}{\sqrt{w\,(2+v)}}\biggl]\right)~,
\end{equation}
where we have defined the quantities
\begin{equation}
v\equiv-1-\frac{\vec{x}\cdot\vec{x}_p}{\ell^2}+\sqrt{1+\frac{r^2}{\ell^2}}\sqrt{1+\frac{r_p^2}{\ell^2}}\quad\text{and }
w\equiv-1+\frac{\vec{x}\cdot\vec{x}_p}{\ell^2}+\sqrt{1+\frac{r^2}{\ell^2}}\sqrt{1+\frac{r_p^2}{\ell^2}}~.
\end{equation}
For large $\ell$ we find 
\begin{equation}
v=\frac{\left(\vec{x}-\vec{x}_p\right)^2}{2\ell^2}+\mathcal{O}\left(\ell^{-4}\right)~,\quad w=\frac{\left(\vec{x}+\vec{x}_p\right)^2}{2\ell^2}+\mathcal{O}\left(\ell^{-4}\right)~,
\end{equation}
and
\begin{equation}
A_\tau=\frac{q}{4\pi|\vec{x}-\vec{x}_p|}+\mathcal{O}\left(\ell^{-2}\right)~.
\end{equation}
We have chosen our normalization such that we get the correct result in the $\ell\rightarrow\infty$ limit, this is why our conventions differ by a factor of $4$ in $F(u)$ from those used in~\cite{D'Hoker:1999jc}.

\begin{figure}[t]
\centering{
\includegraphics[width=0.49\textwidth]{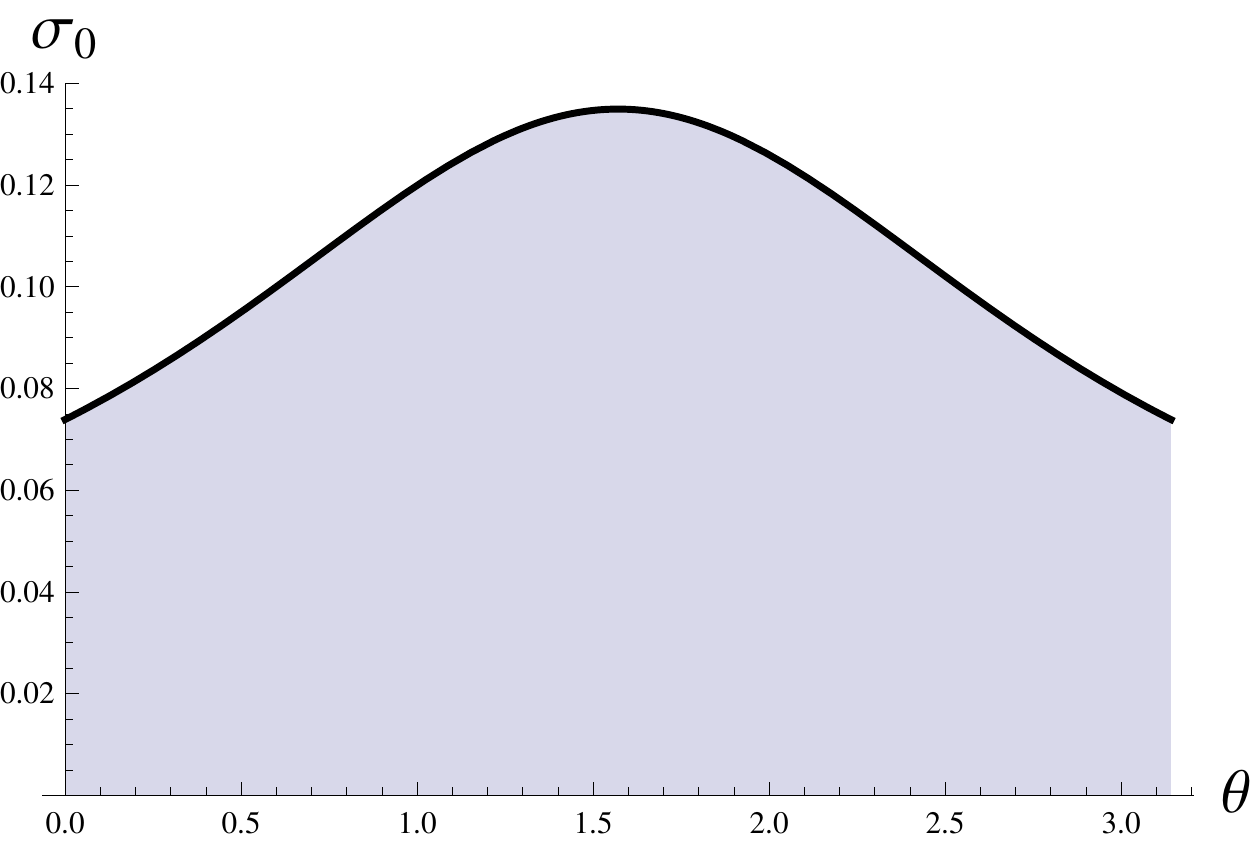}
\includegraphics[width=0.49\textwidth]{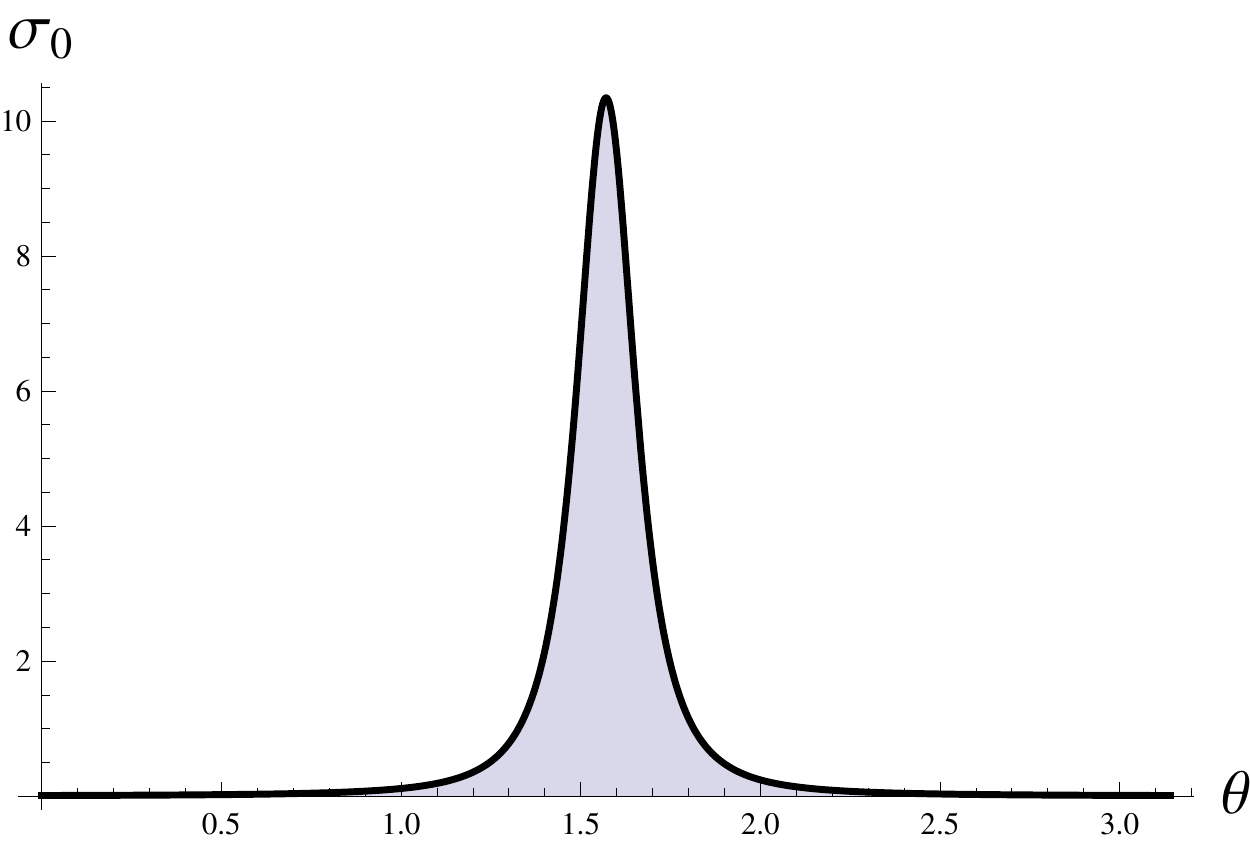}
}
\caption{Charge density $\sigma_0$ for a point charge with $q=1$ induced on the conformal sphere in units where $\ell=1$. We take $\phi=0$. \textbf{Left}: the charge is located at $\vec{x_p}=(0.4,0,0)$. \textbf{Right}: the charge is located at $\vec{x_p}=(8,0,0)$.\label{fig:inducedchargedensity}}
\end{figure}

The charge density induced on the conformal sphere is given simply by $\sigma_0=\lim_{r\rightarrow\infty} r^2 F^{tr} $. We provide some plots of this charge density in figure~\ref{fig:inducedchargedensity}. We have checked that our expression correctly gives $q$ when integrated over the $S^2$.

In order to obtain the $U(1)$  currents induced by a magnetic charge, as explained in the main text, it is not possible to dualize the field strength formed by $A_\tau$ as the corresponding magnetic field would not obey the correct Dirichlet conditions on the boundary sphere.  The currents are obtained by dualizing the field strength obtained from 
\begin{equation}
A^{\rm mag}_\tau=\frac{p}{4\pi^2\ell}\left(\frac{2+w-v}{\sqrt{v\,(2+w)}}\arctan\biggl[\frac{\sqrt{v\,(2+w)}}{v}\biggl]-\frac{2+v-w}{\sqrt{w\,(2+v)}}\arctan\biggl[\frac{w}{\sqrt{w\,(2+v)}}\biggl]\right)~,
\end{equation}
in which case $(j^\theta,j^\phi)=\lim_{r\rightarrow\infty}r^2(\tilde{F}_{\rm mag}^{r\theta},\tilde{F}_{\rm mag}^{r\phi})$.

\section{Static charged particles in black hole backgrounds} \label{app:particlesinBHBG}

In this appendix we work out approximate expressions for the electrostatic field produced by a point particle in a general planar black hole background with non-constant electromagnetic coupling constant.

The history of electrically charged static point particles in a black hole background dates back almost a century. In 1927, Whittaker wrote down an infinite series expansion for the electric field of a charged particle in a Schwarzschild background \cite{whittaker} and subsequently Copson \cite{copson1}, in the same year, found the analytic re-summed solution with the aid of Hadamard's  ``elementary solution'' to general second order partial differential equations \cite{hadamard}. Fifty years later, Copson wrote down the analytic solution of the electric field of a charged particle in an asymptotically flat Reissner-Nordstrom background \cite{copson2}. Considerations of the electric field of a charged particle in Rindler space began with the work of Bradbury in 1962 \cite{brad} (see also \cite{rindler} for a more complete historical account).

As far as we know, the problem of the electric field of a static charged particle in a  charged AdS black hole/brane background with non-constant scalar couplings has not been addressed. In this appendix, we discuss a simple WKB approximation to the problem for the charged black branes considered in the main text. 

\subsection*{General Setup}

Consider an action governing the dynamics of a $U(1)$ gauge field $A_\mu$ in of the form:
\begin{equation}
S = -\int d^4x \sqrt{-g} \; \frac{\sigma(z)}{4} \; F_{\mu\nu} F^{\mu\nu}~ + Q \int A_\mu dx^\mu \, ,
\end{equation}
where we have assumed a background:
\begin{equation}
ds^2 = - g_{tt}(z) dt^2 + g_{zz}(z) dz^2 + g_{xx}(z) \left( dx^2 + dy^2 \right)~,
\end{equation}
and an $z$-dependent coupling $\sigma(z)$. We assume that the horizon is located at $z=z_h$ where $g_{tt}(z_h) = 0$ and the asymptotic boundary of the space is at $z = z_b$ where $z_b < z_h$.

We are interested in an electrostatic problem, and so we set $A_x = A_y = 0$. The equation of motion governing a time independent $A_t$ is given by:
\begin{equation}
\partial_z \left( \frac{\sigma \; g_{xx}}{\sqrt{g_{zz} g_{tt}}}  \partial_z A_t \right) + \sigma \sqrt{ \frac{g_{zz}}{g_{tt}} } (\partial_x^2 + \partial_y^2)A_t = {Q}\,\delta(z-z_p)\delta(x)\delta(y)~.
\end{equation}
We have included a time independent delta function source at $(r_p,0,0)$. Thus we are led to solve an ordinary differential equation of general form (going to Fourier space in the $(x,y)$-coordinates):
\begin{equation}
\partial_z \left( \alpha(z) \partial_z A_t \right) - \beta(z) k^2 A_t = {Q}\,\delta(z-z_p)~.
\end{equation}
The effect of the delta function comes in the boundary conditions between the $z<z_p$ and $z>z_p$ solutions. It is convenient to define $\gamma(z) \equiv \beta(z)/\alpha(z)$ and $\zeta(z) \equiv  \alpha(z) \beta(z)$. In terms of the original metric variables:
\begin{equation}
\gamma(z) = \frac{g_{zz}}{g_{xx}}~, \quad \quad \zeta(z) = \sigma^2 \; \frac{g_{xx}}{g_{tt}}~.
\end{equation}
We now propose a WKB approximation to solve the equations of $A_t$:
\begin{equation}
A_t (z) = \exp {\frac{1}{\lambda}\left[ W_0(z) + \lambda W_1(z) + \ldots \right]}~, \qquad \partial_z \to \lambda \partial_z \, ,  
\end{equation}
where $\lambda$ is a formal small parameter used to keep track of the expansion and then set to one (analogous to $\hbar$ in quantum mechanics). The equation obeyed by $W_0$ is given by:
\begin{equation}
(\partial_z W_0)^2  = k^2 \gamma(z)~,
\end{equation} 
from which it follows that $W_0(z) = \pm k \int dz \sqrt{\gamma(z)}$. It is convenient to define a ``flat'' coordinate 
\begin{equation}
   \upsilon \equiv \int dz \sqrt{\gamma(z)} \, ,
\end{equation}   
in terms of which $W_0 = \pm k \upsilon + {\rm constant}$. 
From this solution one can readily find that:
\begin{equation}
{W_1}(z) = -\frac{1}{4} \log \zeta(z)~. 
\end{equation}
Validity of the WKB approximation requires $W_0' \gg W_1'$, that is $k \sqrt{\gamma} \gg \zeta'/\zeta$. The general solution is the linear combination:
\begin{equation}\label{WKB}
A_t (z) = \zeta(z)^{-1/4} \left( c_k \;  e^{k \upsilon(z)} + d_k \;  e^{-k \upsilon(z)} \right)~.
\end{equation}
We denote the coefficients of the solution in the $z>z_p$ region by $\left(c^{(-)}_k,d^{(-)}_k \right)$ and those in the $z<z_p$ region by $\left(c^{(+)}_k,d^{(+)}_k \right)$. Similarly $\upsilon^{(+)}(z) = \int_{z_b}^z dz \sqrt{\gamma(z)}$ and $\upsilon^{(-)}(z) = \int_z^{z_h} dz\sqrt{\gamma(z)}$. Notice that $\upsilon^{(-)}(z) + \upsilon^{(+)}(z)= \upsilon^{(+)}(z_h) = \upsilon^{(-)}(z_b)$. Also note that $\upsilon^{(+)}(z)$ is monotonically increasing with increasing $z$.

\subsection*{Boundary conditions near AdS bondary}

To fully specify the solution we must impose appropriate boundary conditions. We now assume we are in an asymptotically AdS space and that the boundary lies at $z_b = 0$ and the horizon at $z=z_h \gg 1$. Naturally, the point charge lies in the interval $0<z_p<z_h$. 

For $z < z_p$ one requires that the solution is fast-falling near the AdS boundary. So our boundary condition at $z = 0$ leads to the following $z < z_p$ solution:
\begin{equation}
A^{(+)}_t = c^{(+)}_k \zeta(z)^{-1/4} \sinh  k \upsilon^{(+)}(z)~.
\end{equation}  

\subsection*{Matching at $z=z_p$}

We must also impose continuity at $z=z_p$. In addition to continuity, the delta function source imposes a condition on the first derivative of $A_t$ at $z=z_p$:
\begin{equation}
\lim_{\epsilon \to 0^+} \left( \partial_z A_t(z_p+\epsilon) - \partial_z A_t(z_p-\epsilon) \right) = \sqrt{\frac{\gamma(z_p)}{\zeta(z_p)}} \, Q 
~.
\end{equation}
The above conditions at $z_p$ fix the remaining coefficients to:
\begin{equation}\label{matching}
{d_k^{(-)}} = \frac{e^{k \upsilon^{(+)}_h}}{2}\left( c_k^{(+)} - \frac{Q \, e^{-k {\upsilon^{(+)}_p}} }{ k \, {\zeta_p}^{1/4}} \right), \quad 
{c_k^{(-)}} =  \frac{e^{-k \upsilon^{(+)}_h}}{2}\left( - c_k^{(+)} + \frac{Q \, e^{k {\upsilon^{(+)}_p}} }{ k \, {\zeta_p}^{1/4}} \right)~,
\end{equation}
where we have defined $\upsilon^{(+)}_p \equiv \upsilon^{(+)}(z_p)$ and $\zeta_p \equiv \zeta(z_p)$.

\subsection*{Boundary conditions near the horizon}

Let us assume that the metric very near the horizon is Rindler space, i.e. we are dealing with a non-extremal black brane. For our purposes here we may take the Rindler horizon to be at $z=1$, with $\sigma(z) \equiv \sigma_h$ constant, and
\begin{equation}
ds^2 = -(1-z) \, dt^2 + \frac{dz^2}{(1-z)} + dx^2 + dy^2~,
\end{equation}
where $z$ ranges from $-\infty$ to 1. For this geometry we have $\gamma(z) = 1/(z-1)$ and $\zeta(z) = \sigma_h/(1-z)$. Validity of the WKB approximation requires $k \sqrt{\gamma} \gg \zeta'/\zeta$, that is $k \gg 1/\sqrt{1-z}$. This is satisfied asymptotically for $z \to -\infty$ but breaks down when $z \to 1$, when the horizon is approached. Fortunately, we can obtain the analytic Fourier modes in the Rindler near horizon region and match to the WKB ansatz at the asymptotic boundary of the Rinder region, where it becomes reliable. This will allow us to obtain the proper boundary conditions on the WKB modes.\footnote{This is analogous to how in quantum mechanics the proper WKB boundary conditions at turning points are obtained from matching to the asymptotics of the exact solution of the Schr\"odinger equation in a linear potential.}

As a boundary condition at the horizon we impose that the gauge field vanishes at the horizon $z = 1$. This means that the black hole horizon is an equipotential. In the Rindler region we can solve the equation for $A_t$ exactly and find:
\begin{equation}
A^{(h)}_t(z) \propto \sqrt{1-z} \; I_1 \left(2 k \sqrt{1-z} \right)~,
\end{equation}
where $I_\nu(z)$ is the modified Bessel function of the first kind. For $z \to - \infty$ we can expand our solution and find: 
\begin{equation}
A^{(h)}_t(z) \propto ({1-z})^{1/4} \; \sinh \left( 2 k \sqrt{1-z} \right) 
\end{equation}
The above takes a WKB form which fixes the coefficients of the $z>z_p$ WKB solution (\ref{WKB}) to:
\begin{equation}\label{rmatch}
c^{(-)}_k =- d^{(-)}_k~.
\end{equation}
We can now combine (\ref{matching}) with (\ref{rmatch}) to solve for $c^{(+)}_k$:
\begin{equation}
c^{(+)}_k =  \frac{Q}{\zeta_p^{1/4} k} \; \frac{\sinh  k \left(\upsilon^{(+)}_h - \upsilon^{(+)}_p \right) }{\sinh k \upsilon^{(+)}_h}~.
\end{equation}

\subsection*{Boundary CFT charge density}

We would like to obtain the holographic charge density corresponding to the probe field. In general this is given by $j^t = \sigma \partial_z A_z|_{z=0}$. In Fourier space, this is given by:\footnote{Here and in what follows we drop the $(+)$ index from $\upsilon$.}
\begin{equation}
j^t(k) =  \sigma_b \, \partial_z A^{(+)}_z(0)  = \frac{Q \sigma_b^{1/2}}{\zeta_p^{1/4}} \; \frac{\sinh  k \left(\upsilon_h - \upsilon_p \right) }{\sinh k \upsilon_h}~.
\end{equation}
Consistent with the WKB approximation scheme, we ignored the contribution from the $z$-dependence of the normalization factor, since this is a higher order term in the WKB expansion, which we have been neglecting. 

To obtain its expression in the $(x,y)$-coordinates, we must Fourier transform $j^t(k)$:
\begin{equation}
j^t(r) = \frac{1}{(2\pi)^2} \int_0^{2\pi} d\theta \int_0^\infty dk \;  k \; e^{-i k r \cos\theta} j^t(k)~.
\end{equation}
This integral can be performed by series expanding $j^t(k)$ in powers of $e^{-k \upsilon_h}$, integrating over $k$ and then over $\theta$:
\begin{equation} \label{jseries}
 j^t = \frac{Q}{2\pi} \, \eta_p \, \sum_{n=0}^\infty 
\left( \frac{\upsilon_p+2 n \upsilon_h}{(r^2+(\upsilon_p+2 n \upsilon_h)^2)^{3/2}}
 - \frac{2 (n+1) \upsilon_h-\upsilon_p}{(r^2+(2 (n+1) \upsilon_h-\upsilon_p)^2)^{3/2}} \right) \, ,
\end{equation}
where
\begin{equation}
 \eta_p = \frac{\sigma_b^{1/2}}{\zeta_p^{1/4}} =  \sqrt{\frac{\sigma_b}{\sigma_p}} \left.\frac{g_{tt}^{1/4}}{g_{xx}^{1/4}}\right|_{z_p} \, .
\end{equation}
In the large $v_h$ limit, the $n=0$ term in (\ref{jseries}) dominates and we can write
\begin{equation} \label{approxjt}
 j^t(r) = \frac{Q}{2 \pi} \, \eta_p \left( 
 \frac{\upsilon_p}{(r^2+\upsilon_p^2)^{3/2}}
 - \frac{2 \upsilon_h-\upsilon_p}{(r^2+(2 \upsilon_h-\upsilon_p)^2)^{3/2}} \right) \, .
\end{equation}
Notice that the second term can be thought of as due to an image charge behind the horizon, or equivalently due to an induced charge on the horizon, mimicking an image charge behind the horizon. As $z_p \to z_h$ the charge and its image cancel each other and the profile goes to zero. As we saw in the pure AdS case (cf.\ eq.\ (\ref{electricpot})), the first term can furthermore be thought of as being due to the combination of a charge and an image charge reflected across the AdS boundary. The higher $n$ corrections in (\ref{jseries}) can be interpreted as contributions from further image charges, obtained by sequences of mirroring across the two conducting boundary surfaces $z=0$ and $z=z_h$. These are suppressed by inverse powers of $\upsilon_h^{(+)}$, the ``optical distance'' between the boundary of AdS and the horizon.

As a check, notice that for $\sigma_b=\sigma_p=\frac{1}{4 g^2}$, $g_{tt}=g_{zz}=g_{xx}=1/z^2$, $z_h=\infty$, we recover the Poincar\'e AdS result (\ref{AdS4jt}).

\end{document}